\documentclass[aps,pre,reprint,twocolumn]{revtex4-2}
\usepackage{graphicx}
\usepackage{dcolumn}
\usepackage{bm}
\usepackage{graphicx}
\usepackage{dcolumn}
\usepackage{bm}
\usepackage{dcolumn}
\usepackage{amsmath}
\usepackage{graphicx}
\usepackage{dcolumn}
\usepackage{bm}
\usepackage{easyReview} 
\usepackage{multirow}
\usepackage{xcolor}
\newcommand{\ro}  { {\bf r}}
\usepackage[caption=false]{subfig}
\usepackage[normalem]{ulem}
\usepackage[range-units=single,per-mode=reciprocal]{siunitx}
\begin{document}
	\title{Thermodynamic Insights into Polyelectrolyte Complexation: A Theoretical Framework}
	\author{Souradeep Ghosh}
	\affiliation{Department of Physical Sciences and Centre for Advanced Functional Materials, Indian Institute of Science Education and Research
		Kolkata, Mohanpur 741246, India}
		
	\begin{abstract}
		In this study, we propose a theoretical framework to investigate the interactions between flexible polymer chains, specifically polyelectrolytes (PEs). By calculating the system's free energy while considering position-dependent mutual interactions and chain conformations, we gain insights into the local dielectricity as PEs overlap. Our analysis reveals that the thermodynamic drive for complex coacervation is influenced by factors such as the number of ions bound to the polymer backbone and the entropy associated with free ions, challenging earlier assumptions about the relationship between entropy gain and electrostatic temperature. We demonstrate that global thermodynamic behavior is strongly influenced by local factors like dielectric constant, providing clarity on discrepancies between experimental and computational studies. Additionally, we found that entropy gain is inversely proportional to the local dielectric constant, assuming a constant electrostatic temperature. Our findings highlight the importance of considering polymer-specific parameters when exploring the thermodynamic behavior of charged polymer complexation.
	\end{abstract}
	
	\maketitle
	\section{Introduction}
	There has recently been an increased interest in studying complexation between oppositely charged polyelectrolytes, due to their potential applications as molecular cages and controlled delivery agents. Pioneering experimental works have been carried out to explore and develop these proposed uses for polyelectrolyte complexes and coacervates, with a focus on the complexation of oppositely charged polyelectrolytes as the first step of coacervate formation.
	
	The foundational theory of polyelectrolyte coacervation, put forth by Voorn and Overbeek,\cite{voorn1957} begins with a physical picture where the chain connectivity correlation is not explicitly included and treats the backbone charges as disconnected free ions. However, this theory is not able to fully account for the electrostatic correlations present in flexible, deformable polyelectrolytes, which continues to attract considerable theoretical\cite{borue1988,borue1990,mahdi2000,joanny2001,delacruz2003,delacruz2004,kudlay2004,oskolkov2007,perry2015,salehi2016,sing2017,lytle2017,adhikari2018,potemkin2017, panyukov2018,lytle2019,zhang2018,ylitalo2021,wang-zhao2019,rumyantsev2018,rumyantsev2019,chen-yang2021,knoerdel2021,sayko2021,mitra2023,Ghosh2023,chen2022-PNAS}, experimental\cite{record1978,kabanov1985,dautzenberg2002,cousin2005,gummel2007,porcel2009,spruijt2010macro,chollakup2010,gucht2011,tirrell2012,lemmers2012,gucht2012,tirrell2013,tirrell2014,		vitorazi2014,perry2014,salehi2015,lutkenhaus2015,kayitmazer2015,fu2016,meka2017,schlenoff2017,ali2018,dePablo2018,tirrell2018,		ali2019,wang2019,huang2019,pde2020,tirrell2020,meng2020,tirrell2020,tirrell2021,chen2021,neitzel2021,priftis2012,priftis2012-softmatter,subbotin2021,friedowitz2021,ma2021,lalwani2021,digby2022}, and simulational \cite{hayashi2002,winkler2002,hayashi2003,hayashi2004,zhang2005,zhaoyang2006,fredrickson2007,larson2009,semenov2012,fredrickson2012,juarez2015,		peng2015,lytle2016,dzubiella2016,fredrickson2017,chang2017,radhakrishna2017,whitmer2018,whitmer2018macro,rumyantsev2019macro,shakya2020,neitzel2021,		sayko2021,bobbili2022,chen2022,chen2022-PNAS}. The thermodynamic drive of the complex formation process has been identified as the entropy gain from the released counterions\cite{michaels1965,tianaka1980,dautzenberg2002,zhaoyang2006,gummel2007,larson2009,beltran2012,lemmers2012,tirrell2012,gucht2012,		semenov2012,perry2015,peng2015,dzubiella2016,salehi2016,fu2016,muthu2017,chang2017,radhakrishna2017,lytle2017,schlenoff2017,meka2017,tirrell2018,adhikari2018,whitmer2018,		whitmer2018macro,lytle2019,wang2019,tirrell2020,tirrell2021,mitra2023,Ghosh2023}, as observed in numerous experiments\cite{michaels1965,tianaka1980,dautzenberg2002,gummel2007,beltran2012,lemmers2012,gucht2012,tirrell2012,tirrell2013,tirrell2014,fu2016,meka2017,schlenoff2017,tirrell2018,wang2019,tirrell2020,tirrell2021}, simulations\cite{zhaoyang2006,larson2009,semenov2012,peng2015,dzubiella2016,chang2017,radhakrishna2017,whitmer2018,whitmer2018macro}, and theories\cite{perry2015,salehi2016,muthu2017,lytle2017,adhikari2018,lytle2019,mitra2023,Ghosh2023}.
	
	A key challenge in the theoretical study of polyelectrolytes is the proper description of electrostatic correlations and their consequences on the conformation and thermodynamics. The physical origin of electrostatic correlation is the preferential interaction between opposite charges, and the long-range nature of these interactions leads to nontrivial correlation effects in polyelectrolyte systems. Here, we attempt to elucidate a clearer picture via analytical and numerical techniques by examining the equilibrium complexation behavior of oppositely charged flexible polyelectrolyte in solutions. Emphasizing the essential difference between the two-body interaction of soft, interpenetrating macromolecules from that of a pair of rigid molecules. 
	Although mean-field theories and coarse-grained simulations have shown that complex formation between oppositely charged polymers can be either endothermic or exothermic,\cite{zhaoyang2006,whitmer2018macro,mitra2023,Ghosh2023} there remains a significant divergence from experimental interpretations of enthalpy data.\cite{priftis2012,fu2016,schlenoff2017,wang2019,tirrell2020,tirrell2021} A common explanation for this discrepancy points to the inability of mean-field theories to account for the polar solvent reorganization entropy that occurs before and after complexation.\cite{priftis2012,chen2022-PNAS,Muthukumar2023-book} In this context, we propose a new perspective that focuses on the role of local dielectricity. Our approach is to taking into account the molecular nature of the complex via a variational calculation of the free energy, as a function of the center-of-mass separation between the two chains, which systematically captures the coupling between the position-dependent mutual interaction of the individual chains and their consequences on the conformation and thermodynamics. Consequently, it provides insight into the spatial variation of local dielectricity as the two polyelectrolytes undergo mutual overlap. 
	
	The rest of this article is organized as follows. In Section II  we introduce a model to study the equilibrium behavior of the complexes of oppositely charged polyelectrolytes, we present a full derivation of the interaction of oppositely charged polyelectrolyte pairs to study flexible charged chains, and demonstrate how the self-consistent procedure and single chain averages can be approximated with a variational procedure. In Section III we present results and discuss the effective interaction, polyelectrolyte self-energy,  and thermodynamic drive of the binding process. We further discuss the importance of local dielectric environment in determining the total energy and thermodynamics. Finally, in Section IV we conclude with a summary of the key results and future outlook.
	
	\section{Model}
	All charges are monovalent, and the PA and PC consist of $N_1$ and $N_2$ monomers, respectively. $N_{c1}$ and $N_{c2}$ ionizable monomers are uniformly distributed along the contour of the polymer chains. During complexation, a single pair of ionizable monomers from either chain forms the intermediate complex, resulting in one neutral monomer. At any given instant, $n$ ionizable monomers from PC form the intermediate complex, along with an equal number of ionizable monomers from PA. The remaining ionizable monomers in the uncomplexed parts of PA and PC are $N_{c1}-n$ and $N_{c2}-n$, respectively. The maximum degree of ionization for PA and PC are defined as $f_{m1}=N_{c1}/N_1$ and $f_{m2}=N_{c2}/N_2$, respectively.
	
	The partition sum of the system can be denoted by $Z$ and is written as $Z=Z_1 Z_2 Z_3$, where $Z_1$ represents the combinatorial factor arising from randomly choosing $M$ repeat units out of $N$ units for adsorbing counterions, $Z_2$ represents the combinatorial factor associated with the indistinguishability of the uncondensed counterions and dissociated salt ions ($N_{i}-M_{i}+n_{+/-}$), and $Z_3$ represents all other contributions arising from chain conformations and distributions of all species subject to the various potential interactions in the system.
	
	To determine the entropic contribution arising from the various distributions of the adsorbed counterions and coions, we note that there are $N_i$ monomers and $M_i$ adsorbed monovalent counterions. In terms of the progressive monomer overlapping units, the partition function is given by:
	
	\begin{align}
		\label{z1}
		Z_1=\left[\prod_{i=1}^2 \frac{\left(N_i-n_i\right) !}{\left(N_i-n_i-M_i\right) ! M_{i} !}\right]
	\end{align}
	
	where $n_1$ is the number of PA monomers complexed with PC and $n_2$ is the number of PC monomers complexed with PA, with $n_1=n_2=n$.
	
	To determine the translational entropy of the unadsorbed ions which are distributed in the bulk volume $\Omega$, we count as mobile ions: $N_1-M_1+n_{-}$ monovalent counterions from poly-cation molecule and $N_2-M_2+n_{+}$ monovalent counterions from polyanion molecule. Therefore, the partition function related to the translational entropy in volume $\Omega$ is:
	
	\begin{widetext}
		\begin{align}
			\label{z2}
			Z_2 &=\left[\frac{\left(\Omega / \ell^3\right) \sum_{i=1}^2\left(N_i-n_i-M_i+n_i\right)+n_{+}+n_{-}}{\left(N_1-n_1-M_1+n_1+n_{+}\right) !\left(N_2-n_2-M_2+n_2+n_{-}\right) !}\right].
		\end{align}

	The free energy of the system is given by $F_0=-k_B T \ln( Z_{1} Z_{2})$, where $k_B$ is the Boltzmann constant and $T$ is the temperature.
	
In reality, the process of adsorption of counterions on the polymer is cooperative and not random. However, in the present model, the cooperativity of counterion adsorption is not included. The chain connectivity, self and inter-chain interactions are written in the Edwards path integral representation, where the contour length of a chain is $L=N \ell$. By integrating over the positions of counterions, salt ions, and solvent molecules, the Helmholtz free energy, $F$, of the system is obtained as
\begin{align}
	\label{partition-sum}
	e^{-\beta F}&=Z_1 Z_2 \exp \left(\frac{\Omega \kappa^3}{12 \pi}\right) \int \prod_{i} \mathcal{D} {\bf R}(s_{i})\exp \left[-\beta \left( H+U_{a d}\right)\right].
\end{align}
where $\beta=1/k_{B}T$ and the integral $\mathcal{D} {\bf R}(s_{i})$ is a conformational integral, $ {\bf R}(s_{i})$ is the position vector of the $ i $-th chain at the arc length variable $s_{i}$ where $s_{i}(0\le s_{i}\le \mathcal{N}_i)$. The inverse Debye length $\kappa$ is defined as
\begin{align}\label{kappa}
	\tilde{\kappa}^{2}=&4 \pi  \tilde{\ell}_{B} \sum_{p=+,-} z_{p}^{2} n_{p} / \Omega
	=4 \pi \tilde{\ell}_{B} \left[\sum_{i=1}^{2}\widetilde{\rho}_{i}\left(f_{m i}-\frac{n}{N_{i}}\right)\left(1-\alpha_{i}\right)+\frac{\widetilde{\rho}_{i} n}{N_{i}}+\widetilde{c}_{s}\right].
\end{align}
The dimensionless parameters are defined as follows: $\widetilde{\rho}_{i}=N_{i}/(\Omega/\ell^3)$, which represents the monomer densities of the respective chains in the solution ($i=1,2$), and $\widetilde{c}_{s}=c_s \ell^3$, where $\ell$ represents the size of a monomer as well as a counterion (from either chain), and is also the Kuhn length for these flexible polyelectrolytes. Additionally, $\widetilde{\kappa}=\kappa \ell$, $n_p$ and $z_p$ represent the number and valency of the dissociated ions of the $p^{\text{th}}$ species, respectively.
\end{widetext}

The energy of the three types of ion-pairs, negatively charged monomer-counter-cation, positively charged monomer-counter-anion, and oppositely charged monomers, is of the Coulombic form $-e^2 / 4 \pi \epsilon_0 \epsilon_l d$, where $d$ is the dipole length of the pair and $\epsilon_l$ the local dielectric constant in the vicinity of the ion-pairs. For PA and PC, respectively, we get the degree of ionization of the dangling parts of the chains as (considering $M_1$ counter-cations and $M_2$ counter-anions remain condensed)
\begin{equation}
	f_1 = \frac{N_1 f_{m1} - M_1 - n}{N_1 - (n/f_{m1})}, \qquad f_2 = \frac{N_2 f_{m2} - M_2 - n}{N_2 - (n/f_{m2})},
	\label{doidef}
\end{equation}
where $n_i=n/f_{mi}$ is the total number of monomers from chain $i$ which form part of the intermediate complex. Summing over all ion-pairs of three types for the complex and the dangling chains, the electrostatic free energy of the ion-pairs is obtained as
	\begin{align}
	\frac{U_{ad}(\epsilon_{local})}{k_B T} &=-\sum_i N_i\left\{\left(f_{m i}-\frac{n}{N_i}\right)-\left(1-\frac{n_i}{N_i}\right) f_i\right\} \frac{\tilde{\ell}_B}{\tilde{d}} \delta \nonumber\\
	&-n \widetilde{\ell}_B \delta
	\label{uad}
\end{align}
where the dielectric mismatch parameter, $\delta(\equiv\epsilon_{bulk}/\epsilon_{local})$, represents the disparity between the local dielectric constant, $\epsilon_{local}$, and the bulk dielectric constant, $\epsilon_{bulk}$. The bound pair energy is proportional to the product of $\delta$ and the Bjerrum length, $\widetilde{\ell}_B/\tilde{d}$, which is considered the electrostatic or Coulomb strength of the system. In the equation \eqref{uad}, the term in the sum is represented by $-M_i \widetilde{\ell}_B \delta/\tilde{d}$.

	Additionally, we note that the Coulomb energy for bound monomer pairs can be represented by \(\delta_{12}\widetilde{\ell}_{B}\). We can rewrite the Coulomb term as follows:
	\begin{align}
		\label{gamma-def}
		\Gamma = \delta\widetilde{\ell}_{B} = \frac{\epsilon \ell}{\epsilon_{local} d} \frac{e^2}{4 \pi \epsilon_0 \ell \epsilon k_B T} = \frac{e^2}{4 \pi \epsilon_0 \epsilon_{l12} d k_B T},
	\end{align}
	where \(\epsilon_{local}\) represents the local dielectric constant, and the dipole length is given by \(d\). This formulation indicates that the effective Coulomb strength can be modulated by changing the local dielectric constant. The Coulomb strength of chemically symmetrical poly-ions with opposite charges can be calculated assuming their estimated size is $\ell\sim 0.25$nm, and the dimensionless Bjerrum length at room temperature is constant at $\widetilde{\ell}_{B}=3$.
	
\begin{figure}
		\centering
		\fbox{\includegraphics[width=0.99\linewidth]{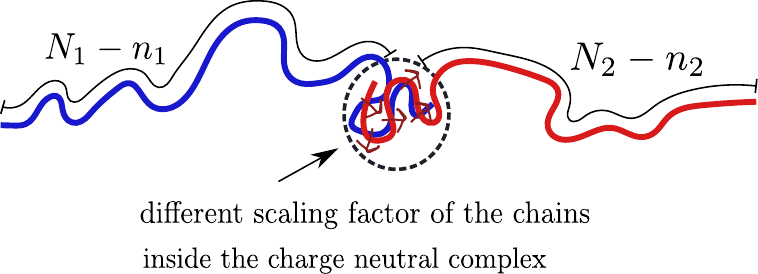}}
		\caption{A schematic representation of partially charged asymmetric polyelectrolyte (PE) chains complexing: In the formation of a polyelectrolyte complex (PEC) between oppositely charged polyelectrolyte (PE) chains, the chains overlap and slide along each other as their monomers form bound ion pairs, releasing their counterions in a cooperative manner. The remaining, ``dangling'' portions of the chains retain partial compensation by counterions. It is important to note that the complexed and dangling portions of each chain will have different scaling factors relative to one another.\cite{mitra2023,Ghosh2023} This implies that, in order to fully understand the behavior of the PEC, one must consider the distinct properties of not only the two complexed chains, but also the two dangling portions of the original PE chains.}\label{schematic}
	\end{figure}

	The interaction energy between monomers within a polyion and between different polyions is modeled using interaction potentials, resulting in a Hamiltonian, $H$. At an intermediate state of complexation, both polyions (PA and PC) will have two distinct parts: one that is dangling and one that is complexed with the other, oppositely charged polyion. As the size scaling and monomer distributions in these two parts are generally different, they are treated as separate chains to calculate interactions and apply the Gaussian trial Hamiltonian. In this way, we consider four polymer chains (indicated by indices 1 to 4), with chains 3 and 4 forming the complexed part of the original polyions. The free energy, $F_{pol}$, resulting from the Hamiltonian at a particular degree of overlap (at which $n_i$ molecules from chain $i$ form the complexed part of the chains), is a function of the charges ($f_i$) of the dangling parts of the polyions, the configurational distribution of monomers in all four chains (in terms of the size expansion factors $\widetilde{\ell}_{1i}$, $i$=1 to 4), as well as parameters such as chain lengths ($N_i$), salt concentration ($\widetilde{c}_s$), Bjerrum length ($\widetilde{\ell}_B$), and those related to excluded volume interactions. In terms of a general Hamiltonian, it will be given by 
	\begin{align}\label{Partition_function}
		e^{- \beta F_{pol}}= \prod_{i=1}^{4}\int \mathcal{D} {\bf R}(s_{i}) \exp(-\beta H),
	\end{align}
	where indices $i$=1, 2 stand for the dangling parts of the polyanion(PA) and polycation(PC), and $i$=3, 4 the complexed part, respectively (Fig. \ref{schematic}). The parameter $\mathcal{N}_i$ is the number of monomers in chains $i$=1 to 4, with $\mathcal{N}_i=N_i-n_i$ for the dangling parts (chains $i=1,2$), $\mathcal{N}_3=n_1$, and $\mathcal{N}_4=n_2$. Here, $n_i=0$ represents the uncomplexed polyions PA and PC. The right-hand side of Eq.\eqref{Partition_function} represents the conformal integral of the canonical partition function of the polyions, which are now considered to comprise four independent chains. It is worth noting that the degrees of freedom of the counterions and the Coulomb energy of counterion-monomer ion-pairs have already been accounted for in partition sum \eqref{partition-sum} using Eqs. \eqref{z1}, \eqref{z2}, and \eqref{uad}.

	The Hamiltonian, $H$, can be further broken down into two contributions: $H_{0}$ and $H_{int}$. $H_{0}$ arises from the connectivity of the monomers individually in each of the four chains, and is given by
	\begin{align}\label{H_0_appendix}
		{\beta H_{0}}&=\frac{3}{2 \ell^2}\sum_{i=1}^{4}\int_{0}^{\mathcal{N}_{i}} d s_{i}\left(\frac{\partial {\bf R}\left(s_{i}\right)}{\partial s_{i}}\right)^{2}
	\end{align}
	where $\ell$ is the Kuhn length. Note that the limits of integrals in the last two terms can be alternatively and equivalently written as 0 to $n_i$. The interaction term, $H_{int}$, in the Hamiltonian can be separated into two contributions: (i) $H_{ex}$, the non-electrostatic excluded volume interactions and (ii) $H_{el}$, electrostatic interactions between uncompensated charged monomers (which neither have their counterion nor another oppositely charged monomer adsorbed on them). The contribution from excluded volume interaction, with strength $w_{ij}$ (between a pair of monomers from the $i$-th and $j$-th chains, respectively), is given by:
\begin{align}\label{eq:H_ex}
	H_{ex}&=\sum_{i=1}^{4}\omega_{ii}\ell^{3} \int_{0}^{\mathcal{N}_{i}} ds_{i} \int_{0}^{\mathcal{N}_{i}} ds_{i}^\prime \delta( \mathbf{R}(s_{i})-\mathbf{R}(s_{i}^\prime))\\ \nonumber
	&+2\omega_{34}\ell^{3} \int_{0}^{\mathcal{N}_{3}} ds_{3} \int_{0}^{\mathcal{N}_{4}} ds_{4} \delta(\mathbf{R}(s_{3})-\mathbf{R}(s_{4})),
\end{align}
	where the arguments of the $\delta$-functions within the above integrals are the difference in contour vectors corresponding to the monomer pair involved in the excluded volume interaction. The energy contribution from the pairwise excluded volume interactions among monomers within individual PE molecules is given by the first four terms on the right-hand side of Eq.\eqref{eq:H_ex}, corresponding to interaction strengths $w_{11}, w_{22}, w_{33},$ and $w_{44}$. These terms will be sensitively dependent on the configurations of the individual chains. The energy due to the mutual excluded volume interaction between the two polyions is given by the fifth term, corresponding to the interaction strength $w_{34}$. The other contributions corresponding to $w_{12}, w_{13}, w_{14}, w_{23}$, and $w_{24}$ are neglected, as those pairs of chains in our four-chain model are separated from each other and do not have any overlapping monomers (Fig. \ref{schematic}).
	It is well known that the Coulomb interaction between any two charges separated by a distance $ \mathbf{r-r^{\prime}} $ in an electrically neutral system is screened by the presence of other charges to the familiar Debye–Hu\"ckel form,
	\begin{align}\label{Yukawa-Uij}
		U_{ij}^{el}\left(\mathbf{r},\mathbf{r}^{\prime}\right)=\frac{f_{i}f_{j}\ell_{B}}{2} \frac{\exp \left[-\kappa \left|\mathbf{r}-\mathbf{r}^{\prime}\right|\right]}{\left|\mathbf{r}-\mathbf{r}^{\prime}\right|},
	\end{align}
	where $ f_{i},f_{j} $ given by Eq.\eqref{doidef} and $\kappa$ is the inverse Debye length which is dependent on the charge density of the system as described in Eq.\eqref{kappa}.
	
	The electrostatic interaction between intra- and inter-chain segments is given by,
	\begin{align}\label{H_el}
		\beta H_{el}=&\int_{0}^{\mathcal{N}_{1}} ds_{1} \int_{0}^{\mathcal{N}_{1}} ds_{1}^\prime   U_{11}^{el}({\bf R}(s_1),{\bf R}(s_1^\prime))\nonumber\\
		&+\int_{0}^{\mathcal{N}_{2}} ds_{2} \int_{0}^{\mathcal{N}_{2}} ds_{2}^\prime  U_{12}^{el}({\bf R}(s_2),{\bf R}(s_2^\prime) )\nonumber\\
		&+2 \int_{0}^{\mathcal{N}_{1}} ds_{1} \int_{0}^{\mathcal{N}_{2}} ds_{2}   U_{12}^{el}({\bf R}(s_1),{\bf R}(s_2)),
	\end{align}
	where $ (N_{i}-n_{i}) $ is the number of uncomplexed monomers from the dangling parts of chain $i$ and $ f_{i} $ is the degree of ionization of the dangling part of chain $i$ given by Eq. \ref{doidef}. It is worth noting that chains 3 and 4 (as depicted in Figure \ref{schematic}) are assumed to be uncharged, as it is assumed that the charged monomers within these chains have formed ion pairs as part of the complex. As previously stated, the complexed portion, composed of monomer-monomer ion pairs, is assumed to be charge neutral ($ f_{m1}n_{1}=f_{m2}n_{2} $) and does not affect the electrostatic interactions in the uncomplexed portions of the polyions. The complex has $ n $ bound monomer pairs, resulting in $ n_{i}(=n/f_{mi}) $ total monomers from polyion $i$.
It is quite challenging to directly evaluate the partition sum in Eq. \ref{Partition_function} using the Hamiltonian discussed earlier (Eqs. \ref{H_0_appendix}, \ref{eq:H_ex}, and \ref{H_el}). A common approach is to introduce a trial Hamiltonian, such as the one proposed by Edwards \cite{edwards1979}, which takes the form of:
\begin{align}\label{trial-Hamiltonian}
	\beta H_{trial}=\sum_{i=1}^{4}\frac{3}{2 \ell\ell_{1i}}\int_{0}^{\mathcal{N}_{i}} d s_{i}\left(\frac{\partial {\bf R}\left(s_{i}\right)}{\partial s_{i}}\right)^{2}.
\end{align}
Here, $\ell_{1i}$ is a variational parameter that gives the effective expansion factor of the polyion relative to its Gaussian size. The total Hamiltonian can then be separated into the trial Hamiltonian and the deviation from it, as given by:
\begin{align}\label{variational-Hamiltonian}
	H=& H_{trial}+\left(H-H_{trial}\right) \nonumber\\
	=& H_{trial}+\delta H.
\end{align}
By minimizing the free energy with respect to $\ell_{1i}$, we can obtain the optimal value of $\ell_{1i}$. Hence, the free energy of the chain ($F_5$) given in Eq. \eqref{Partition_function} can be rewritten as (by adding and subtracting the term $ H_{trial}$ to the original Hamiltonian in Eq. \eqref{Partition_function}, as in Eq.\eqref{variational-Hamiltonian})
\begin{align}\label{step-1-variational}
	&e^{-\beta F_{pol}}= \prod_{i=1}^{4}\int D{\bf R}(s_{i}) \exp(-\beta H_{trial}-\beta H+\beta H_{trial})\\ \nonumber
	&=\prod_{i=1}^{4}\int D{\bf R}(s_{i}) \exp(-\beta H_{trial}-X-\beta H_{ex}-\beta H_{el}),
\end{align}
where
\begin{align}\label{X}
X&=\frac{3}{2 \ell} \sum_{i=1}^{4} \int_{0}^{\mathcal{N}_{i}} d s_{i} \left(\frac{1}{\ell} -\frac{1}{ \ell_{1i}}\right) \left(\frac{\partial {\bf R}\left(s_{i}\right)}{\partial s_{i}}\right)^{2},
\end{align}
and $ H_{ex}$ and $H_{el} $ are given by Eq. \eqref{eq:H_ex} and Eq. \eqref{H_el}, respectively. We note that the function of the form $ e^{- x} $ in Eq. \eqref{step-1-variational} is being averaged over the configurations of the chains which are assumed to have Gaussian distributions, albeit with the expansion factors $\ell_{1i}$, given by the trial Hamiltonian (Eq. \eqref{trial-Hamiltonian}). We recall the Gibbs-Bogoliubov inequality, which states that for any function $e^{-x}$ averaged over some distribution one may write
\begin{align}\label{Jensen-Peierls-inequality}
	\langle e^{-x}\rangle_{0} \ge  e^{-\langle x\rangle_{0}},
\end{align}
where the angular brackets with subscript zero indicate the averaging over the specific probability distribution. Using this we obtain from Eq. \eqref{step-1-variational},
\begin{align}\label{avg-inequality}
	\langle e^{-X-\beta H_{ex}-\beta H_{el}} \rangle_{0}\ge e^{-\langle X \rangle_{0}-\langle \beta H_{ex} \rangle_{0}-\langle \beta H_{el} \rangle_{0}},
\end{align}
where for the polyions the distribution function is over the segment positions ${\bf R} (s_i)$, and is given by $\exp(-\beta H_{trial} [{\bf R} (s_i)])$ (Eq. \eqref{trial-Hamiltonian}).
\begin{widetext}
By using Eq. \eqref{avg-inequality} and Eq. \eqref{step-1-variational} the resulting expression can be obtained as
\begin{align}\label{step-2}
	e^{-\beta F_{pol}}&\ge \exp{(-\langle X \rangle_{0}-\langle \beta H_{ex} \rangle_{0}-\langle \beta H_{el} \rangle_{0})} \prod_{i=1}^{4}\int D {\bf R}(s_{i}) \exp(-\beta H_{trial})\equiv e^{-\beta\widetilde{F}_{pol}},
\end{align}
The average is defined as,
	\begin{equation}\label{avgexample}
		\langle g [\mathbf{R} (s_1),\mathbf{R} (s_2)] \rangle_0 = \frac{\prod_{i=1}^{2}\int \mathcal{D}\mathbf{R}(s_{i}) g [\mathbf{R} (s_1),\mathbf{R} (s_2)] e^{-\beta H_{trial}}}
		{\prod_{i=1}^{2}\int \mathcal{D}\mathbf{R}(s_{i}) e^{-\beta H_{trial} }}.
	\end{equation}
Since, $ H_{trial} $ has a general Gaussian form, these averages (given in Eq. \eqref{step-2}) can be calculated explicitly. First of all, we have, for a general function $ g[\mathbf{R}(s_{i}),\mathbf{R}(s^{\prime}_{i})] $, we write, using \eqref{avgexample}
	\begin{align}\label{avgexample-FT}
	\langle g[\mathbf{R}(s_{i}),\mathbf{R}(s^{\prime}_{i})]\rangle_{0}&=\left \langle \int ds \int ds^\prime \int \frac{d^3 k}{(2\pi)^3}g(\mathbf{k}) e^{-i\mathbf{k}\cdot \left|\mathbf{R}(s_{i})-\mathbf{R}(s^{\prime}_{i})\right|}\right\rangle_{0}= \int ds \int ds^\prime \int \frac{d^3 k}{(2\pi)^3}g(\mathbf{k})\left \langle e^{-i \mathbf{k}\cdot \left|\mathbf{R}(s_{i})-\mathbf{R}(s^{\prime}_{i})\right|}\right\rangle_{0}
\end{align}
The details of the average calculation is shown in the appendix \ref{structure-factor}. Using the result of Eq. \eqref{Exp-avg} derived using the definitions of Eqs.\eqref{avgexample-FT},\eqref{avgexample} it follows that, in general,
	\begin{align}\label{avg-of-delta-H}
		&\langle g[\mathbf{R}(s_{i}),\mathbf{R}(s^{\prime}_{i})]\rangle_{0}=\int ds \int ds^\prime \int \frac{d^3 k}{(2\pi)^3}g(\mathbf{k})~\exp \left[-k^2 \int_{-\infty}^{\infty} \frac{d q}{2 \pi} g(q) \sin ^2\left(\frac{q\left|s-s^{\prime}\right|}{2}\right) \right],
	\end{align}
where $ g(\mathbf{k}) $ is the Fourier conjugate of $ g[\mathbf{R}(s_{i}),\mathbf{R}(s^{\prime}_{i})] $ (See appendix \ref{structure-factor} for details). 
\end{widetext}
Hence, the free energy can be obtained by using the Eq. \eqref{step-2} as
\begin{align}
	\widetilde{F}_{pol}=\left\langle\beta\left(H_0-H_{trial}\right)\right\rangle_{0}+\left\langle\beta H_{ex}\right\rangle_{0}+\left\langle\beta H_{el}\right\rangle_{0}.
	\label{F5-avg}
\end{align}
where the average of Eqs. \eqref{eq:H_ex} and \eqref{H_el} can be obtained using Eqs.\eqref{avgexample},\eqref{avgexample-FT}, \eqref{Exp-avg} and \eqref{avg-of-delta-H}, and can be written as follows 
		\begin{align}\label{H_ex_avg}
		&\left\langle\beta H_{ex}\right\rangle_{0}=\nonumber\\
		&\sum_{i=1}^{4}\omega_{ii}\ell^{3} \int_{0}^{\mathcal{N}_{i}} ds_{i} \int_{0}^{\mathcal{N}_{i}} ds_{i}^\prime \int \frac{d^3 k}{(2\pi)^3} \langle e^{-i \mathbf{k}\cdot \left| \mathbf{R}(s_{i})-\mathbf{R}(s_{i}^\prime)\right| } \rangle_{0}\nonumber\\
		&+2\omega_{34}\ell^{3} \int_{0}^{\mathcal{N}_{3}} ds_{3} \int_{0}^{\mathcal{N}_{4}} ds_{4} \int \frac{d^3 k}{(2\pi)^3} \langle e^{-i \mathbf{k}\cdot \left|\mathbf{R}(s_{3})-\mathbf{R}(s_{4})\right|}\rangle_{0},
	\end{align}
	and
	\begin{align}\label{H_el_avg}
		&\left\langle\beta H_{el}\right\rangle_{0}=\nonumber\\
		&\int_{0}^{\mathcal{N}_{1}} ds_{1} \int_{0}^{\mathcal{N}_{1}} ds_{1}^\prime \int \frac{d^3 k}{(2\pi)^3} \hat{U}_{11}^{el}(\mathbf{k}) \langle e^{-i \mathbf{k}\cdot \left|\mathbf{R}(s_{1})-\mathbf{R}(s_{1}^\prime)\right| } \rangle_{0}\nonumber\\
		&+\int_{0}^{\mathcal{N}_{2}} ds_{2} \int_{0}^{\mathcal{N}_{2}} ds_{2}^\prime \int \frac{d^3 k}{(2\pi)^3} \hat{U}_{22}^{el}(\mathbf{k}) \langle e^{-i \mathbf{k}\cdot \left|\mathbf{R}(s_{2})-\mathbf{R}(s_{2}^\prime)\right|}\rangle_{0}\nonumber\\
		&+2\int_{0}^{\mathcal{N}_{1}} ds_{1} \int_{0}^{\mathcal{N}_{2}} ds_{2} \int \frac{d^3 k}{(2\pi)^3} \hat{U}_{12}^{el}(\mathbf{k}) \langle e^{-i\mathbf{k}\cdot \left|\mathbf{R}(s_{1})-\mathbf{R}(s_{2})\right|}\rangle_{0}.
	\end{align}

Note that $\widetilde{F}_{5}$, in Eq.\eqref{F5-avg}, is the extrema of the free energy $ F_{5}$. Eventually, the optimal values of the parameters $ f_{1} $,$ f_{2} ,\ell_{11}, \text{ and }\ell_{12} $  are obtained from the conditions
	\begin{align}
		\frac{\partial \widetilde{F}_{pol}}{\partial \ell_{11}}=\frac{\partial \widetilde{F}_{pol}}{\partial \ell_{12}}=\frac{\partial \widetilde{F}_{pol}}{\partial f_{1}}=\frac{\partial \widetilde{F}_{pol}}{\partial f_{2}}=0.
	\end{align}
	Using the optimal values of $\ell_{1 i}$, the dimensionless mean square end-to-end distance is defined as, $\left\langle \widetilde{R}_{i}^{2}\right\rangle \equiv R_i^2/\ell^2=\mathcal{N}_{i} \ell_{1 i} \ell/\ell^2 \equiv \mathcal{N}_{i} \widetilde{\ell}_{1 i}$. With the number of monomers in the $i$-th chain being $ \mathcal{N}_{i} $, within the approximation of uniform expansion of the polyelectrolyte chains, the average dimensionless radii of gyration of the chains is defined as $ \widetilde{R}_{gi}^{2}={\mathcal{N}_{i} \widetilde{\ell}_{1 i}}/{6} $. The total free energy has a closed form as follows,
		\begin{align}\label{F5}
			&\beta F_{pol}=\nonumber\\
			&\sum_{i=1}^{4}\frac{3}{2}\left(\widetilde{\ell}_{1i}-1-\log \widetilde{\ell}_{1i}\right)+\sum_{i=1}^{2}2\sqrt{\frac{6}{\pi}}{f_{1}}^{2} \widetilde{\ell}_{B} \frac{\mathcal{N}_{i}^{3/2}}{\sqrt{\ell_{1i}}}  \Theta_{s}\left(\widetilde{\kappa},a_{i}\right)\nonumber\\
			&+\frac{4}{3}\left(\frac{3}{2\pi}\right)^{3/2} \left[\frac{w_{ii}{(N_{i}-n_{i})}^{1/2}}{\widetilde{\ell}_{1i}^{3/2}}\right]+w_{34}n_{1}n_{2}\Theta_{ex}(\widetilde{R})\\ \nonumber
			&- {f_{1}} {f_{2}} \mathcal{N}_{1}\mathcal{N}_{2}\widetilde{\ell}_{B}\Theta_{m}(\widetilde{\kappa},\widetilde{R},a).\\\nonumber
		\end{align}
	In the above equation $\widetilde{\ell}_{11}$, $\widetilde{\ell}_{12}$, $\widetilde{\ell}_{13}$ and $\widetilde{\ell}_{14}$ are the effective expansion factors for the mean square end-to-end distance of the danging part of PA and PC and complexed part of PA and PC respectively. The number of monomers that participate in the charge-neutral complex is given by $n_1$ and $n_2$, and the degree of ionization of the dangling part of the chains is represented by $f_1$ and $f_2$. The strengths of the excluded volume interaction are given by $ w_{11},w_{22},w_{33},w_{44}\text{ and } w_{12} $ for monomers of individual chains PA, PC in the dangling part and in the complex and between the monomers of the PA-PC chains respectively. The dimensionless separation distance between two chains is assumed to be, $ \widetilde{R}={\left|\ro_{1}-\ro_{2}\right|}/{\ell}\equiv \widetilde{R}_{g1}+\widetilde{R}_{g2}+2\widetilde{R}_{g4}$, where the dimensionless radius of gyration $ \widetilde{R}_{gi} $. 
	\begin{widetext}
	In the Eq. \eqref{F5}, $\Theta_{s}\left(\widetilde{\kappa},a_{i}\right)$ is given by (See appendix \ref{appendix-section:F5-derivation} for details)
	\begin{align}
		\Theta_{s}\left(\widetilde{\kappa},a_{i}\right)=& {\left[-\frac{\pi}{a^{3/2}}  e^a \text{erfc}\left(\sqrt{a}\right)+\frac{\pi}{a^{1/2}}+\frac{\pi}{a^{3/2}}-\frac{2 \sqrt{\pi}}{a}\right]}
		\label{Theta_s}
	\end{align}
	where $a_{i}=\widetilde{\kappa}^2 \widetilde{R}_{gi}^{2}/3=\widetilde{\kappa}^2 (N_{i}-n_{i})\widetilde{\ell}_{1i}/18$. In case of two symmetric chains (same charge, $ N_{c1}=N_{c2} $ and length, $ N_{1}=N_{2} $) the interaction term can be calculated analytically and is given by
	\begin{align}\label{Theta_m_muthu}
		\Theta_{m}(\widetilde{\kappa},\widetilde{R},a_{12})=&\frac{\pi  N_{1}^2}{24 \widetilde{\kappa}^4 \widetilde{R} \widetilde{R}_{g}^4} \left[6 \widetilde{\kappa}^2 \widetilde{R}\left(\widetilde{R}\text{ erfc}\left(\frac{\widetilde{R}}{2 \widetilde{R}_{g}}\right)-2 \sqrt{\frac{6}{\pi }} \widetilde{R}_{g} e^{-\frac{\widetilde{R}^2}{4 \widetilde{R}_{g}^2}}\right)+6 e^{-\widetilde{\kappa} \left(\widetilde{\kappa} \widetilde{R}_{g}^2+\widetilde{R}\right)} \text{erfc}\left(\widetilde{\kappa} \widetilde{R}_{g}+\frac{\widetilde{R}}{2 \widetilde{R}_{g}}\right)\right.\nonumber\\
		&\left.+\frac{3}{\sqrt{2}} e^{-\pi \widetilde{\kappa}\widetilde{R}} \left\{\frac{4 \pi}{5} \left(1-e^{\pi \widetilde{\kappa} \widetilde{R}} \text{erfc}\left(\sqrt{2} \widetilde{\kappa} \widetilde{R}_{g}+\frac{\widetilde{R}}{2 \widetilde{R}_{g}}\right)\right)-\frac{\pi}{8}\left(1- \text{erfc}\left(\sqrt{2} \widetilde{\kappa} \widetilde{R}_{g}-\frac{\widetilde{R}}{2 \widetilde{R}_{g}}\right)\right)\right\}\right.\nonumber\\
	&\left.+12 \widetilde{\kappa}^2 \widetilde{R}_{g}^2 \left(e^{-\widetilde{\kappa}\widetilde{R}}-\text{erfc}\left(-\frac{\widetilde{R}}{2 \widetilde{R}_{g}}\right)\right)-6 e^{-\pi \widetilde{\kappa}\widetilde{R}}\right],
	\end{align}
\end{widetext}
and 
\begin{align}
	\Theta_{ex}(\widetilde{R}_{34})=&\frac{\sqrt{\pi} N_{1}^2 \exp\left({-\frac{\widetilde{R}_{34}^2}{4 \widetilde{R}_{g}^2}}\right)}{2 \widetilde{R}_{g}^3}\nonumber\\
	&-N_{1}^2 \left(\frac{\pi \widetilde{R}_{34}}{4 \widetilde{R}_{g}^4}+\frac{\pi }{2 \widetilde{R}_{34} \widetilde{R}_{g}^2}\right) \text{erfc}\left(\frac{\widetilde{R}_{34}}{2 \widetilde{R}_{g}}\right)
\end{align}
	where $ \widetilde{R}_{34}=\left|{\bf r}_{3}-{\bf r}_{4}\right|/\ell$.

	We note that the analytical calculation of the interaction between two asymmetric chains using the method described can be challenging and may require numerical techniques.
	
	However, an alternative method to calculate $\langle\beta H_{ex}\rangle$ and $\langle\beta H_{el}\rangle$ of equation \eqref{F5-avg} is to use a Flory-type approach\cite{flory1950,podgornik1993,Muthukumar2012}. To do so, we first introduce the monomer density $\rho(\mathbf{r})$, which is related to the number concentration and is defined as
	\begin{align}
		\rho(\mathbf{r})=\int_0^N \mathrm{~d} s \delta(\mathbf{R(s)}-\mathbf{r})\label{rho} \\
		\text { with } \int \mathrm{d}^3 \mathbf{r} \rho(\mathbf{r})=L.
	\end{align}
	The total length of the polymer is denoted by $L$, and the integral of the monomer density over all space gives $L$. We further assume $\rho(\mathbf{r})$ to have a Gaussian distribution and for i-th chain we obtained the expression as (detailed in appendix \ref{appendix-section:rho-fourier})
	\begin{align}
		&\rho_{i}(\ro)=\mathcal{N}_{i}\left(\frac{3}{4 \pi R_{gi}^{2}}\right)^{3 / 2} \exp \left[-\frac{3 (\left|\ro-\ro_{i}\right|)^{2}}{2 R_{gi}^{2}}\right].
		\label{rho}
	\end{align}
	
	 Using this definition, the interaction terms in Eq.\eqref{F5-avg} can be expressed in terms of $\rho(\mathbf{r})$. Details of this transformation is given in Appendix \ref{appendix-section:rho-fourier}, \ref{appendix-section:Hex}, and \ref{appendix-section:Hel}.
	
	The excluded volume interaction Eq.\eqref{eq:H_ex}, using \eqref{inter-chain-H-ex} and \eqref{intra-chain-H-ex}, can be written as,
	\begin{align}\label{H_ex_rho}
	 \beta H_{ex} &=\omega_{11}\ell^{3}\int \frac{d^{3} \mathbf{k}}{(2 \pi)^{3}} \left|\hat{\rho}_{1}(\mathbf{k})\right|^{2}+\omega_{22}\ell^{3}\int \frac{d^{3} \mathbf{k}}{(2 \pi)^{3}} \left|\hat{\rho}_{2}(\mathbf{k})\right|^{2}\nonumber\\
		&+\omega_{33}\ell^{3}\int \frac{d^{3} \mathbf{k}}{(2 \pi)^{3}} \left|\hat{\rho}_{3}(\mathbf{k})\right|^{2}+\omega_{44}\ell^{3}\int \frac{d^{3} \mathbf{k}}{(2 \pi)^{3}} \left|\hat{\rho}_{4}(\mathbf{k})\right|^{2}\nonumber\\
		&+2\omega_{34}\ell^{3}\int \frac{d^{3} \mathbf{k}}{(2 \pi)^{3}} \hat{\rho}_{3}(\mathbf{k})\hat{\rho}_{4}(-\mathbf{k})e^{i\mathbf{k}.{\mathbf{ R}_{ex}}},
	\end{align}
	where $ \hat{\rho}_{i}(\mathbf{k}) $ is the Fourier conjugate of $ \rho_{i}(\ro) $.
	
	The electrostatic interaction given in Eq.\eqref{H_el}, using Eqs. \eqref{inter-chain-H-el} and \eqref{intra-chain-H-el}, can be written as,
	\begin{align}\label{H_el_kspace}
		\beta H_{el}&=\int \frac{d^{3} \mathbf{k}}{(2 \pi)^{3}} \left|\hat{\rho}_{1}(\mathbf{k})\right|^{2}\hat{U}_{11}^{el}(\mathbf{k})+\int \frac{d^{3} \mathbf{k}}{(2 \pi)^{3}} \left|\hat{\rho}_{2}(\mathbf{k})\right|^{2}\hat{U}_{22}^{el}(\mathbf{k})\nonumber\\
		&+2\int \frac{d^{3} \mathbf{k}}{(2 \pi)^{3}} \hat{\rho}_{1}(\mathbf{k})\hat{U}_{12}^{el}(\mathbf{k})\hat{\rho}_{2}(-\mathbf{k})e^{i\mathbf{k}.{\mathbf{ R}}},
	\end{align}
	where $ \hat{U}_{ij}(\mathbf{k}) $ is Fourier conjugate of Eq. \eqref{Yukawa-Uij} and is given by Eq. \eqref{conjugate-Uij}.
	
	We note that the average of a general function depending on polymer coordinate can be written in the form of convolution,\cite{podgornik1993}
	\begin{align}
		\int_0^N\langle f(\mathbf{R(s)})\rangle_{H_0} d s=\left\langle\int d^3 \mathbf{r} f(\mathbf{r}) \rho(\mathbf{r})\right\rangle_{\mathbf{r_{1}},\mathbf{r_{2}}},
	\end{align}
	where the avg in the rhs is over CM of the PEs. In the absence of any an-isotropic electric field that breaks the translation symmetry of the problem, the systems CM dependency vanishes. If we use this approximation, as detailed in Ref. \cite{podgornik1993}, instead of the variational calculation	used here, the $ \Theta_{ex}$, $\Theta_s$ and $\Theta_{m}$ of Eq. \eqref{F5} are	replaced, respectively
	\begin{align}
	\Theta_{ex}(\widetilde{R})=\left(\frac{3}{4\pi \widetilde{R}_{g0}^2}\right)^{3/2}\exp\left(-\frac{3\widetilde{R}_{34}^2}{4 \widetilde{R}_{g0}^2}\right),
	\end{align}
	\begin{align}\label{Theta_s}
		&\Theta_{s}\left(\widetilde{\kappa},a_{i}\right)=\frac{2}{\pi}\left[\sqrt{\frac{\pi\widetilde{\kappa}^{2}}{4 a_{i}}}-\frac{\widetilde{\kappa} \pi}{2} \exp{\left(a_{i}\right)} \text{erfc}\left(\sqrt{a_{i}}\right)\right],
	\end{align}
	and
	\begin{align}\label{Theta_m}
		\Theta_{m}\left(\widetilde{\kappa}, \widetilde{R},a_{12}\right)=&\frac{e^{a_{12}}}{ \widetilde{R}}\left[e^{-\widetilde{\kappa}  \widetilde{R}} \text{erfc}\left(\sqrt{a_{12}}-\frac{\widetilde{\kappa}  \widetilde{R}}{2 \sqrt{a_{12}}}\right)\right.\nonumber\\
		&\left.-e^{\widetilde{\kappa}\widetilde{R}} \text{erfc}\left(\sqrt{a_{12}}+\frac{\widetilde{\kappa}  \widetilde{R}}{2 \sqrt{a_{12}}}\right)\right],
	\end{align}
	where $a_{i}=\widetilde{\kappa}^2 \widetilde{R}_{gi}^{2}/3=\widetilde{\kappa}^2 (N_{i}-n_{i})\widetilde{\ell}_{1i}/18$, $a_{12}=\sum_{i=1}^{2}\widetilde{\kappa}^{2}\widetilde{R}_{gi}^{2}/6=\sum_{i=1}^{2}\widetilde{\kappa}^{2}\widetilde{\ell}_{1i}(N_{i}-n_{i})/6$, $ \widetilde{R}_{g0}^{2}=({n_{1}
		\widetilde{\ell}_{1 3}+n_{2}\widetilde{\ell}_{1 4}})/{6}$, and $ \widetilde{R}_{34}=\left|{\bf r}_{3}-{\bf r}_{4}
	\right|/\ell \equiv 0 $. .
	
	In the limit of $\widetilde{\kappa}\to \infty$, the electrostatic interaction terms given in Eqs. \eqref{Theta_m} and \eqref{Theta_m_muthu} exhibit similar behavior. Both terms can be approximated as
	\begin{align}
		\Theta_{m}(\widetilde{\kappa},\widetilde{R},a)\sim \exp\left({-\frac{\widetilde{R}^2}{4 \widetilde{R}_{g}^2}}\right),
	\end{align}
	which indicates a decay with distance comparable to the excluded volume interaction. This decay arises from short-range interactions. In the absence of salt, both calculations yield a $1/\widetilde{R}$ -type decay, indicating a long-range interaction behavior.
	
\section{result and discussion}
	This study considers a two-state model to determine the equilibrium conformations, effective expansion factors of oppositely charged polyions and their effective charge. The free energy is self-consistently minimized with respect to six variables, namely the charge $f_{i}$ and size $\tilde{\ell}_{1 i}$ of the PEs. At an intermediate step of complexation, the PEs are divided into dangling and complexed parts. The complexed part (chains 3 and 4 in Fig. 1) is assumed to take Gaussian conformations, and if excluded volume interactions are ignored, the size parameters $\tilde{\ell}_{13}$ and $\tilde{\ell}_{14}$ can be taken as $ 1 $. This leads to a four-variable minimization with respect to $f_{1}, f_{2}, \tilde{\ell}_{11}, \tilde{\ell}_{12}$. For the symmetric fully ionizable case, where both polyions have the same number of monomers ($N_{1}=N_{2}=N$) and are fully ionizable, the minimization reduces to effectively two variables $f$ and $\tilde{\ell}_1$. The resulting equilibrium size and charge of the polyions, PA and PC, are determined for the completely separated ($n=0$) and fully complexed ($n=N_{c1}$) chains.

\begin{figure}[!htbp]%
	\centering%
	\subfloat[]{\includegraphics[width=0.7\linewidth]{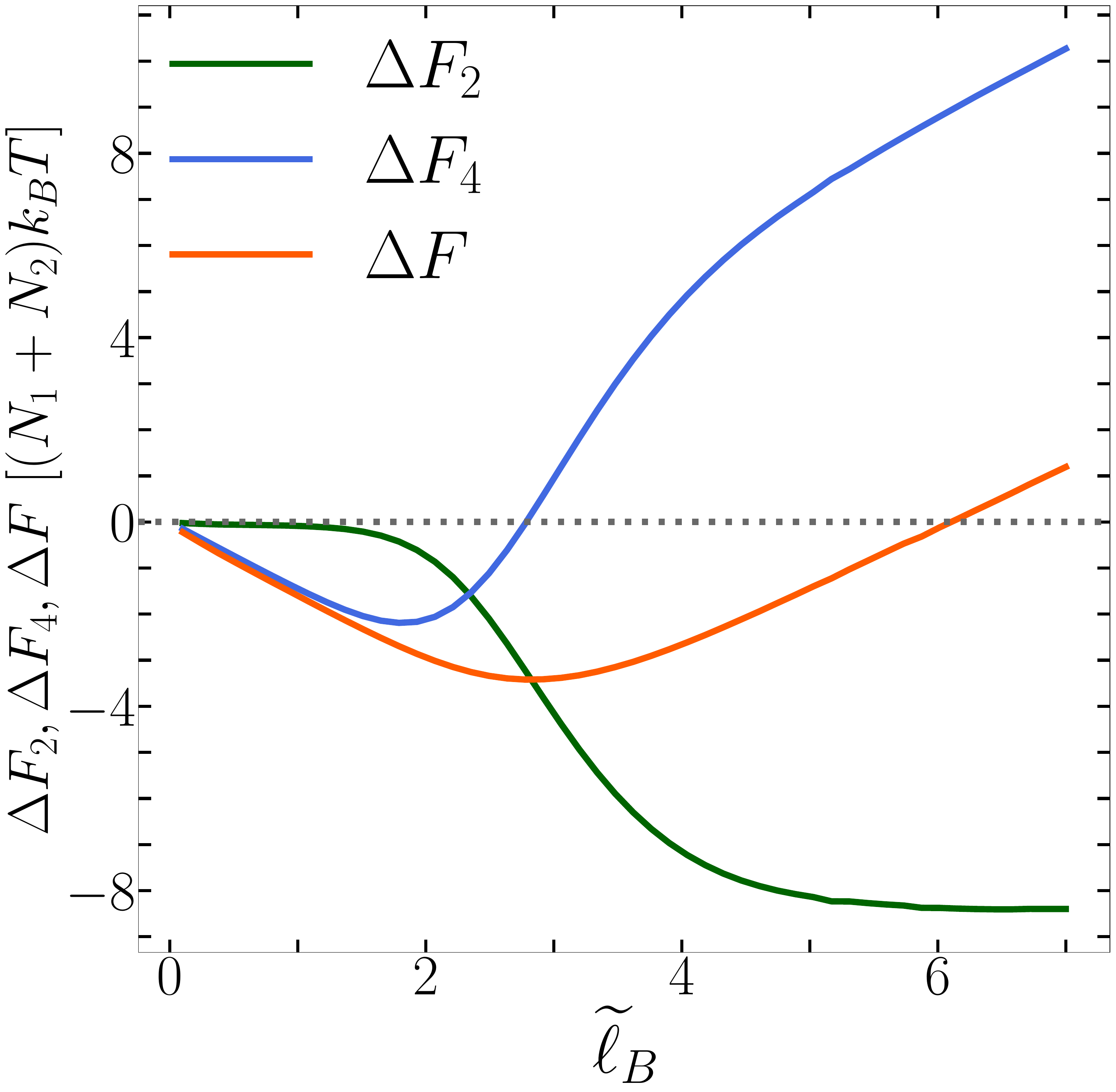}}
	\caption{\textbf{Relative Thermodynamic Drive:} (a) Variation in the free energy components of complexation, \(\Delta F_2\) (free ion entropy) and \(\Delta F_4\) (ion-pair energy), in units of \((N_1 + N_2) k_B T\), as functions of \(\delta\), plotted for \(N_1 = N_2=N= 1000\). Other parameters are: \(\tilde{c}_s = 0.0\), \(w_{ij} = 0.0\), and \(\tilde{\rho}_i = 0.0005\). The crossover strength, \(\Gamma_c\), divides the complexation process into two regimes: enthalpy-dominated on the left and entropy-dominated on the right. At very high \(\delta\) values, the complexation is disfavored due to enthalpy loss.}
	\label{fig:f2f4vslb}
\end{figure}

In this study, excluded volume parameters for all interactive pairs - monomer pair within PA, monomer pair within PC, and monomer pair with one monomer from each PA and PC - are assumed to be zero. This is due to the assumption that electrostatic interactions will overwhelm excluded volume interactions, especially under good solvent conditions. Thus, $w_{11}=w_{22}=w_{12}=0.0$ or $w_{ij}=0.0$ for the rest of the paper. To estimate the monomer densities of the individual chains, the dimensionless volume of the system is set to $\Omega/\ell^3=2\times10^6$. For the $i$-th polyion of length $N_i$ (here, $i=1,2$ corresponding to the full PA and PC), the monomer density is given by $\tilde{\rho}_i=N_i/(\Omega/\ell^3)$. For $N_i=1000$, the monomer density is approximately $0.0005$.

\begin{figure}[!htbp]%
	\centering
	\fbox{\includegraphics[width=0.99\linewidth]{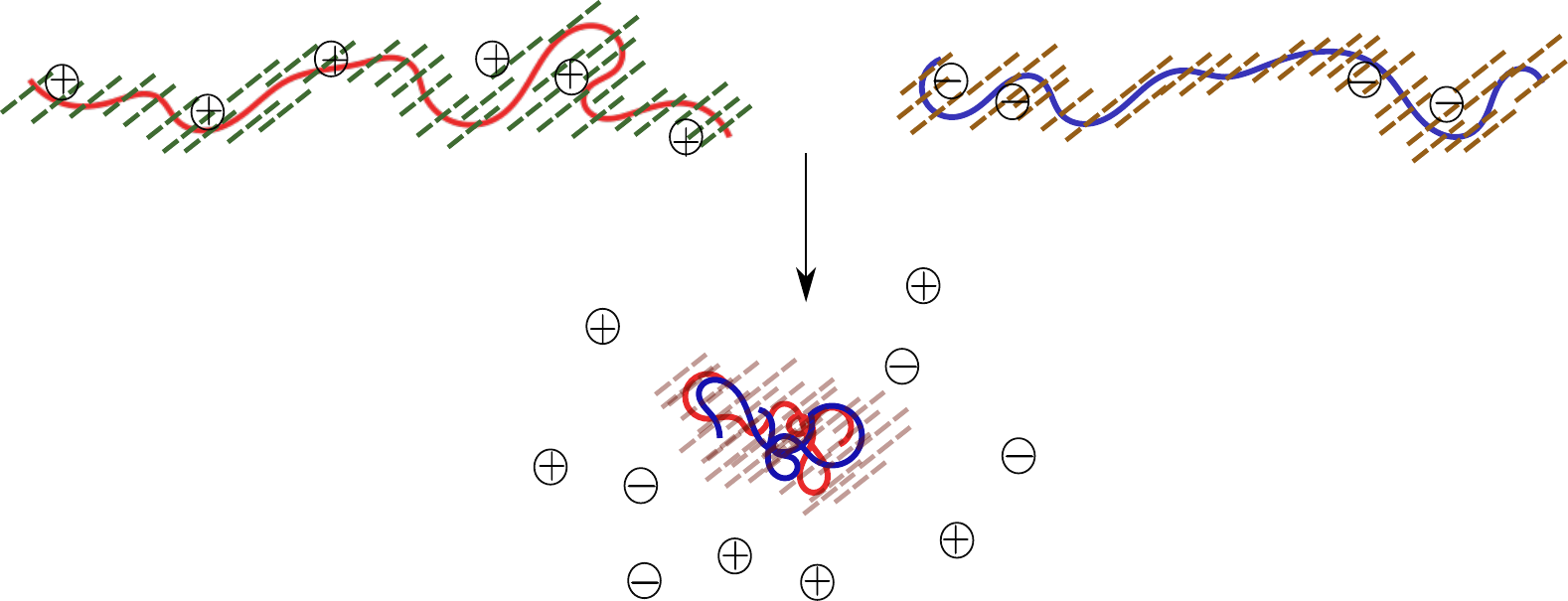}}
	\caption{The adsorption of counterions on the backbone of polyelectrolyte chains can result in the formation of temporary dipoles, causing a localized reduction of the dielectric constant in the vicinity of the chain, as shown in the shaded area of the figure. The dielectric behavior of the system changes upon the complexation of two polyelectrolyte chains, leading to an overall reduction in the local dielectric constant. It is crucial to keep in mind that the local dielectric constant of each individual chain is dependent on its specific polymer structure, emphasizing the significance of considering polymer specificity when evaluating the thermodynamic driving forces of polyelectrolyte complexation.}
	\label{fig:schematic-2}
\end{figure}

	We first benchmark the general problem of complexation of two oppositely but partially ionizable PEs of different length by the known results for fully ionizable and symmetric chains (of same length and number of monomers).  The correlation between counterion adsorption, release and pairwise electrostatic attraction at various ambient conditions has already been investigated in details\cite{zhaoyang2006,mitra2023,Ghosh2023}. Here, we briefly note the key results, obtained by the minimization of the free energy (Eq. \eqref{partition-sum}) for the symmetric, fully ionizable case ($f_{m1}=f_{m2}=1$, $f_{T1}=f_{T2}=f_T$, and $\tilde{\ell}_{11}=\tilde{\ell}_{12}=\tilde{\ell}_{1}$), rendering the minimization to be effectively with respect to two variables $f$ and $\tilde{\ell}_{1}$.

The relationship between enthalpy, entropy, and electrostatic temperature($\widetilde{\ell}_{B}$)is well-documented in the context of complex formation among polyions\cite{zhaoyang2006,mitra2023,Ghosh2023,chen2022-PNAS}. Here We first benchmark the general problem of counterion condensation by the known results for a fully ionizable PE, taking the counterion size equal to the monomer size as is traditionally done (See Fig. (2)). The mean-field implicit solvent model reveals that counterion condensation, initially on single chains, can lead to enthalpy loss while charge neutral complex formation, a phenomenon discussed in earlier literature. The initial state involves both chains being in equilibrium, separated by a distance in a dilute condition. The final state, in the case of symmetric PEs, involves both chains forming a charge neutral complex. In general, the mean-field model predicts that complex formation among polyions at moderate temperatures is primarily driven by counterion release entropy (See Fig. (2) and the schematic in Fig. (3)). Although the formation of pairwise complexes between two oppositely charged polyions can exhibit three distinct regimes depending on the thermodynamic drive. At low Coulomb strengths ($\delta \ell_B$, see Eq. \eqref{gamma-def}), the complex formation is primarily driven by the enthalpy gain from monomer-monomer ion pairs.\cite{zhaoyang2006,mitra2023,Ghosh2023} At moderate Coulomb strengths, the drive is dominated by the entropy gain of free counterions, with minimal changes in enthalpy (See Fig. (2)).\cite{zhaoyang2006,priftis2012,mitra2023,chen2022-PNAS} Finally, at high electrostatic strengths, the complexation is still driven by entropy gain from release of counterions, but this is significantly offset by enthalpy loss (See Fig. (2)).\cite{zhaoyang2006,priftis2012,dzubiella2016,chen2022-PNAS,mitra2023,Ghosh2023} Consequently, The role of ion pairing becomes a crucial factor in determining the thermodynamic drive for the formation of charge complexes between oppositely charged polyions.\cite{zhaoyang2006,chen2022-PNAS,mitra2023,Ghosh2023}

\begin{figure}[!htbp]
	\centering
	\subfloat[Change in enthalpy as a function of varying local dielectric constant]{%
		\includegraphics[width=0.51\linewidth]{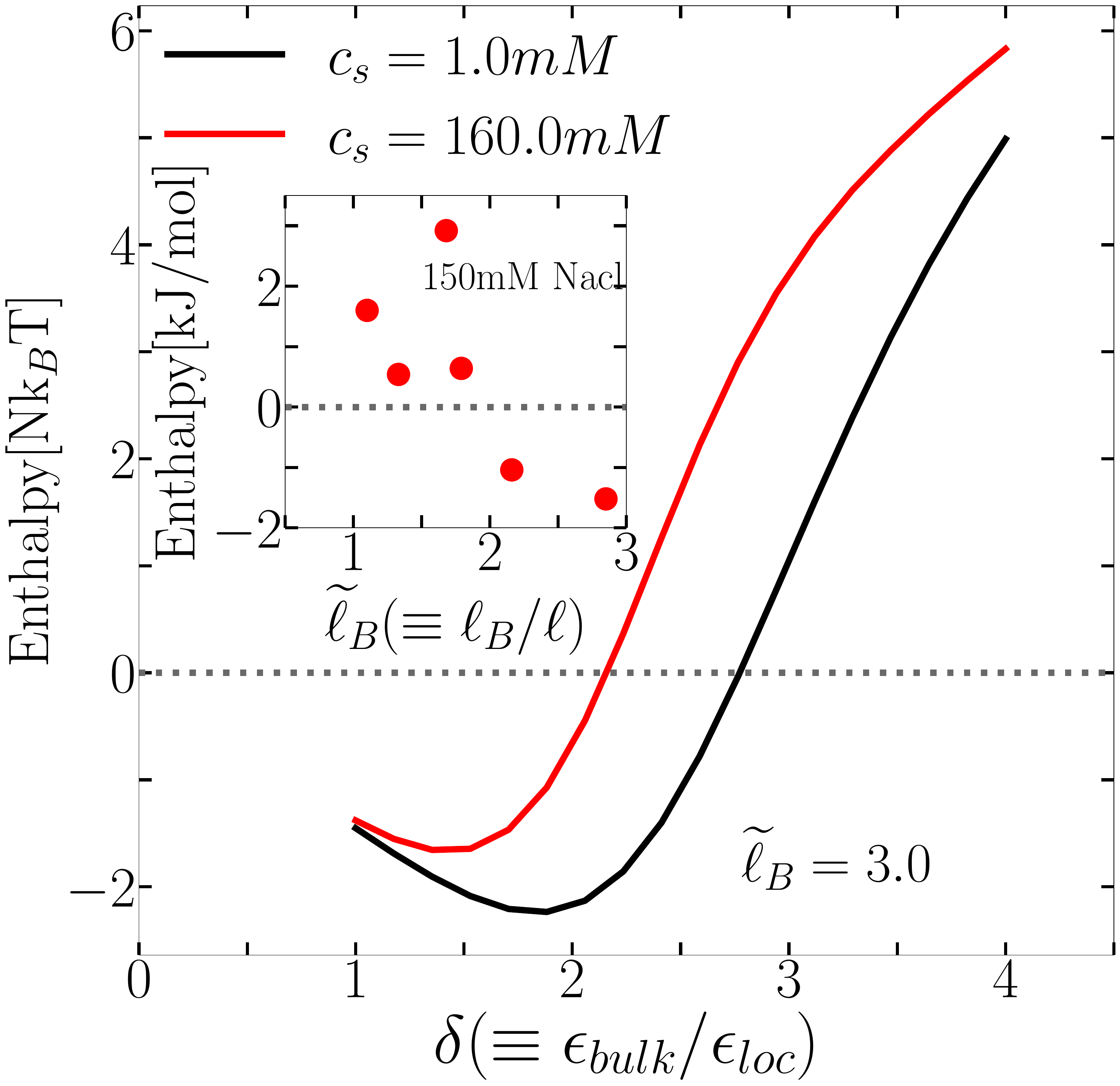}
	}
	\subfloat[Change in enthalpy as a function of temperature]{%
		\includegraphics[width=0.5\linewidth]{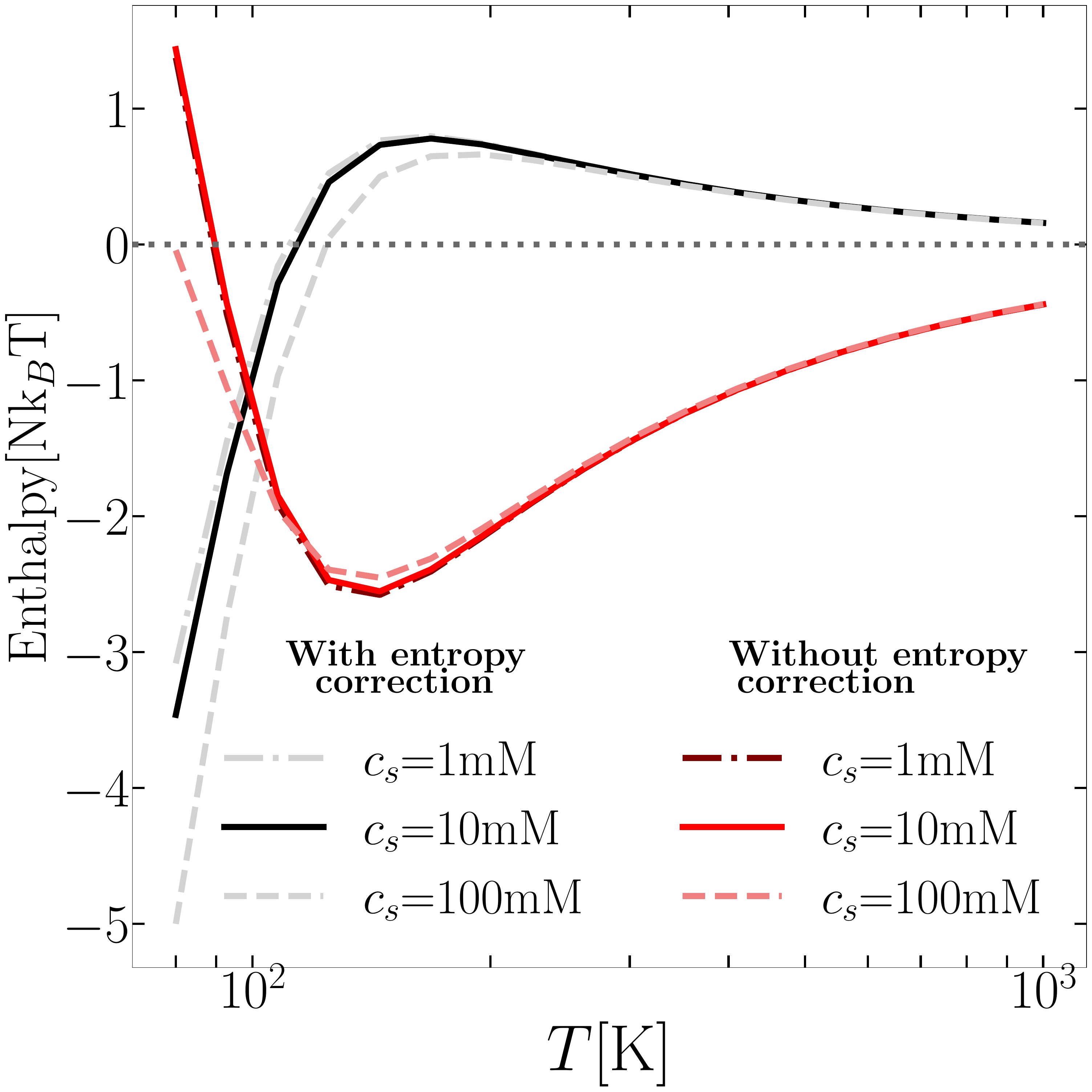}
	}
	\qquad
	\subfloat[Change in enthalpy as a function of salt concentration]{%
		\includegraphics[width=0.9\linewidth]{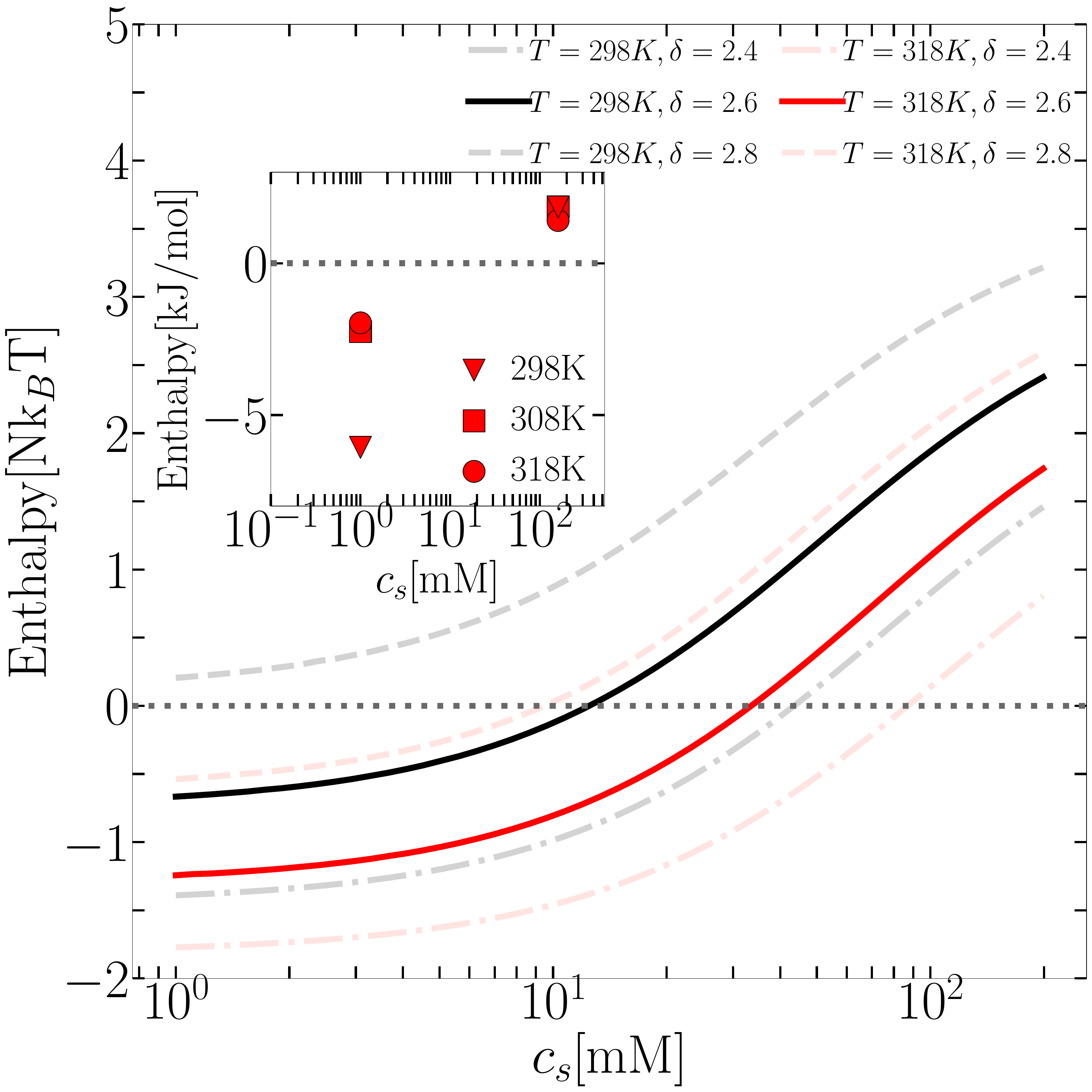}
	}
	\caption{
		Change in enthalpy for the formation of a charge-neutral complex as a function of (a) local dielectric constant and (b) temperature, and (c) salt concentration. The enthalpy data points derived from the experimental $\Delta H_{PEC}$ values from the isothermal titration calorimetry (ITC) data of Laugel et al.,\cite{Laugel2006} for a range of complexing polyelectrolytes in 0.15 M NaCl, show significant variation. While mean-field theories account for both endothermic and exothermic complexation enthalpies, the scaling of predicted $\Delta H_{PEC}$ differs substantially from the experimental results (as seen in the inset of Figure (a)). However, our results (Figures (a-c)) demonstrate that global thermodynamic behavior is significantly influenced by local factors such as the dielectric constant, which can explain both endothermic and exothermic complexation enthalpies. The parameters are: $N_1=N_2=N=1000,~w_{i j}=0.0,$ and $\tilde{\rho}_i=0.0005$.
	}
	\label{fig:E}
\end{figure}

We note that the monomers in a polyelectrolyte chain are strongly correlated topologically and electrostatically, surrounded by a neutralizing background of charged ions in a highly polarizable solvent. As a result, the polar solvent is attached to the nonpolar backbone, resulting in a significantly lower local dielectric constant $\epsilon_l$ in the vicinity of the sulfonate group (in case of NaPss). This effect has been recognized in early investigations\cite{Mehler1984,Lamm1997,Rouzina1998}. Theoretical formulations of the polarizability effect in crowded environments are still a challenge in polyelectrolyte physics.\cite{muthu2017} The local dielectric environment of the polyions influences the degree of counterion condensation.\cite{muthu2004,muthu2017,Ghosh20231} In regions with a low dielectric constant, counterion binding is stronger. Consequently, at a given salt concentration (as shown in Fig. 4(a)), enthalpy shifts from negative to positive in environments with lower dielectric constants. Increasing salt concentration makes counterion binding more energetically favorable, thereby shifting the transition point of enthalpy toward higher dielectric constants.

Hence, variations in the local dielectric behavior among different polyions lead to diverse enthalpy changes: polyions with lower local dielectric constants may exhibit positive enthalpy changes (enthalpy loss), while those with relatively higher local dielectric constants tend to exhibit negative enthalpy changes. Previous attempts to explain why experimental observations of enthalpy changes can be both positive and negative (see inset of Fig. 4(a) for fixed salt concentration and Fig. 4(c) for fixed temperature) focused on the idea of solvent reorganization entropy. Additionally, in the proposed correction for the origin of entropy due to implicit solvent effects, it is hypothesized that solvent reorganization incurs an additional entropy cost.\cite{chen2022-PNAS} With this correction, the enthalpy change tends to remain positive at moderate physiological temperatures(see Fig. 4(b)). However, the corrected theory overlooks potential dielectric heterogeneity within the polyions. However, we hypothesis that these variation in enthalpy could be due to the specific characteristics of the polyions that are forming the pairs.\cite{Laugel2006, fu2016}

However in an attempt to look deeper into the electrostatic entropy contribution, it can be estimated using the traditional form of the Born equation of solvation, which relates the electrostatic solvation energy to the dielectric constant of the solvent. The electrostatic entropy contribution can be calculated as $TS=(T/\epsilon)\left(\partial \epsilon / \partial T\right) U_{ad}$, where $\epsilon$ represents the dielectric constant of the solvent and $\partial \epsilon / \partial T$ represents the temperature dependence of the dielectric constant. Approximately, this expression can be used to estimate the electrostatic entropy contribution in polyelectrolyte systems, where the solvent is highly polarizable and attached to the nonpolar backbone, and explicit solvent molecules are absent.

Although the Born equation is a useful tool for predicting the free energy of solvation, it is not very sensitive to solvent properties when the dielectric constant of the bulk is used instead of the local dielectric constant. Therefore, it falls short in accurately describing the entropy of solvation.\cite{Dinpajooh2016} Several studies have highlighted this limitation and emphasized the importance of considering the local dielectric behavior in accurately estimating the electrostatic entropy contribution\cite{Roux1990,Rick1994,LyndenBell1997,Rajamani2004,Dinpajooh2016}s

For water, with a bulk dielectric constant of $\epsilon_{bulk}=80$, the entropy can be approximately calculated as $TS=\left(\partial \epsilon_{bulk} / \partial T\right)(298/\epsilon_{bulk})U_{ad}(\epsilon_{bulk})$, the bulk dielectric constant variation ($ \epsilon_{bulk} $) was estimated to be $\left(\partial \epsilon_{bulk} / \partial T\right) = -0.36 \mathrm{~K}^{-1}$\cite{Muthukumar2023-book}, hence $TS=-0.36(298/\epsilon_{bulk})U_{ad}(\epsilon_{bulk})\simeq-1.36 U_{ad}(\epsilon_{bulk})$, where $U_{ad}(\epsilon_{bulk})$ is the ion pair formation energy given by Eq.\eqref{uad}(note that, $\epsilon_{l12}\text{ and }\epsilon_{l}$ corresponds to $\epsilon_{bulk}$) \cite{Dinpajooh2016,chen2022-PNAS}.  In this case, the entropy contribution has a significant impact, indicating the contribution from solvent reorientation during complex formation, even at higher temperatures or lower values of the Bjerrum length $\ell_{B}$.\cite{chen2022-PNAS} 

It is important to note that the conditions in polyelectrolytes or proteins are somewhat different than in the ionic situation in the case of a single ion, it is completely accessible to the polar media or water, whereas in polyelectrolytes or proteins, the bulk of the macromolecule and the fractal nature of it limit the accessibility of water and ions. This limited accessibility results in a slower rate of increase of $\epsilon_{l}$ with distance.\cite{Mehler1984,Lamm1997}
The exact value of the local dielectric constant depends on several factors such as local density, temperature, electric field, and dipole moment of ion pairs. However, it is challenging to obtain a precise understanding of the spatial variation of the dielectric constant embedded in the parameter $\delta$. Despite this challenge, we can use our knowledge of the monomer density profiles\cite{Kumar2012} to make a qualitative estimation of the local dielectric constant. The chain conformations and charges provide information about the spatial dielectric behavior near the polyelectrolyte interior and interface with the solvent, allowing us to estimate the local dielectric constant in a phenomenological manner. Following the approach outlined in \cite{Grant2001}, the local dielectric constant can be estimated by writing
\begin{align}
	\epsilon(\mathbf{r})=\rho(\mathbf{r}) \epsilon_{local}^{\prime}+\left(1-\rho(\mathbf{r})\right) \epsilon_{bulk},
	\label{epsilon}
\end{align}
where $\epsilon_{local}^{\prime}=4$, $\epsilon_{bulk}=80$, and $\rho(\mathbf{r})$ is given by Eqs. \eqref{rho}. To fully determine Eq. \eqref{rho} (See appendix\ref{appendix-section:rho-fourier}), we need to minimize the total free energy with respect to the size and charge of the polymers, as detailed in ref \cite{mitra2023,Ghosh2023}.

Equation \eqref{epsilon} gives the dielectric distribution function $\epsilon(\mathbf{r})$, which modulates the formation of counterions or monomer pairs based on its spatial variation. To simplify the calculation, $\epsilon(\mathbf{r})$ can be replaced by the local dielectric constant ($\epsilon_{local}$), which is defined as the average of the continuous dielectric function $\epsilon(\mathbf{r})$ over the volume of the sphere
\begin{align}
	\epsilon_{local} = \frac{1}{V} \int_V \epsilon(\mathbf{r})d^3 \mathbf{r}.
	\label{local-dielectricity}
\end{align}
This average is a distance-weighted average of the dielectric constant calculated within the radius of gyration of the polymer.

Previously, the variation of $ \epsilon_{local} $ with temperature was estimated to be $\left(\partial \epsilon_{local} / \partial T\right) \simeq-0.8 \times 10^{-4} \mathrm{~K}^{-1}$, hence, using empirical knowledge of $ \epsilon_{local} $, we can calculate the entropy contribution as $TS=-0.8 \times10^{-4}(298/\epsilon_{local})U_{ad}(\epsilon_{local})\simeq-8\times10^{-4}U_{ad}(\epsilon_{local})$. \cite{Roux1990,Dinpajooh2016} Hence, considering the impact of local dielectric behavior significantly lowers the estimated electrostatic entropy cost, suggesting it could be negligible. However, a detailed examination of the unique properties of polyions is crucial to confirm this assertion. Existing evidence from previous studies and the above phenomenological derivation indicates that the local dielectric environment plays a pivotal role in these processes.

\begin{figure}[!htbp]%
	\centering%
	\subfloat[]{\includegraphics[width=0.7\linewidth]{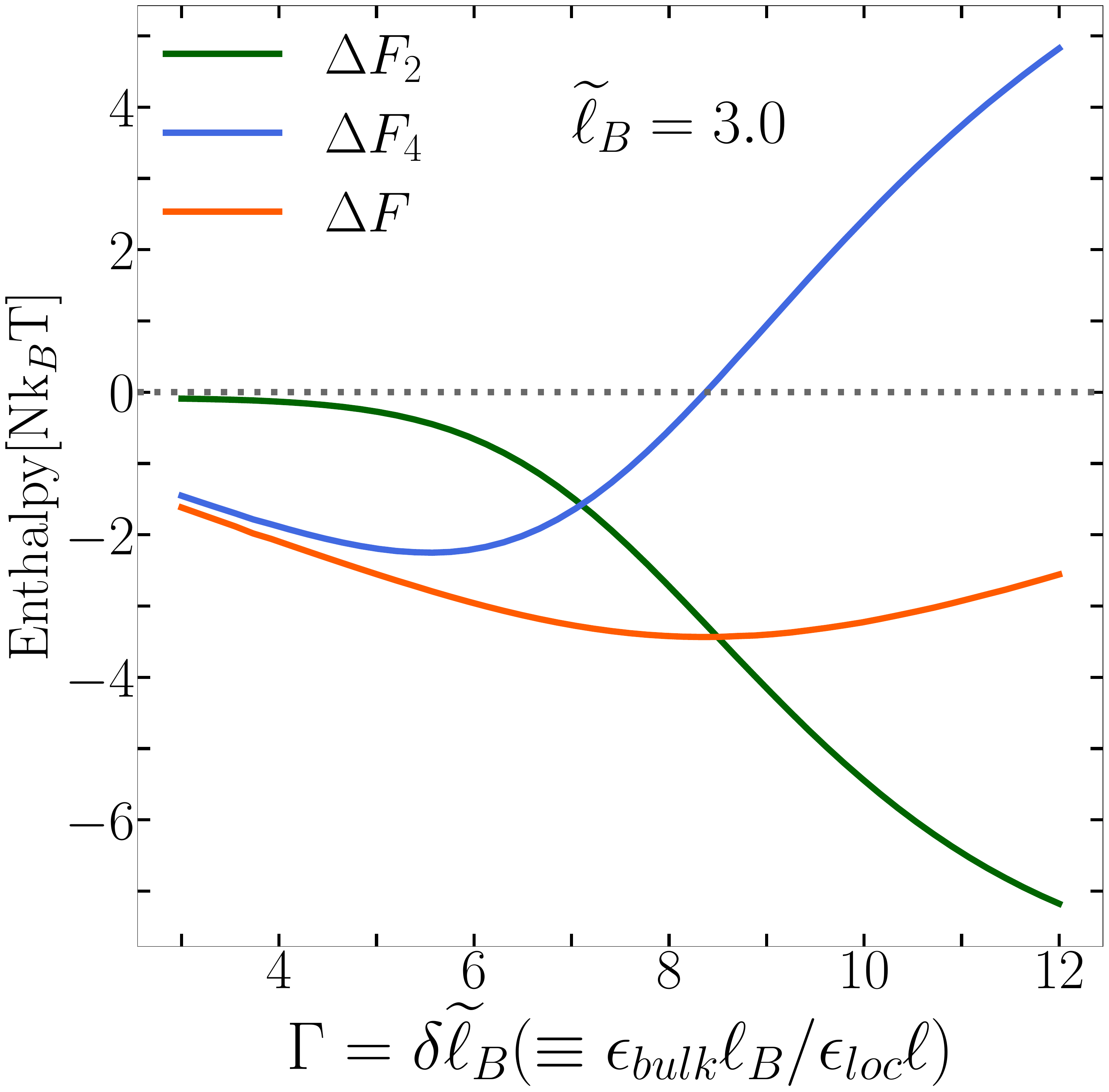}}
	\qquad
	\hspace*{0.8cm}\subfloat[]{\includegraphics[width=0.88\linewidth]{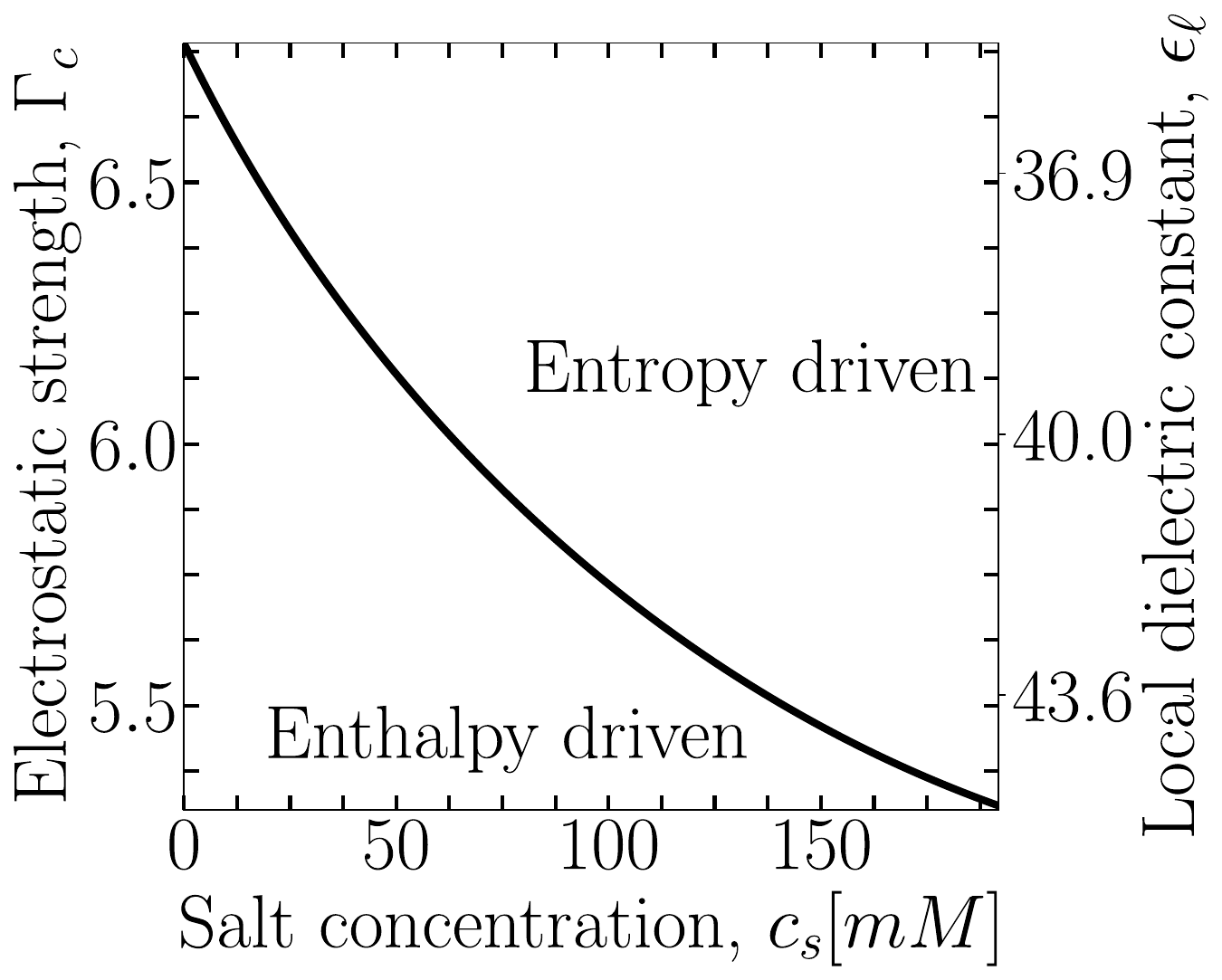}}
	\caption{\textbf{Relative Thermodynamic Drive:} (a) Variation in the free energy components of complexation, \(\Delta F_2\) (free ion entropy) and \(\Delta F_4\) (ion-pair energy), in units of \((N_1 + N_2) k_B T\), as functions of \(\delta\), plotted for \(N_1 = N_2 =N= 1000\). The crossover point, \(\Gamma_c\), where \(\Delta F_2 = \Delta F_4\), is identified. Other parameters are: \(\tilde{c}_s = 0.0\), \(w_{ij} = 0.0\), and \(\tilde{\rho}_i = 0.0005\). The crossover strength, \(\Gamma_c\), divides the complexation process into two regimes: enthalpy-dominated on the left and entropy-dominated on the right. At very high \(\delta\) values, the complexation is disfavored due to enthalpy loss. (b) The crossover Coulomb strength, \(\Gamma_c(= \delta \widetilde{\ell}_{B})\), at which entropy and enthalpy are equal, is plotted against salt concentration, with temperature constant at room temperature. This line indicates the boundary between enthalpic and entropic driving forces for complex formation. The other parameters are: \(\tilde{\ell}_B = 3.0, ~\tilde{c}_s = 0.0, ~w_{ij} = 0.0\), and \(\tilde{\rho}_i = 0.0005\).}
	\label{fig:size-asy}
\end{figure}

It is worth noting that when the electrostatic interaction strength is weak, the use of \(\epsilon_{bulk}\) in the electrostatic energy calculations implies that entropy becomes the primary thermodynamic driving force (as illustrated in Fig. 4(b)). However, if the local dielectric behavior near non-polar polymers in polar solvents is considered, the thermodynamic drive could remain enthalpic (as seen in Figs. 4(b) and 5(a)). This suggests that the ability of a polyion to attract ions from the solution largely determines its thermodynamic characteristics. Given the complex interplay between the binding behavior of different polymers under various counterion and solvent conditions, it is challenging to ascertain the exact nature of the thermodynamic driving force.


To further illustrate the point, experimental data shows a clear qualitative shift in enthalpy from negative to positive as salt concentration increases (see the inset plot in Fig. 4(c)).

The mean-field implicit solvent model, which accounts for local binding constants, indicates that the overall thermodynamic drive is unique to each pair of polyions. The model qualitatively explains the temperature dependence of enthalpy as found in the experiments.\cite{Laugel2006} At a given temperature, lower dielectric constants result in greater enthalpy loss. For instance, at $ 298 $ K, the enthalpy loss with a dielectric constant of $ 2.8 $ is greater than with a dielectric constant of $ 2.6 $. Note that Figs. 4(a) and 4(c) complement each other in illustrating this point.

We evaluated enthalpy and entropy as functions of \(\delta\), a measure of the local dielectric constant, at a fixed electrostatic temperature. At lower values of \(\delta\), where the local dielectric constant is closer to that of the bulk, enthalpy predominantly drives complex formation. In contrast, at higher \(\delta\) values, when the local dielectric constant is lower relative to the bulk, entropy from counterion release becomes the driving force behind complex formation. Between these extremes, there is a point where enthalpy and entropy are balanced, indicating a thermodynamic state boundary. The crossover point is evaluated for various salt concentrations and is depicted in Fig. 5(b).

Earlier studies indicated that entropy gain is inversely proportional to electrostatic temperature. Our findings, however, suggest that entropy gain is also inversely proportional to the local dielectric constant, assuming a constant electrostatic temperature.

The crossover point becomes increasingly dependent on salt concentration and local dielectric constant. At higher salt concentrations, the thermodynamic drive can be entropic if the local dielectric constant of the polyions is below $ 44 $ (at $\sim 150 $ mM). At lower salt concentrations, this threshold for entropic drive drops to a dielectric constant of 37 (at $\sim 0$ mM). This complexity underscores the challenge of explaining experimental outcomes without a thorough understanding of local dielectric behavior in polyions.

Note that the value of $\Gamma$ at crossover point is quite narrow; within the range of salt concentrations explored, it is approximately  $6.0\pm 0.6$ . Additionally, the estimated local dielectric constant at the crossover point is about $40.0\pm 4$.

Further note that, the anticipated discrepancies in interpreting experimental data\cite{Laugel2006,fu2016,chen2022-PNAS} in cases with fixed salt concentrations (inset plot of Fig. 4(a)) can be attributed to replacing the local parameter (\(\epsilon_{local}\)) with a global one (\(\epsilon_{bulk}\)). However, the results (Figs. 4(a-c),5(a-b)) demonstrate that global thermodynamic behavior is significantly influenced by local parameters like dielectric constant, underscoring the importance of understanding these factors for accurate predictions and interpretations of experimental results.
	
	\section{Conclusions}
	In this study, a theoretical framework was proposed for investigating the interactions between flexible polymer chains, specifically polyelectrolytes. The free energy of the system was calculated taking into account the position-dependent mutual interactions and conformations of the individual chains. An approximate analytical formula was derived for the inter-chain pair potential, based on the self-regulating degree of ionization and radius of gyration, as a function of the distance between the centers of mass of the chains.
	
	The thermodynamic driving force for complex-coacervation formation, a process in which oppositely charged polyelectrolytes form a complex, was found to be influenced by the number of ions bound to the chain backbone and the volume entropy of free ions. Earlier research suggested that entropy gain is inversely proportional to electrostatic temperature, but our analysis indicates that entropy gain is also inversely proportional to the local dielectric constant, given a constant electrostatic temperature. This insight sheds light on a significant discrepancy: while experimental studies often suggest that entropy is the dominant factor in complex coacervation, computational simulations with implicit solvents typically show considerable energetic contributions, particularly in systems with weak to moderate electrostatic strengths. Another explanation for the discrepancies involves solvent reorganization entropy, but this approach often overestimates the effect by using global parameters like \(\epsilon_{bulk}\) instead of local parameters like \(\epsilon_{local}\); however, our results (Figs. 4(a-c),5(a-b)) show that global thermodynamic behavior is significantly influenced by local factors like the dielectric constant.
	
	The unique characteristics of poly-ions/polyelectrolytes, which are organic molecules with oily backbones, resulting in a higher electrostatic attraction near the backbone were highlighted. The optimal value of the internal protein dielectric constant, which is dependent on the macromolecule's structure and sequence specificity, is still a topic of ongoing debate in the literature. We point out how the thermodynamic drive is modified by the local environment of the polyions and how modifications in Coulomb interactions ultimately determine the nature of the thermodynamic drive and gives a new perspective. Our findings provide a new perspective on the nature of the thermodynamic drive, highlighting the crucial role played by electrostatic correlations.
	
	However, it is important to acknowledge that the results of this study are not entirely rigorous as it does not account for certain factors such as self-consistent equations for multiple chains and solvent orientation-dependent interactions. These limitations suggest that further research is necessary to fully understand the underlying mechanisms and to account for the missing factors. Despite these limitations, the study provides new insights into the thermodynamic driving force of complex-coacervation formation and the unique properties of polyelectrolytes, which can be useful for future studies.
	
	\section{Acknowledgment}
	The author would like to express their gratitude to Arindam Kundagrami, Soumik Mitra, Aritra Chowdhury, Benjamin Schuler, and Neelanjana Sengupta for their valuable contributions in the intellectual discussions regarding this work. The author also extends appreciation to Raveena Gupta for assistance with the manuscript. The research was supported by the Indian Institute of Science Education and Research Kolkata and the Ministry of Education, Government of India.
	
	\appendix
	\begin{widetext}
	\section{Derivation of the structure factor}\label{structure-factor}
	
		Defining the Fourier transform of $\mathbf{R}(s)$ as
		\begin{align}\label{rs2rq}
				\mathbf{R}(s)=\int_{-\infty}^{\infty} \frac{d q}{2 \pi} \mathbf{R}(q) \exp (i q s).
		\end{align}
	We take the trial Hamiltonian of a chain and representing it using the Fourier transform (given in Eq.\eqref{rs2rq}), we obtain
	\begin{align}
		\beta H_{0}(\{{\bf R}\left(q_{1}\right)\})=& \int_{-\infty}^{\infty} \frac{d q_{1}}{2 \pi} \frac{R_{q_{1}}^2}{g(q_{1})}
	\end{align}
	with $ g(q)=\frac{2 \ell_{11}}{3 q_{1}^{2}} $. Here, we follow the exact steps as outlined in Muthukumar's 1987 paper \cite{muthu1987}. Using the definitions in Eqs. \eqref{avgexample} and \eqref{rs2rq}, we can discretize the notation and obtain
		\begin{align}\label{Exp-avg}
			&\left\langle\exp\left\{i \mathbf{k} \cdot \int_{-\infty}^{\infty} \frac{d q^{\prime}}{2 \pi} \mathbf{R}\left(q^{\prime}\right)\left[\exp \left(i q^{\prime} s\right)-\exp \left(i q^{\prime} s^{\prime}\right)\right]\right\}\right\rangle_0 \\
			=& \frac{\int \prod_p d \mathbf{R}(p) \exp \left\{-\sum_{p=-\infty}^{\infty} \frac{\mathbf{R}^2(p)}{L g}+\frac{i \mathbf{k}}{L} \cdot \sum_{p=-\infty}^{\infty} \mathbf{R}(p)\left[\exp \left(\frac{2 \pi i p s}{L}\right)-\exp \left(\frac{2 \pi i p s^{\prime}}{L}\right)\right]\right\}}{\int \prod_p d \mathbf{R}(p) \exp \left[-\sum_{p=-\infty}^{\infty} \frac{\mathbf{R}^2(p)}{L g}\right]} \\
			=& \prod_p \frac{\int d \mathbf{R}(p) \exp \left\{-\frac{\mathbf{R}^2(p)}{L g}+\frac{i \mathbf{k}}{L} \cdot \mathbf{R}(p)\left[\exp \left(\frac{2 \pi i p s}{L}\right)-\exp \left(\frac{2 \pi i p s^{\prime}}{L}\right)\right]\right\}}{2 \pi(p) \exp \left[-\mathbf{R}^2(p) / L g\right]} \\
			=& \exp \left[-k^2 \int_{-\infty}^{\infty} \frac{d q}{2 \pi} g(q) \sin ^2\left(\frac{q\left|s-s^{\prime}\right|}{2}\right) \right].
		\end{align}
	\end{widetext}
	\section{$ \rho(\mathbf{r}) $}\label{appendix-section:rho-fourier}
	The segment density profile of the $i$-th chain, $\rho_{i}(\mathbf{r})$, is defined as $\int d s_{i} \delta\left(\mathbf{R}(s_{i})-\mathbf{r}\right)$ in Eq.\eqref{rho-ito-Rsi}. We assume that $\rho_{i}(\mathbf{r})$ can be approximated by a Gaussian distribution\cite{flory1950,tanford1961physical}, and at a position $\mathbf{r}$, the monomer density is given by:
	\begin{align}\label{rho-def-determination-of-rho}
		\rho_{i}(\mathbf{r}) &=A e^{-a \mathbf{(\left|\ro-\ro_{i}\right|)}^{2}},
	\end{align}
	where $A$ and $a$ are determined based on the definitions $ \mathcal{N}_{i} $ and $ R_{gi} $:
	\begin{align}\label{N-def-determination-of-rho}
		&\mathcal{N}_{i}=\int_{0}^{\infty} 4 \pi \mathbf{r}^{2} \rho_{i}(\mathbf{r})d\mathbf{r}=\frac{\pi^{3 / 2} A}{a^{3 / 2}},
	\end{align}
	\begin{align}\label{Rg-def-determination-of-rho}
		&R_{gi}^{2}=\frac{\int_{0}^{\infty} 4 \pi \mathbf{r}^{4} \rho_{i}(\mathbf{r}) d \mathbf{r}}{\int 4 \pi \mathbf{r}^{2} \rho_{i}(\mathbf{r}) d \mathbf{r}}=\frac{3 \pi^{3 / 2} A}{2 \mathcal{N}_{i} a^{5 / 2}},
	\end{align}
	respectively. $\mathcal{N}_{i}$ denotes the number of monomers inside the segmental density profiles. Solving Eq.\eqref{N-def-determination-of-rho} and\eqref{Rg-def-determination-of-rho}, we get
	\begin{align}
		&a=\frac{3}{2 R_{gi}^{2}},\qquad
		A=\mathcal{N}_{i}\left(\frac{3}{2 \pi R_{gi}^{2}}\right)^{3 / 2}.
	\end{align}
	Therefore, by putting these values of $a$ and $A$ back into Eq.\eqref{rho-def-determination-of-rho}, we obtain:
	\begin{align}
		&\rho_{i}(\ro)=\mathcal{N}_{i}\left(\frac{3}{4 \pi R_{gi}^{2}}\right)^{3 / 2} \exp \left[-\frac{3 (\left|\ro-\ro_{i}\right|)^{2}}{2 R_{gi}^{2}}\right].
		\label{rho-appendix}
	\end{align}
	The Fourier conjugate of $ \rho_{i}(\ro) $ is given by 
	\begin{align}\label{conjugate-rho}
		\mathcal{F}[\rho_{i}(\mathbf{r})]=\hat{\rho}_{i}(\mathbf{k})=\mathcal{N}_{i} \exp{\left(-\frac{k^{2} R_{gi}^{2}}{6}\right)} .
	\end{align}
	The monomer density distribution of the dangling parts of the chains is given by $ \hat{\rho}_{1}(\mathbf{k}) $ and $ \hat{\rho}_{2}(\mathbf{k}) $ respectively for PA and PC. The detailed form is given by
	\begin{align}\label{rho-def-each-component-1}
		\hat{\rho}_{1}(\mathbf{k})&=(N_{1}-n_{1}) \exp{\left(-i \mathbf{k}\cdot \mathbf{r}_{1}-\frac{\mathbf{k}^{2} R_{g1}^{2}}{6}\right)},
	\end{align}
	and
	\begin{align}\label{rho-def-each-component-2}
		\hat{\rho}_{2}(\mathbf{k})&=(N_{2}-n_{2}) \exp{\left(-i \mathbf{k}\cdot \mathbf{r}_{2}-\frac{\mathbf{k}^{2} R_{g2}^{2}}{6}\right)}.
	\end{align}
	\begin{widetext}
		\section{$ H_{ex} $}\label{appendix-section:Hex}
		To calculate the averaged excluded volume interaction, we need to move to the polymer coordinate system. We start by considering the inter-chain interaction term (last term) in Eq. \eqref{eq:H_ex},
		%
		\begin{align}
			^{12}H_{ex}\beta&=2\omega_{12}\ell^{3} \int_{0}^{N_{1}} ds_{1} \int_{0}^{N_{2}} ds_{2} \delta(\mathbf{R}(s_{1})-\mathbf{R}(s_{2}))\nonumber\\
			&=w_{12}\ell^{3}\int d s_{1} \int d s_{2} \left[\int d^{3} \mathbf{r} \int d^{3} \mathbf{r}^{\prime} \delta\left(\mathbf{R}\left(s_{1}\right)-\mathbf{r}\right) \delta\left(\mathbf{r}-\mathbf{r}^{\prime}\right)\delta\left(\mathbf{R}\left(s_{2}\right)-\mathbf{r}^{\prime}\right)\right]\nonumber\\
			&=w_{12}\ell^{3}\int d^{3} \mathbf{r} \int d^{3} \mathbf{r}^{\prime} \left[\int d s_{1} \delta(\mathbf{R}(s_{1})-\mathbf{r}) \delta(\mathbf{r}-\mathbf{r}^{\prime})\int d s_{2} \delta(\mathbf{R}(s_{2})-\mathbf{r}^\prime)\right].
		\end{align}
		In the above equation, we define, 
		\begin{align}\label{rho-ito-Rsi}
			\rho_{1}(\mathbf{r})=\int d s_{1} \delta\left(\mathbf{R}(s_{1})-\mathbf{r}\right) \text{ and } \rho_{2}(\mathbf{r}^{\prime})=\int d s_{2} \delta(\mathbf{R}(s_{2})-\mathbf{r}^{\prime}).
		\end{align}
		where $\rho_{1}(\mathbf{r})$ and $\rho_{2}(\mathbf{r})$ are the monomer density of PA and PC at position $\mathbf{r}$ and so on. Hence we write 
		\begin{align}
			^{12}H_{ex}\beta	&=w_{12}\ell^{3}\int d^{3} \mathbf{r} \int d^{3} \mathbf{r}^{\prime} \rho_{1}(\mathbf{r}) \delta\left(\mathbf{r}-\mathbf{r}^{\prime}\right) \rho_{2}\left(\mathbf{r}^{\prime}\right) =w_{12}\ell^{3}\int d^{3} \mathbf{r} \int d^{3} \mathbf{r}^{\prime} \rho_{1}(\mathbf{r})\left[\int \frac{d^{3} \mathbf{k}}{(2 \pi)^{3}} e^{-i \mathbf{k} \cdot\left(\mathbf{r}-\mathbf{r}^{\prime}\right)}\right] \rho_{2}\left(\mathbf{r}^{\prime}\right),
		\end{align}
		where the detailed form of $ \rho_{1}(\mathbf{r}) \text{ and }\rho_{2}(\mathbf{r})  $  is calculated assuming Gaussian segment distribution in the appendix. The Fourier conjugates of $ \rho_{1}(\mathbf{r}) \text{ and }\rho_{2}(\mathbf{r})  $ are given by Eqs. \eqref{rho-def-each-component-1} and \eqref{rho-def-each-component-2}. Centers of mass of the two chains are separated by a distance $ \mathbf{R}_{ex}=\mathbf{r}_{1}-\mathbf{r}_{2} $. Hence the integral becomes,
		\begin{align}\label{inter-chain-H-ex}
			^{12}H_{ex}\beta&=w_{12}\ell^{3}\int \frac{d^{3} \mathbf{k}}{(2 \pi)^{3}} \quad \hat{\rho}_{1}(-\mathbf{k}) \hat{\rho}_{2}(\mathbf{k}) e^{-i \mathbf{k} \cdot \mathbf{R}_{ex}}.
		\end{align}
		Therefore, in the case of self-excluded volume interaction of the polymers, the above equation becomes,
		\begin{align}\label{intra-chain-H-ex}
			&^{11}H_{ex}\beta=\omega_{11}\ell^{3}\int \frac{d^{3} \mathbf{k}}{(2 \pi)^{3}} \left|\hat{\rho}_{1}(\mathbf{k})\right|^{2}\text{, } ^{22}H_{ex}\beta=\omega_{22}\ell^{3}\int \frac{d^{3} \mathbf{k}}{(2 \pi)^{3}} \left|\hat{\rho}_{2}(\mathbf{k})\right|^{2}.
		\end{align}
		\section{$ H_{el} $}\label{appendix-section:Hel}
		Now that we have an explicit knowledge of the monomer density profile (given by Eq. \eqref{rho}), we use that to rewrite the electrostatic interaction. The Fourier conjugate of $ 	U_{ij}\left(\mathbf{r}-\mathbf{r}^{\prime}\right) $ can be obtained as,
		\begin{align}\label{conjugate-Uij}
			\hat{U}_{ij}^{el}(\mathbf{k})=\frac{f_{i}f_{j}\ell_{B}}{2}\frac{4\pi}{\mathbf{k}^2+\kappa^2}.
		\end{align}
		Using Eqs. \eqref{Yukawa-Uij}, \eqref{rho} and assuming $ \widetilde{R}_{gi} $ contains the excluded volume effect for the $ i $-th chain, we write the inter-chain electrostatic interaction (third term of Eq. \eqref{H_el}) as,
		\begin{align}\label{inter-chain-H-el}
			^{12}H_{el}(\mathbf{R})\beta=&\int d^{3} \mathbf{r}^{\prime} \int d^{3} \mathbf{r} \quad \rho_{1}\left(\mathbf{r}^{\prime}\right) U_{12}^{el}\left(\mathbf{r}-\mathbf{r}^{\prime}\right) \rho_{2}(\mathbf{r}) =\int d^{3} \mathbf{r}^{\prime} \int d^{3} \mathbf{r} \rho_{1}\left(\mathbf{r}^{\prime}\right)\left[\int \frac{d^{3} \mathbf{k}}{(2 \pi)^{3}} \hat{U}_{12}^{el}(\mathbf{k}) e^{-i \mathbf{k} \cdot\left(\mathbf{r}-\mathbf{r}^{\prime}\right)}\right] \rho_{2}(\mathbf{r})\nonumber\\
			&=\int \frac{d^{3} \mathbf{k}}{(2 \pi)^{3}} \hat{\rho}_{1}(\mathbf{k}) \hat{U}_{12}^{el}(\mathbf{k}) \hat{\rho}_{2}(-\mathbf{k}) e^{-i \mathbf{k}\cdot \mathbf{R}}.
		\end{align}	
		Therefore the intra-chain electrostatic interaction (first and second term of Eq. \eqref{H_el}) can be obtained as
		\begin{align}\label{intra-chain-H-el}
			^{11}H_{el}(\mathbf{R})\beta=\int \frac{d^{3} \mathbf{k}}{(2 \pi)^{3}} \left|\hat{\rho}_{1}(\mathbf{k})\right|^{2}\hat{U}_{11}^{el}(\mathbf{k})\text{ and } ^{22}H_{el}(\mathbf{R})\beta=\int \frac{d^{3} \mathbf{k}}{(2 \pi)^{3}} \left|\hat{\rho}_{2}(\mathbf{k})\right|^{2}\hat{U}_{22}^{el}(\mathbf{k}).
		\end{align}
	\section{$ F_{5} $ Derivation, uniform expansion method}\label{appendix-section:F5-derivation}
	The free energy formalism developed by Muthukumar \cite{muthu2004}and the Edwards-Singh\cite{edwards1979} method are two equivalent approaches for calculating the effective free energy of a polyelectrolyte chain. The variational equation used in the Edwards-Singh method directly estimates the derivative of the free energy with respect to the radius of gyration ($\partial F/\partial R_g$), which can be used to estimate the effective free energy of the chain through the method of thermodynamic integration\cite{kundu2014}. The calculations for the electrostatic interaction term are standard and detailed in \cite{muthu1987}, are outlined below
	\begin{align}
		&\int_{0}^{\mathcal{N}_{1}} ds_{1} \int_{0}^{\mathcal{N}_{1}} ds_{1}^\prime \int \frac{d^3 k}{(2\pi)^3} \hat{U}_{11}^{el}(\mathbf{k}) \langle e^{-i \mathbf{k}\cdot \left|\mathbf{R}(s_{1})-\mathbf{R}(s_{1}^\prime)\right| } \rangle_{0}\nonumber\\
		&=\frac{f_{1}^2 \ell_{B}}{2}\int_{0}^{\mathcal{N}_{1}} ds_{1} \int_{0}^{\mathcal{N}_{1}} ds_{1}^\prime \int \frac{4\pi k^2 dk}{(2\pi)^3}\frac{4\pi e^{-{k}^2\left|s_{1}-s_{1}^\prime\right|\ell_{11}/6}}{{k}^2+\kappa^2}\nonumber\\
		&=\frac{f_{1}^2 \ell_{B}}{2}\int_{0}^{\mathcal{N}_{1}} ds_{1} \int_{0}^{\mathcal{N}_{1}} ds_{1}^\prime \int_{0}^{\infty} \frac{d k}{(2\pi)^3}\left[4\pi k^2 e^{-{k}^2\left|s_{1}-s_{1}^\prime\right|\ell_{11}/6}\int_{0}^{\infty}e^{-x({k}^2+\kappa^2)}dx\right]\nonumber\\
		&=\frac{f_{1}^2 \ell_{B}}{2}\int_{0}^{\mathcal{N}_{1}} ds_{1} \int_{0}^{\mathcal{N}_{1}} ds_{1}^\prime\frac{\sqrt{\pi}}{2} \left[\sqrt{\frac{6}{\ell_{11}(s_{1}-s_{1}^\prime)}}-\sqrt{\pi } \kappa \exp{\left(\frac{\kappa^2 \ell_{11} (s_{1}-s_{1}^\prime)}{6}\right)} \text{erfc}\left(\sqrt{\frac{\kappa^2 \ell_{11} (s_{1}-s_{1}^\prime)}{6}}\right)\right].
	\end{align}
	After integrating over $ s_{1} \text{ and } s_{1}^{\prime} $, leads to the following result: 
	\begin{align}
		\beta F_{5}=2\sqrt{\frac{6}{\pi}}{f_{1}}^{2} \widetilde{\ell}_{B} \frac{\mathcal{N}_{1}^{3/2}}{\sqrt{\ell_{11}}} {\left[-\frac{\pi}{a_{1}^{3/2}}  \exp(a_{1}) \text{erfc}\left(\sqrt{a_{1}}\right)+\frac{\pi}{a_{1}^{1/2}}+\frac{\pi}{a_{1}^{3/2}}-\frac{2 \sqrt{\pi}}{a_{1}}\right]},
	\end{align}
	where $ a_{1}=\kappa^2 \ell_{11} \mathcal{N}_1/6 $.
	\end{widetext}

		\bibliography{manuscript2chain}

\begin{thebibliography}{117}%
\makeatletter
\providecommand \@ifxundefined [1]{%
 \@ifx{#1\undefined}
}%
\providecommand \@ifnum [1]{%
 \ifnum #1\expandafter \@firstoftwo
 \else \expandafter \@secondoftwo
 \fi
}%
\providecommand \@ifx [1]{%
 \ifx #1\expandafter \@firstoftwo
 \else \expandafter \@secondoftwo
 \fi
}%
\providecommand \natexlab [1]{#1}%
\providecommand \enquote  [1]{``#1''}%
\providecommand \bibnamefont  [1]{#1}%
\providecommand \bibfnamefont [1]{#1}%
\providecommand \citenamefont [1]{#1}%
\providecommand \href@noop [0]{\@secondoftwo}%
\providecommand \href [0]{\begingroup \@sanitize@url \@href}%
\providecommand \@href[1]{\@@startlink{#1}\@@href}%
\providecommand \@@href[1]{\endgroup#1\@@endlink}%
\providecommand \@sanitize@url [0]{\catcode `\\12\catcode `\$12\catcode
  `\&12\catcode `\#12\catcode `\^12\catcode `\_12\catcode `\%12\relax}%
\providecommand \@@startlink[1]{}%
\providecommand \@@endlink[0]{}%
\providecommand \url  [0]{\begingroup\@sanitize@url \@url }%
\providecommand \@url [1]{\endgroup\@href {#1}{\urlprefix }}%
\providecommand \urlprefix  [0]{URL }%
\providecommand \Eprint [0]{\href }%
\providecommand \doibase [0]{https://doi.org/}%
\providecommand \selectlanguage [0]{\@gobble}%
\providecommand \bibinfo  [0]{\@secondoftwo}%
\providecommand \bibfield  [0]{\@secondoftwo}%
\providecommand \translation [1]{[#1]}%
\providecommand \BibitemOpen [0]{}%
\providecommand \bibitemStop [0]{}%
\providecommand \bibitemNoStop [0]{.\EOS\space}%
\providecommand \EOS [0]{\spacefactor3000\relax}%
\providecommand \BibitemShut  [1]{\csname bibitem#1\endcsname}%
\let\auto@bib@innerbib\@empty
\bibitem [{\citenamefont {Overbeek}\ and\ \citenamefont
  {Voorn}(1957)}]{voorn1957}%
  \BibitemOpen
  \bibfield  {author} {\bibinfo {author} {\bibfnamefont {J.~T.~G.}\
  \bibnamefont {Overbeek}}\ and\ \bibinfo {author} {\bibfnamefont {M.~J.}\
  \bibnamefont {Voorn}},\ }\bibfield  {title} {\bibinfo {title} {Phase
  separation in polyelectrolyte solutions. theory of complex coacervation},\
  }\href {https://doi.org/10.1021/ja0276321} {\bibfield  {journal} {\bibinfo
  {journal} {J. {C}ell. {C}omp. {P}hysiol.}\ }\textbf {\bibinfo {volume}
  {49}},\ \bibinfo {pages} {7} (\bibinfo {year} {1957})}\BibitemShut {NoStop}%
\bibitem [{\citenamefont {Borue}\ and\ \citenamefont
  {Erukhimovich}(1988)}]{borue1988}%
  \BibitemOpen
  \bibfield  {author} {\bibinfo {author} {\bibfnamefont {V.~Y.}\ \bibnamefont
  {Borue}}\ and\ \bibinfo {author} {\bibfnamefont {I.~Y.}\ \bibnamefont
  {Erukhimovich}},\ }\bibfield  {title} {\bibinfo {title} {A statistical theory
  of weakly charged polyelectrolytes: fluctuations, equation of state and
  microphase separation},\ }\href {https://doi.org/10.1021/ma00189a019}
  {\bibfield  {journal} {\bibinfo  {journal} {Macromolecules}\ }\textbf
  {\bibinfo {volume} {21}},\ \bibinfo {pages} {3240} (\bibinfo {year}
  {1988})}\BibitemShut {NoStop}%
\bibitem [{\citenamefont {Borue}\ and\ \citenamefont
  {Erukhimovich}(1990)}]{borue1990}%
  \BibitemOpen
  \bibfield  {author} {\bibinfo {author} {\bibfnamefont {V.~Y.}\ \bibnamefont
  {Borue}}\ and\ \bibinfo {author} {\bibfnamefont {I.~Y.}\ \bibnamefont
  {Erukhimovich}},\ }\bibfield  {title} {\bibinfo {title} {A statistical theory
  of globular polyelectrolyte complexes},\ }\href
  {https://doi.org/10.1021/ma00217a015} {\bibfield  {journal} {\bibinfo
  {journal} {Macromolecules}\ }\textbf {\bibinfo {volume} {23}},\ \bibinfo
  {pages} {3625} (\bibinfo {year} {1990})},\ \Eprint
  {https://arxiv.org/abs/https://doi.org/10.1021/ma00217a015}
  {https://doi.org/10.1021/ma00217a015} \BibitemShut {NoStop}%
\bibitem [{\citenamefont {Mahdi}\ and\ \citenamefont {de~la
  Cruz}(2000)}]{mahdi2000}%
  \BibitemOpen
  \bibfield  {author} {\bibinfo {author} {\bibfnamefont {K.~A.}\ \bibnamefont
  {Mahdi}}\ and\ \bibinfo {author} {\bibfnamefont {M.~O.}\ \bibnamefont {de~la
  Cruz}},\ }\bibfield  {title} {\bibinfo {title} {Phase diagrams of salt-free
  polyelectrolyte semidilute solutions},\ }\href
  {https://doi.org/10.1021/ma000142d} {\bibfield  {journal} {\bibinfo
  {journal} {Macromolecules}\ }\textbf {\bibinfo {volume} {33}},\ \bibinfo
  {pages} {7649} (\bibinfo {year} {2000})}\BibitemShut {NoStop}%
\bibitem [{\citenamefont {Castelnovo}\ and\ \citenamefont
  {Joanny}(2001)}]{joanny2001}%
  \BibitemOpen
  \bibfield  {author} {\bibinfo {author} {\bibfnamefont {M.}~\bibnamefont
  {Castelnovo}}\ and\ \bibinfo {author} {\bibfnamefont {J.-F.}\ \bibnamefont
  {Joanny}},\ }\bibfield  {title} {\bibinfo {title} {Complexation between
  oppositely charged polyelectrolytes: Beyond the random phase approximation},\
  }\href {https://doi.org/10.1007/s10189-001-8051-7} {\bibfield  {journal}
  {\bibinfo  {journal} {EPJE}\ }\textbf {\bibinfo {volume} {6}},\ \bibinfo
  {pages} {377} (\bibinfo {year} {2001})}\BibitemShut {NoStop}%
\bibitem [{\citenamefont {Ermoshkin}\ and\ \citenamefont {Olvera de~la
  Cruz}(2003)}]{delacruz2003}%
  \BibitemOpen
  \bibfield  {author} {\bibinfo {author} {\bibfnamefont {A.~V.}\ \bibnamefont
  {Ermoshkin}}\ and\ \bibinfo {author} {\bibfnamefont {M.}~\bibnamefont {Olvera
  de~la Cruz}},\ }\bibfield  {title} {\bibinfo {title} {A modified random phase
  approximation of polyelectrolyte solutions},\ }\href
  {https://doi.org/10.1021/ma034148p} {\bibfield  {journal} {\bibinfo
  {journal} {Macromolecules}\ }\textbf {\bibinfo {volume} {36}},\ \bibinfo
  {pages} {7824} (\bibinfo {year} {2003})},\ \Eprint
  {https://arxiv.org/abs/https://doi.org/10.1021/ma034148p}
  {https://doi.org/10.1021/ma034148p} \BibitemShut {NoStop}%
\bibitem [{\citenamefont {Kudlay}\ \emph {et~al.}(2004)\citenamefont {Kudlay},
  \citenamefont {Ermoshkin},\ and\ \citenamefont {Olvera de~la
  Cruz}}]{delacruz2004}%
  \BibitemOpen
  \bibfield  {author} {\bibinfo {author} {\bibfnamefont {A.}~\bibnamefont
  {Kudlay}}, \bibinfo {author} {\bibfnamefont {A.~V.}\ \bibnamefont
  {Ermoshkin}},\ and\ \bibinfo {author} {\bibfnamefont {M.}~\bibnamefont
  {Olvera de~la Cruz}},\ }\bibfield  {title} {\bibinfo {title} {Complexation of
  oppositely charged polyelectrolytes: Effect of ion pair formation},\ }\href
  {https://doi.org/10.1021/ma048519t} {\bibfield  {journal} {\bibinfo
  {journal} {Macromolecules}\ }\textbf {\bibinfo {volume} {37}},\ \bibinfo
  {pages} {9231} (\bibinfo {year} {2004})},\ \Eprint
  {https://arxiv.org/abs/https://doi.org/10.1021/ma048519t}
  {https://doi.org/10.1021/ma048519t} \BibitemShut {NoStop}%
\bibitem [{\citenamefont {Kudlay}\ and\ \citenamefont {Olvera de~la
  Cruz}(2004)}]{kudlay2004}%
  \BibitemOpen
  \bibfield  {author} {\bibinfo {author} {\bibfnamefont {A.}~\bibnamefont
  {Kudlay}}\ and\ \bibinfo {author} {\bibfnamefont {M.}~\bibnamefont {Olvera
  de~la Cruz}},\ }\bibfield  {title} {\bibinfo {title} {Precipitation of
  oppositely charged polyelectrolytes in salt solutions},\ }\href
  {https://doi.org/10.1063/1.1629271} {\bibfield  {journal} {\bibinfo
  {journal} {JCP}\ }\textbf {\bibinfo {volume} {120}},\ \bibinfo {pages} {404}
  (\bibinfo {year} {2004})},\ \Eprint
  {https://arxiv.org/abs/https://doi.org/10.1063/1.1629271}
  {https://doi.org/10.1063/1.1629271} \BibitemShut {NoStop}%
\bibitem [{\citenamefont {Oskolkov}\ and\ \citenamefont
  {Potemkin}(2007)}]{oskolkov2007}%
  \BibitemOpen
  \bibfield  {author} {\bibinfo {author} {\bibfnamefont {N.~N.}\ \bibnamefont
  {Oskolkov}}\ and\ \bibinfo {author} {\bibfnamefont {I.~I.}\ \bibnamefont
  {Potemkin}},\ }\bibfield  {title} {\bibinfo {title} {Complexation in
  asymmetric solutions of oppositely charged
  polyelectrolytes:{\hspace{0.167em}} phase diagram},\ }\href
  {https://doi.org/10.1021/ma0709304} {\bibfield  {journal} {\bibinfo
  {journal} {Macromolecules}\ }\textbf {\bibinfo {volume} {40}},\ \bibinfo
  {pages} {8423} (\bibinfo {year} {2007})}\BibitemShut {NoStop}%
\bibitem [{\citenamefont {Perry}\ and\ \citenamefont {Sing}(2015)}]{perry2015}%
  \BibitemOpen
  \bibfield  {author} {\bibinfo {author} {\bibfnamefont {S.~L.}\ \bibnamefont
  {Perry}}\ and\ \bibinfo {author} {\bibfnamefont {C.~E.}\ \bibnamefont
  {Sing}},\ }\bibfield  {title} {\bibinfo {title} {Prism-based theory of
  complex coacervation: Excluded volume versus chain correlation},\ }\href
  {https://doi.org/10.1021/acs.macromol.5b01027} {\bibfield  {journal}
  {\bibinfo  {journal} {Macromolecules}\ }\textbf {\bibinfo {volume} {48}},\
  \bibinfo {pages} {5040} (\bibinfo {year} {2015})},\ \Eprint
  {https://arxiv.org/abs/https://doi.org/10.1021/acs.macromol.5b01027}
  {https://doi.org/10.1021/acs.macromol.5b01027} \BibitemShut {NoStop}%
\bibitem [{\citenamefont {Salehi}\ and\ \citenamefont
  {Larson}(2016)}]{salehi2016}%
  \BibitemOpen
  \bibfield  {author} {\bibinfo {author} {\bibfnamefont {A.}~\bibnamefont
  {Salehi}}\ and\ \bibinfo {author} {\bibfnamefont {R.~G.}\ \bibnamefont
  {Larson}},\ }\bibfield  {title} {\bibinfo {title} {A molecular thermodynamic
  model of complexation in mixtures of oppositely charged polyelectrolytes with
  explicit account of charge association/dissociation},\ }\href
  {https://doi.org/10.1021/acs.macromol.6b01464} {\bibfield  {journal}
  {\bibinfo  {journal} {Macromolecules}\ }\textbf {\bibinfo {volume} {49}},\
  \bibinfo {pages} {9706} (\bibinfo {year} {2016})},\ \Eprint
  {https://arxiv.org/abs/https://doi.org/10.1021/acs.macromol.6b01464}
  {https://doi.org/10.1021/acs.macromol.6b01464} \BibitemShut {NoStop}%
\bibitem [{\citenamefont {Sing}(2017)}]{sing2017}%
  \BibitemOpen
  \bibfield  {author} {\bibinfo {author} {\bibfnamefont {C.~E.}\ \bibnamefont
  {Sing}},\ }\bibfield  {title} {\bibinfo {title} {Development of the modern
  theory of polymeric complex coacervation},\ }\href
  {https://doi.org/https://doi.org/10.1016/j.cis.2016.04.004} {\bibfield
  {journal} {\bibinfo  {journal} {Advances in Colloid and Interface Science}\
  }\textbf {\bibinfo {volume} {239}},\ \bibinfo {pages} {2 } (\bibinfo {year}
  {2017})},\ \bibinfo {note} {complex Coacervation: Principles and
  Applications}\BibitemShut {NoStop}%
\bibitem [{\citenamefont {Lytle}\ and\ \citenamefont {Sing}(2017)}]{lytle2017}%
  \BibitemOpen
  \bibfield  {author} {\bibinfo {author} {\bibfnamefont {T.~K.}\ \bibnamefont
  {Lytle}}\ and\ \bibinfo {author} {\bibfnamefont {C.~E.}\ \bibnamefont
  {Sing}},\ }\bibfield  {title} {\bibinfo {title} {Transfer matrix theory of
  polymer complex coacervation},\ }\href {https://doi.org/10.1039/c7sm01080j}
  {\bibfield  {journal} {\bibinfo  {journal} {Soft Matter}\ }\textbf {\bibinfo
  {volume} {13}},\ \bibinfo {pages} {7001} (\bibinfo {year}
  {2017})}\BibitemShut {NoStop}%
\bibitem [{\citenamefont {Adhikari}\ \emph {et~al.}(2018)\citenamefont
  {Adhikari}, \citenamefont {Leaf},\ and\ \citenamefont
  {Muthukumar}}]{adhikari2018}%
  \BibitemOpen
  \bibfield  {author} {\bibinfo {author} {\bibfnamefont {S.}~\bibnamefont
  {Adhikari}}, \bibinfo {author} {\bibfnamefont {M.~A.}\ \bibnamefont {Leaf}},\
  and\ \bibinfo {author} {\bibfnamefont {M.}~\bibnamefont {Muthukumar}},\
  }\bibfield  {title} {\bibinfo {title} {Polyelectrolyte complex coacervation
  by electrostatic dipolar interactions},\ }\href
  {https://doi.org/10.1063/1.5029268} {\bibfield  {journal} {\bibinfo
  {journal} {JCP}\ }\textbf {\bibinfo {volume} {149}},\ \bibinfo {pages}
  {163308} (\bibinfo {year} {2018})}\BibitemShut {NoStop}%
\bibitem [{\citenamefont {Rumyantsev}\ and\ \citenamefont
  {Potemkin}(2017)}]{potemkin2017}%
  \BibitemOpen
  \bibfield  {author} {\bibinfo {author} {\bibfnamefont {A.~M.}\ \bibnamefont
  {Rumyantsev}}\ and\ \bibinfo {author} {\bibfnamefont {I.~I.}\ \bibnamefont
  {Potemkin}},\ }\bibfield  {title} {\bibinfo {title} {Explicit description of
  complexation between oppositely charged polyelectrolytes as an advantage of
  the random phase approximation over the scaling approach},\ }\href
  {https://doi.org/10.1039/C7CP05300B} {\bibfield  {journal} {\bibinfo
  {journal} {PCCP}\ }\textbf {\bibinfo {volume} {19}},\ \bibinfo {pages}
  {27580} (\bibinfo {year} {2017})}\BibitemShut {NoStop}%
\bibitem [{\citenamefont {Rubinstein}\ \emph {et~al.}(2018)\citenamefont
  {Rubinstein}, \citenamefont {Liao},\ and\ \citenamefont
  {Panyukov}}]{panyukov2018}%
  \BibitemOpen
  \bibfield  {author} {\bibinfo {author} {\bibfnamefont {M.}~\bibnamefont
  {Rubinstein}}, \bibinfo {author} {\bibfnamefont {Q.}~\bibnamefont {Liao}},\
  and\ \bibinfo {author} {\bibfnamefont {S.}~\bibnamefont {Panyukov}},\
  }\bibfield  {title} {\bibinfo {title} {Structure of liquid coacervates formed
  by oppositely charged polyelectrolytes},\ }\href
  {https://doi.org/10.1021/acs.macromol.8b02059} {\bibfield  {journal}
  {\bibinfo  {journal} {Macromolecules}\ }\textbf {\bibinfo {volume} {51}},\
  \bibinfo {pages} {9572} (\bibinfo {year} {2018})}\BibitemShut {NoStop}%
\bibitem [{\citenamefont {Lytle}\ \emph {et~al.}(2019)\citenamefont {Lytle},
  \citenamefont {Chang}, \citenamefont {Markiewicz}, \citenamefont {Perry},\
  and\ \citenamefont {Sing}}]{lytle2019}%
  \BibitemOpen
  \bibfield  {author} {\bibinfo {author} {\bibfnamefont {T.~K.}\ \bibnamefont
  {Lytle}}, \bibinfo {author} {\bibfnamefont {L.-W.}\ \bibnamefont {Chang}},
  \bibinfo {author} {\bibfnamefont {N.}~\bibnamefont {Markiewicz}}, \bibinfo
  {author} {\bibfnamefont {S.~L.}\ \bibnamefont {Perry}},\ and\ \bibinfo
  {author} {\bibfnamefont {C.~E.}\ \bibnamefont {Sing}},\ }\bibfield  {title}
  {\bibinfo {title} {Designing electrostatic interactions via polyelectrolyte
  monomer sequence},\ }\href {https://doi.org/10.1021/acscentsci.9b00087}
  {\bibfield  {journal} {\bibinfo  {journal} {{ACS} Central Science}\ }\textbf
  {\bibinfo {volume} {5}},\ \bibinfo {pages} {709} (\bibinfo {year}
  {2019})}\BibitemShut {NoStop}%
\bibitem [{\citenamefont {Zhang}\ \emph {et~al.}(2018)\citenamefont {Zhang},
  \citenamefont {Alsaifi}, \citenamefont {Wu},\ and\ \citenamefont
  {Wang}}]{zhang2018}%
  \BibitemOpen
  \bibfield  {author} {\bibinfo {author} {\bibfnamefont {P.}~\bibnamefont
  {Zhang}}, \bibinfo {author} {\bibfnamefont {N.~M.}\ \bibnamefont {Alsaifi}},
  \bibinfo {author} {\bibfnamefont {J.}~\bibnamefont {Wu}},\ and\ \bibinfo
  {author} {\bibfnamefont {Z.-G.}\ \bibnamefont {Wang}},\ }\bibfield  {title}
  {\bibinfo {title} {Polyelectrolyte complex coacervation: Effects of
  concentration asymmetry},\ }\href {https://doi.org/10.1063/1.5028524}
  {\bibfield  {journal} {\bibinfo  {journal} {JCP}\ }\textbf {\bibinfo {volume}
  {149}},\ \bibinfo {pages} {163303} (\bibinfo {year} {2018})}\BibitemShut
  {NoStop}%
\bibitem [{\citenamefont {Ylitalo}\ \emph {et~al.}(2021)\citenamefont
  {Ylitalo}, \citenamefont {Balzer}, \citenamefont {Zhang},\ and\ \citenamefont
  {Wang}}]{ylitalo2021}%
  \BibitemOpen
  \bibfield  {author} {\bibinfo {author} {\bibfnamefont {A.~S.}\ \bibnamefont
  {Ylitalo}}, \bibinfo {author} {\bibfnamefont {C.}~\bibnamefont {Balzer}},
  \bibinfo {author} {\bibfnamefont {P.}~\bibnamefont {Zhang}},\ and\ \bibinfo
  {author} {\bibfnamefont {Z.-G.}\ \bibnamefont {Wang}},\ }\bibfield  {title}
  {\bibinfo {title} {Electrostatic correlations and temperature-dependent
  dielectric constant can model {LCST} in polyelectrolyte complex
  coacervation},\ }\href {https://doi.org/10.1021/acs.macromol.1c02000}
  {\bibfield  {journal} {\bibinfo  {journal} {Macromolecules}\ }\textbf
  {\bibinfo {volume} {54}},\ \bibinfo {pages} {11326} (\bibinfo {year}
  {2021})}\BibitemShut {NoStop}%
\bibitem [{\citenamefont {Wang}\ \emph {et~al.}(2019)\citenamefont {Wang},
  \citenamefont {Xu},\ and\ \citenamefont {Zhao}}]{wang-zhao2019}%
  \BibitemOpen
  \bibfield  {author} {\bibinfo {author} {\bibfnamefont {F.}~\bibnamefont
  {Wang}}, \bibinfo {author} {\bibfnamefont {X.}~\bibnamefont {Xu}},\ and\
  \bibinfo {author} {\bibfnamefont {S.}~\bibnamefont {Zhao}},\ }\bibfield
  {title} {\bibinfo {title} {Complex coacervation in asymmetric solutions of
  polycation and polyanion},\ }\href
  {https://doi.org/10.1021/acs.langmuir.9b02787} {\bibfield  {journal}
  {\bibinfo  {journal} {Langmuir}\ }\textbf {\bibinfo {volume} {35}},\ \bibinfo
  {pages} {15267} (\bibinfo {year} {2019})}\BibitemShut {NoStop}%
\bibitem [{\citenamefont {Rumyantsev}\ \emph {et~al.}(2018)\citenamefont
  {Rumyantsev}, \citenamefont {Kramarenko},\ and\ \citenamefont
  {Borisov}}]{rumyantsev2018}%
  \BibitemOpen
  \bibfield  {author} {\bibinfo {author} {\bibfnamefont {A.~M.}\ \bibnamefont
  {Rumyantsev}}, \bibinfo {author} {\bibfnamefont {E.~Y.}\ \bibnamefont
  {Kramarenko}},\ and\ \bibinfo {author} {\bibfnamefont {O.~V.}\ \bibnamefont
  {Borisov}},\ }\bibfield  {title} {\bibinfo {title} {Microphase separation in
  complex coacervate due to incompatibility between polyanion and polycation},\
  }\href {https://doi.org/10.1021/acs.macromol.8b00721} {\bibfield  {journal}
  {\bibinfo  {journal} {Macromolecules}\ }\textbf {\bibinfo {volume} {51}},\
  \bibinfo {pages} {6587} (\bibinfo {year} {2018})}\BibitemShut {NoStop}%
\bibitem [{\citenamefont {Rumyantsev}\ \emph
  {et~al.}(2019{\natexlab{a}})\citenamefont {Rumyantsev}, \citenamefont
  {Jackson}, \citenamefont {Yu}, \citenamefont {Ting}, \citenamefont {Chen},
  \citenamefont {Tirrell},\ and\ \citenamefont {de~Pablo}}]{rumyantsev2019}%
  \BibitemOpen
  \bibfield  {author} {\bibinfo {author} {\bibfnamefont {A.~M.}\ \bibnamefont
  {Rumyantsev}}, \bibinfo {author} {\bibfnamefont {N.~E.}\ \bibnamefont
  {Jackson}}, \bibinfo {author} {\bibfnamefont {B.}~\bibnamefont {Yu}},
  \bibinfo {author} {\bibfnamefont {J.~M.}\ \bibnamefont {Ting}}, \bibinfo
  {author} {\bibfnamefont {W.}~\bibnamefont {Chen}}, \bibinfo {author}
  {\bibfnamefont {M.~V.}\ \bibnamefont {Tirrell}},\ and\ \bibinfo {author}
  {\bibfnamefont {J.~J.}\ \bibnamefont {de~Pablo}},\ }\bibfield  {title}
  {\bibinfo {title} {Controlling complex coacervation via random
  polyelectrolyte sequences},\ }\href
  {https://doi.org/10.1021/acsmacrolett.9b00494} {\bibfield  {journal}
  {\bibinfo  {journal} {{ACS} Macro Letters}\ }\textbf {\bibinfo {volume}
  {8}},\ \bibinfo {pages} {1296} (\bibinfo {year}
  {2019}{\natexlab{a}})}\BibitemShut {NoStop}%
\bibitem [{\citenamefont {Chen}\ \emph
  {et~al.}(2021{\natexlab{a}})\citenamefont {Chen}, \citenamefont {Chen},
  \citenamefont {Shi},\ and\ \citenamefont {Yang}}]{chen-yang2021}%
  \BibitemOpen
  \bibfield  {author} {\bibinfo {author} {\bibfnamefont {X.}~\bibnamefont
  {Chen}}, \bibinfo {author} {\bibfnamefont {E.-Q.}\ \bibnamefont {Chen}},
  \bibinfo {author} {\bibfnamefont {A.-C.}\ \bibnamefont {Shi}},\ and\ \bibinfo
  {author} {\bibfnamefont {S.}~\bibnamefont {Yang}},\ }\bibfield  {title}
  {\bibinfo {title} {Multiphase coacervates driven by electrostatic
  correlations},\ }\href {https://doi.org/10.1021/acsmacrolett.1c00282}
  {\bibfield  {journal} {\bibinfo  {journal} {{ACS} Macro Letters}\ }\textbf
  {\bibinfo {volume} {10}},\ \bibinfo {pages} {1041} (\bibinfo {year}
  {2021}{\natexlab{a}})}\BibitemShut {NoStop}%
\bibitem [{\citenamefont {Knoerdel}\ \emph {et~al.}(2021)\citenamefont
  {Knoerdel}, \citenamefont {McTigue},\ and\ \citenamefont
  {Sing}}]{knoerdel2021}%
  \BibitemOpen
  \bibfield  {author} {\bibinfo {author} {\bibfnamefont {A.~R.}\ \bibnamefont
  {Knoerdel}}, \bibinfo {author} {\bibfnamefont {W.~C.~B.}\ \bibnamefont
  {McTigue}},\ and\ \bibinfo {author} {\bibfnamefont {C.~E.}\ \bibnamefont
  {Sing}},\ }\bibfield  {title} {\bibinfo {title} {Transfer matrix model of
  {pH} effects in polymeric complex coacervation},\ }\href
  {https://doi.org/10.1021/acs.jpcb.1c03065} {\bibfield  {journal} {\bibinfo
  {journal} {JPCB}\ }\textbf {\bibinfo {volume} {125}},\ \bibinfo {pages}
  {8965} (\bibinfo {year} {2021})}\BibitemShut {NoStop}%
\bibitem [{\citenamefont {Sayko}\ \emph {et~al.}(2021)\citenamefont {Sayko},
  \citenamefont {Tian}, \citenamefont {Liang},\ and\ \citenamefont
  {Dobrynin}}]{sayko2021}%
  \BibitemOpen
  \bibfield  {author} {\bibinfo {author} {\bibfnamefont {R.}~\bibnamefont
  {Sayko}}, \bibinfo {author} {\bibfnamefont {Y.}~\bibnamefont {Tian}},
  \bibinfo {author} {\bibfnamefont {H.}~\bibnamefont {Liang}},\ and\ \bibinfo
  {author} {\bibfnamefont {A.~V.}\ \bibnamefont {Dobrynin}},\ }\bibfield
  {title} {\bibinfo {title} {Charged polymers: From polyelectrolyte solutions
  to polyelectrolyte complexes},\ }\href
  {https://doi.org/10.1021/acs.macromol.1c01171} {\bibfield  {journal}
  {\bibinfo  {journal} {Macromolecules}\ }\textbf {\bibinfo {volume} {54}},\
  \bibinfo {pages} {7183} (\bibinfo {year} {2021})}\BibitemShut {NoStop}%
\bibitem [{\citenamefont {Mitra}\ and\ \citenamefont
  {Kundagrami}(2023)}]{mitra2023}%
  \BibitemOpen
  \bibfield  {author} {\bibinfo {author} {\bibfnamefont {S.}~\bibnamefont
  {Mitra}}\ and\ \bibinfo {author} {\bibfnamefont {A.}~\bibnamefont
  {Kundagrami}},\ }\bibfield  {title} {\bibinfo {title} {Polyelectrolyte
  complexation of two oppositely charged symmetric polymers: A minimal
  theory},\ }\href {https://doi.org/10.1063/5.0128904} {\bibfield  {journal}
  {\bibinfo  {journal} {JCP}\ }\textbf {\bibinfo {volume} {158}},\ \bibinfo
  {pages} {014904} (\bibinfo {year} {2023})}\BibitemShut {NoStop}%
\bibitem [{\citenamefont {Ghosh}\ \emph {et~al.}(2023)\citenamefont {Ghosh},
  \citenamefont {Mitra},\ and\ \citenamefont {Kundagrami}}]{Ghosh2023}%
  \BibitemOpen
  \bibfield  {author} {\bibinfo {author} {\bibfnamefont {S.}~\bibnamefont
  {Ghosh}}, \bibinfo {author} {\bibfnamefont {S.}~\bibnamefont {Mitra}},\ and\
  \bibinfo {author} {\bibfnamefont {A.}~\bibnamefont {Kundagrami}},\ }\bibfield
   {title} {\bibinfo {title} {Polymer complexation: Partially ionizable
  asymmetric polyelectrolytes},\ }\bibfield  {journal} {\bibinfo  {journal}
  {The Journal of Chemical Physics}\ }\textbf {\bibinfo {volume} {158}},\ \href
  {https://doi.org/10.1063/5.0147323} {10.1063/5.0147323} (\bibinfo {year}
  {2023})\BibitemShut {NoStop}%
\bibitem [{\citenamefont {Chen}\ and\ \citenamefont
  {Wang}(2022)}]{chen2022-PNAS}%
  \BibitemOpen
  \bibfield  {author} {\bibinfo {author} {\bibfnamefont {S.}~\bibnamefont
  {Chen}}\ and\ \bibinfo {author} {\bibfnamefont {Z.-G.}\ \bibnamefont
  {Wang}},\ }\bibfield  {title} {\bibinfo {title} {Driving force and pathway in
  polyelectrolyte complex coacervation},\ }\bibfield  {journal} {\bibinfo
  {journal} {Proceedings of the National Academy of Sciences}\ }\textbf
  {\bibinfo {volume} {119}},\ \href {https://doi.org/10.1073/pnas.2209975119}
  {10.1073/pnas.2209975119} (\bibinfo {year} {2022})\BibitemShut {NoStop}%
\bibitem [{\citenamefont {Record}\ \emph {et~al.}(1978)\citenamefont {Record},
  \citenamefont {Anderson},\ and\ \citenamefont {Lohman}}]{record1978}%
  \BibitemOpen
  \bibfield  {author} {\bibinfo {author} {\bibfnamefont {M.~T.}\ \bibnamefont
  {Record}}, \bibinfo {author} {\bibfnamefont {C.~F.}\ \bibnamefont
  {Anderson}},\ and\ \bibinfo {author} {\bibfnamefont {T.~M.}\ \bibnamefont
  {Lohman}},\ }\bibfield  {title} {\bibinfo {title} {Thermodynamic analysis of
  ion effects on the binding and conformational equilibria of proteins and
  nucleic acids: the roles of ion association or release, screening, and ion
  effects on water activity},\ }\href
  {https://doi.org/10.1017/s003358350000202x} {\bibfield  {journal} {\bibinfo
  {journal} {Quarterly Reviews of Biophysics}\ }\textbf {\bibinfo {volume}
  {11}},\ \bibinfo {pages} {103} (\bibinfo {year} {1978})}\BibitemShut
  {NoStop}%
\bibitem [{\citenamefont {Kabanov}\ \emph {et~al.}(1985)\citenamefont
  {Kabanov}, \citenamefont {Zezin}, \citenamefont {Izumrudov}, \citenamefont
  {Bronich},\ and\ \citenamefont {Bakeev}}]{kabanov1985}%
  \BibitemOpen
  \bibfield  {author} {\bibinfo {author} {\bibfnamefont {V.~A.}\ \bibnamefont
  {Kabanov}}, \bibinfo {author} {\bibfnamefont {A.~B.}\ \bibnamefont {Zezin}},
  \bibinfo {author} {\bibfnamefont {V.~A.}\ \bibnamefont {Izumrudov}}, \bibinfo
  {author} {\bibfnamefont {T.~K.}\ \bibnamefont {Bronich}},\ and\ \bibinfo
  {author} {\bibfnamefont {K.~N.}\ \bibnamefont {Bakeev}},\ }\bibfield  {title}
  {\bibinfo {title} {Cooperative interpolyelectrolyte reactions},\ }\href
  {https://doi.org/10.1002/macp.1985.020131985111} {\bibfield  {journal}
  {\bibinfo  {journal} {Die Makromolekulare Chemie}\ }\textbf {\bibinfo
  {volume} {13}},\ \bibinfo {pages} {137} (\bibinfo {year} {1985})}\BibitemShut
  {NoStop}%
\bibitem [{\citenamefont {Dautzenberg}\ and\ \citenamefont
  {Jaeger}(2002)}]{dautzenberg2002}%
  \BibitemOpen
  \bibfield  {author} {\bibinfo {author} {\bibfnamefont {H.}~\bibnamefont
  {Dautzenberg}}\ and\ \bibinfo {author} {\bibfnamefont {W.}~\bibnamefont
  {Jaeger}},\ }\bibfield  {title} {\bibinfo {title} {Effect of charge density
  on the formation and salt stability of polyelectrolyte complexes},\ }\href
  {https://doi.org/10.1002/1521-3935(200210)203:14<2095::aid-macp2095>3.0.co;2-9}
  {\bibfield  {journal} {\bibinfo  {journal} {Macromolecular Chemistry and
  Physics}\ }\textbf {\bibinfo {volume} {203}},\ \bibinfo {pages} {2095}
  (\bibinfo {year} {2002})}\BibitemShut {NoStop}%
\bibitem [{\citenamefont {Cousin}\ \emph {et~al.}(2005)\citenamefont {Cousin},
  \citenamefont {Gummel}, \citenamefont {Ung},\ and\ \citenamefont
  {Boue}}]{cousin2005}%
  \BibitemOpen
  \bibfield  {author} {\bibinfo {author} {\bibfnamefont {F.}~\bibnamefont
  {Cousin}}, \bibinfo {author} {\bibfnamefont {J.}~\bibnamefont {Gummel}},
  \bibinfo {author} {\bibfnamefont {D.}~\bibnamefont {Ung}},\ and\ \bibinfo
  {author} {\bibfnamefont {F.}~\bibnamefont {Boue}},\ }\bibfield  {title}
  {\bibinfo {title} {Polyelectrolyte-protein complexes: Structure and
  conformation of each species revealed by sans},\ }\href
  {https://doi.org/10.1021/la0510174} {\bibfield  {journal} {\bibinfo
  {journal} {Langmuir}\ }\textbf {\bibinfo {volume} {21}},\ \bibinfo {pages}
  {9675} (\bibinfo {year} {2005})},\ \bibinfo {note} {pMID: 16207052},\ \Eprint
  {https://arxiv.org/abs/https://doi.org/10.1021/la0510174}
  {https://doi.org/10.1021/la0510174} \BibitemShut {NoStop}%
\bibitem [{\citenamefont {Gummel}\ \emph {et~al.}(2007)\citenamefont {Gummel},
  \citenamefont {Cousin},\ and\ \citenamefont {Bou{\'{e}}}}]{gummel2007}%
  \BibitemOpen
  \bibfield  {author} {\bibinfo {author} {\bibfnamefont {J.}~\bibnamefont
  {Gummel}}, \bibinfo {author} {\bibfnamefont {F.}~\bibnamefont {Cousin}},\
  and\ \bibinfo {author} {\bibfnamefont {F.}~\bibnamefont {Bou{\'{e}}}},\
  }\bibfield  {title} {\bibinfo {title} {Counterions release from electrostatic
  complexes of polyelectrolytes and proteins of opposite
  charge:{\hspace{0.167em}} a direct measurement},\ }\href
  {https://doi.org/10.1021/ja070414t} {\bibfield  {journal} {\bibinfo
  {journal} {JACS}\ }\textbf {\bibinfo {volume} {129}},\ \bibinfo {pages}
  {5806} (\bibinfo {year} {2007})}\BibitemShut {NoStop}%
\bibitem [{\citenamefont {Porcel}\ and\ \citenamefont
  {Schlenoff}(2009)}]{porcel2009}%
  \BibitemOpen
  \bibfield  {author} {\bibinfo {author} {\bibfnamefont {C.~H.}\ \bibnamefont
  {Porcel}}\ and\ \bibinfo {author} {\bibfnamefont {J.~B.}\ \bibnamefont
  {Schlenoff}},\ }\bibfield  {title} {\bibinfo {title} {Compact polyelectrolyte
  complexes: {\textquotedblleft}saloplastic{\textquotedblright} candidates for
  biomaterials},\ }\href {https://doi.org/10.1021/bm900373c} {\bibfield
  {journal} {\bibinfo  {journal} {Biomacromolecules}\ }\textbf {\bibinfo
  {volume} {10}},\ \bibinfo {pages} {2968} (\bibinfo {year}
  {2009})}\BibitemShut {NoStop}%
\bibitem [{\citenamefont {Spruijt}\ \emph {et~al.}(2010)\citenamefont
  {Spruijt}, \citenamefont {Westphal}, \citenamefont {Borst}, \citenamefont
  {Stuart},\ and\ \citenamefont {van~der Gucht}}]{spruijt2010macro}%
  \BibitemOpen
  \bibfield  {author} {\bibinfo {author} {\bibfnamefont {E.}~\bibnamefont
  {Spruijt}}, \bibinfo {author} {\bibfnamefont {A.~H.}\ \bibnamefont
  {Westphal}}, \bibinfo {author} {\bibfnamefont {J.~W.}\ \bibnamefont {Borst}},
  \bibinfo {author} {\bibfnamefont {M.~A.~C.}\ \bibnamefont {Stuart}},\ and\
  \bibinfo {author} {\bibfnamefont {J.}~\bibnamefont {van~der Gucht}},\
  }\bibfield  {title} {\bibinfo {title} {Binodal compositions of
  polyelectrolyte complexes},\ }\href {https://doi.org/10.1021/ma101031t}
  {\bibfield  {journal} {\bibinfo  {journal} {Macromolecules}\ }\textbf
  {\bibinfo {volume} {43}},\ \bibinfo {pages} {6476} (\bibinfo {year}
  {2010})}\BibitemShut {NoStop}%
\bibitem [{\citenamefont {Chollakup}\ \emph {et~al.}(2010)\citenamefont
  {Chollakup}, \citenamefont {Smitthipong}, \citenamefont {Eisenbach},\ and\
  \citenamefont {Tirrell}}]{chollakup2010}%
  \BibitemOpen
  \bibfield  {author} {\bibinfo {author} {\bibfnamefont {R.}~\bibnamefont
  {Chollakup}}, \bibinfo {author} {\bibfnamefont {W.}~\bibnamefont
  {Smitthipong}}, \bibinfo {author} {\bibfnamefont {C.~D.}\ \bibnamefont
  {Eisenbach}},\ and\ \bibinfo {author} {\bibfnamefont {M.}~\bibnamefont
  {Tirrell}},\ }\bibfield  {title} {\bibinfo {title} {Phase behavior and
  coacervation of aqueous poly(acrylic acid)-poly(allylamine) solutions},\
  }\href {https://doi.org/10.1021/ma902144k} {\bibfield  {journal} {\bibinfo
  {journal} {Macromolecules}\ }\textbf {\bibinfo {volume} {43}},\ \bibinfo
  {pages} {2518} (\bibinfo {year} {2010})}\BibitemShut {NoStop}%
\bibitem [{\citenamefont {van~der Gucht}\ \emph {et~al.}(2011)\citenamefont
  {van~der Gucht}, \citenamefont {Spruijt}, \citenamefont {Lemmers},\ and\
  \citenamefont {Stuart}}]{gucht2011}%
  \BibitemOpen
  \bibfield  {author} {\bibinfo {author} {\bibfnamefont {J.}~\bibnamefont
  {van~der Gucht}}, \bibinfo {author} {\bibfnamefont {E.}~\bibnamefont
  {Spruijt}}, \bibinfo {author} {\bibfnamefont {M.}~\bibnamefont {Lemmers}},\
  and\ \bibinfo {author} {\bibfnamefont {M.~A.~C.}\ \bibnamefont {Stuart}},\
  }\bibfield  {title} {\bibinfo {title} {Polyelectrolyte complexes: Bulk phases
  and colloidal systems},\ }\href {https://doi.org/10.1016/j.jcis.2011.05.080}
  {\bibfield  {journal} {\bibinfo  {journal} {Journal of Colloid and Interface
  Science}\ }\textbf {\bibinfo {volume} {361}},\ \bibinfo {pages} {407}
  (\bibinfo {year} {2011})}\BibitemShut {NoStop}%
\bibitem [{\citenamefont {Priftis}\ and\ \citenamefont
  {Tirrell}(2012{\natexlab{a}})}]{tirrell2012}%
  \BibitemOpen
  \bibfield  {author} {\bibinfo {author} {\bibfnamefont {D.}~\bibnamefont
  {Priftis}}\ and\ \bibinfo {author} {\bibfnamefont {M.}~\bibnamefont
  {Tirrell}},\ }\bibfield  {title} {\bibinfo {title} {Phase behaviour and
  complex coacervation of aqueous polypeptide solutions},\ }\href
  {https://doi.org/10.1039/C2SM25604E} {\bibfield  {journal} {\bibinfo
  {journal} {Soft Matter}\ }\textbf {\bibinfo {volume} {8}},\ \bibinfo {pages}
  {9396} (\bibinfo {year} {2012}{\natexlab{a}})}\BibitemShut {NoStop}%
\bibitem [{\citenamefont {Lemmers}\ \emph {et~al.}(2012)\citenamefont
  {Lemmers}, \citenamefont {Spruijt}, \citenamefont {Beun}, \citenamefont
  {Fokkink}, \citenamefont {Leermakers}, \citenamefont {Portale}, \citenamefont
  {Stuart},\ and\ \citenamefont {van~der Gucht}}]{lemmers2012}%
  \BibitemOpen
  \bibfield  {author} {\bibinfo {author} {\bibfnamefont {M.}~\bibnamefont
  {Lemmers}}, \bibinfo {author} {\bibfnamefont {E.}~\bibnamefont {Spruijt}},
  \bibinfo {author} {\bibfnamefont {L.}~\bibnamefont {Beun}}, \bibinfo {author}
  {\bibfnamefont {R.}~\bibnamefont {Fokkink}}, \bibinfo {author} {\bibfnamefont
  {F.}~\bibnamefont {Leermakers}}, \bibinfo {author} {\bibfnamefont
  {G.}~\bibnamefont {Portale}}, \bibinfo {author} {\bibfnamefont {M.~A.~C.}\
  \bibnamefont {Stuart}},\ and\ \bibinfo {author} {\bibfnamefont
  {J.}~\bibnamefont {van~der Gucht}},\ }\bibfield  {title} {\bibinfo {title}
  {The influence of charge ratio on transient networks of polyelectrolyte
  complex micelles},\ }\href {https://doi.org/10.1039/c1sm06281f} {\bibfield
  {journal} {\bibinfo  {journal} {Soft Matter}\ }\textbf {\bibinfo {volume}
  {8}},\ \bibinfo {pages} {104} (\bibinfo {year} {2012})}\BibitemShut {NoStop}%
\bibitem [{\citenamefont {Wang}\ \emph {et~al.}(2012)\citenamefont {Wang},
  \citenamefont {Stuart},\ and\ \citenamefont {van~der Gucht}}]{gucht2012}%
  \BibitemOpen
  \bibfield  {author} {\bibinfo {author} {\bibfnamefont {J.}~\bibnamefont
  {Wang}}, \bibinfo {author} {\bibfnamefont {M.~A.~C.}\ \bibnamefont
  {Stuart}},\ and\ \bibinfo {author} {\bibfnamefont {J.}~\bibnamefont {van~der
  Gucht}},\ }\bibfield  {title} {\bibinfo {title} {Phase diagram of coacervate
  complexes containing reversible coordination structures},\ }\href
  {https://doi.org/10.1021/ma301690t} {\bibfield  {journal} {\bibinfo
  {journal} {Macromolecules}\ }\textbf {\bibinfo {volume} {45}},\ \bibinfo
  {pages} {8903} (\bibinfo {year} {2012})}\BibitemShut {NoStop}%
\bibitem [{\citenamefont {Chollakup}\ \emph {et~al.}(2013)\citenamefont
  {Chollakup}, \citenamefont {Beck}, \citenamefont {Dirnberger}, \citenamefont
  {Tirrell},\ and\ \citenamefont {Eisenbach}}]{tirrell2013}%
  \BibitemOpen
  \bibfield  {author} {\bibinfo {author} {\bibfnamefont {R.}~\bibnamefont
  {Chollakup}}, \bibinfo {author} {\bibfnamefont {J.~B.}\ \bibnamefont {Beck}},
  \bibinfo {author} {\bibfnamefont {K.}~\bibnamefont {Dirnberger}}, \bibinfo
  {author} {\bibfnamefont {M.}~\bibnamefont {Tirrell}},\ and\ \bibinfo {author}
  {\bibfnamefont {C.~D.}\ \bibnamefont {Eisenbach}},\ }\bibfield  {title}
  {\bibinfo {title} {Polyelectrolyte molecular weight and salt effects on the
  phase behavior and coacervation of aqueous solutions of poly(acrylic acid)
  sodium salt and poly(allylamine) hydrochloride},\ }\href
  {https://doi.org/10.1021/ma202172q} {\bibfield  {journal} {\bibinfo
  {journal} {Macromolecules}\ }\textbf {\bibinfo {volume} {46}},\ \bibinfo
  {pages} {2376} (\bibinfo {year} {2013})},\ \Eprint
  {https://arxiv.org/abs/https://doi.org/10.1021/ma202172q}
  {https://doi.org/10.1021/ma202172q} \BibitemShut {NoStop}%
\bibitem [{\citenamefont {Priftis}\ \emph {et~al.}(2014)\citenamefont
  {Priftis}, \citenamefont {Xia}, \citenamefont {Margossian}, \citenamefont
  {Perry}, \citenamefont {Leon}, \citenamefont {Qin}, \citenamefont
  {de~Pablo},\ and\ \citenamefont {Tirrell}}]{tirrell2014}%
  \BibitemOpen
  \bibfield  {author} {\bibinfo {author} {\bibfnamefont {D.}~\bibnamefont
  {Priftis}}, \bibinfo {author} {\bibfnamefont {X.}~\bibnamefont {Xia}},
  \bibinfo {author} {\bibfnamefont {K.~O.}\ \bibnamefont {Margossian}},
  \bibinfo {author} {\bibfnamefont {S.~L.}\ \bibnamefont {Perry}}, \bibinfo
  {author} {\bibfnamefont {L.}~\bibnamefont {Leon}}, \bibinfo {author}
  {\bibfnamefont {J.}~\bibnamefont {Qin}}, \bibinfo {author} {\bibfnamefont
  {J.~J.}\ \bibnamefont {de~Pablo}},\ and\ \bibinfo {author} {\bibfnamefont
  {M.}~\bibnamefont {Tirrell}},\ }\bibfield  {title} {\bibinfo {title}
  {Ternary, tunable polyelectrolyte complex fluids driven by complex
  coacervation},\ }\href {https://doi.org/10.1021/ma500245j} {\bibfield
  {journal} {\bibinfo  {journal} {Macromolecules}\ }\textbf {\bibinfo {volume}
  {47}},\ \bibinfo {pages} {3076} (\bibinfo {year} {2014})},\ \Eprint
  {https://arxiv.org/abs/https://doi.org/10.1021/ma500245j}
  {https://doi.org/10.1021/ma500245j} \BibitemShut {NoStop}%
\bibitem [{\citenamefont {Vitorazi}\ \emph {et~al.}(2014)\citenamefont
  {Vitorazi}, \citenamefont {Ould-Moussa}, \citenamefont {Sekar}, \citenamefont
  {Fresnais}, \citenamefont {Loh}, \citenamefont {Chapel},\ and\ \citenamefont
  {Berret}}]{vitorazi2014}%
  \BibitemOpen
  \bibfield  {author} {\bibinfo {author} {\bibfnamefont {L.}~\bibnamefont
  {Vitorazi}}, \bibinfo {author} {\bibfnamefont {N.}~\bibnamefont
  {Ould-Moussa}}, \bibinfo {author} {\bibfnamefont {S.}~\bibnamefont {Sekar}},
  \bibinfo {author} {\bibfnamefont {J.}~\bibnamefont {Fresnais}}, \bibinfo
  {author} {\bibfnamefont {W.}~\bibnamefont {Loh}}, \bibinfo {author}
  {\bibfnamefont {J.-P.}\ \bibnamefont {Chapel}},\ and\ \bibinfo {author}
  {\bibfnamefont {J.-F.}\ \bibnamefont {Berret}},\ }\bibfield  {title}
  {\bibinfo {title} {Evidence of a two-step process and pathway dependency in
  the thermodynamics of poly(diallyldimethylammonium chloride)/poly(sodium
  acrylate) complexation},\ }\href {https://doi.org/10.1039/c4sm01461h}
  {\bibfield  {journal} {\bibinfo  {journal} {Soft Matter}\ }\textbf {\bibinfo
  {volume} {10}},\ \bibinfo {pages} {9496} (\bibinfo {year}
  {2014})}\BibitemShut {NoStop}%
\bibitem [{\citenamefont {Perry}\ \emph {et~al.}(2014)\citenamefont {Perry},
  \citenamefont {Li}, \citenamefont {Priftis}, \citenamefont {Leon},\ and\
  \citenamefont {Tirrell}}]{perry2014}%
  \BibitemOpen
  \bibfield  {author} {\bibinfo {author} {\bibfnamefont {S.~L.}\ \bibnamefont
  {Perry}}, \bibinfo {author} {\bibfnamefont {Y.}~\bibnamefont {Li}}, \bibinfo
  {author} {\bibfnamefont {D.}~\bibnamefont {Priftis}}, \bibinfo {author}
  {\bibfnamefont {L.}~\bibnamefont {Leon}},\ and\ \bibinfo {author}
  {\bibfnamefont {M.}~\bibnamefont {Tirrell}},\ }\bibfield  {title} {\bibinfo
  {title} {The effect of salt on the complex coacervation of vinyl
  polyelectrolytes},\ }\href {https://doi.org/10.3390/polym6061756} {\bibfield
  {journal} {\bibinfo  {journal} {Polymers}\ }\textbf {\bibinfo {volume} {6}},\
  \bibinfo {pages} {1756} (\bibinfo {year} {2014})}\BibitemShut {NoStop}%
\bibitem [{\citenamefont {Salehi}\ \emph {et~al.}(2015)\citenamefont {Salehi},
  \citenamefont {Desai}, \citenamefont {Li}, \citenamefont {Steele},\ and\
  \citenamefont {Larson}}]{salehi2015}%
  \BibitemOpen
  \bibfield  {author} {\bibinfo {author} {\bibfnamefont {A.}~\bibnamefont
  {Salehi}}, \bibinfo {author} {\bibfnamefont {P.~S.}\ \bibnamefont {Desai}},
  \bibinfo {author} {\bibfnamefont {J.}~\bibnamefont {Li}}, \bibinfo {author}
  {\bibfnamefont {C.~A.}\ \bibnamefont {Steele}},\ and\ \bibinfo {author}
  {\bibfnamefont {R.~G.}\ \bibnamefont {Larson}},\ }\bibfield  {title}
  {\bibinfo {title} {Relationship between polyelectrolyte bulk complexation and
  kinetics of their layer-by-layer assembly},\ }\href
  {https://doi.org/10.1021/ma502273a} {\bibfield  {journal} {\bibinfo
  {journal} {Macromolecules}\ }\textbf {\bibinfo {volume} {48}},\ \bibinfo
  {pages} {400} (\bibinfo {year} {2015})}\BibitemShut {NoStop}%
\bibitem [{\citenamefont {Zhang}\ \emph {et~al.}(2015)\citenamefont {Zhang},
  \citenamefont {Yildirim}, \citenamefont {Antila}, \citenamefont {Valenzuela},
  \citenamefont {Sammalkorpi},\ and\ \citenamefont
  {Lutkenhaus}}]{lutkenhaus2015}%
  \BibitemOpen
  \bibfield  {author} {\bibinfo {author} {\bibfnamefont {Y.}~\bibnamefont
  {Zhang}}, \bibinfo {author} {\bibfnamefont {E.}~\bibnamefont {Yildirim}},
  \bibinfo {author} {\bibfnamefont {H.~S.}\ \bibnamefont {Antila}}, \bibinfo
  {author} {\bibfnamefont {L.~D.}\ \bibnamefont {Valenzuela}}, \bibinfo
  {author} {\bibfnamefont {M.}~\bibnamefont {Sammalkorpi}},\ and\ \bibinfo
  {author} {\bibfnamefont {J.~L.}\ \bibnamefont {Lutkenhaus}},\ }\bibfield
  {title} {\bibinfo {title} {The influence of ionic strength and mixing ratio
  on the colloidal stability of {PDAC}/{PSS} polyelectrolyte complexes},\
  }\href {https://doi.org/10.1039/c5sm01184a} {\bibfield  {journal} {\bibinfo
  {journal} {Soft Matter}\ }\textbf {\bibinfo {volume} {11}},\ \bibinfo {pages}
  {7392} (\bibinfo {year} {2015})}\BibitemShut {NoStop}%
\bibitem [{\citenamefont {Kayitmazer}\ \emph {et~al.}(2015)\citenamefont
  {Kayitmazer}, \citenamefont {Koksal},\ and\ \citenamefont
  {Iyilik}}]{kayitmazer2015}%
  \BibitemOpen
  \bibfield  {author} {\bibinfo {author} {\bibfnamefont {A.~B.}\ \bibnamefont
  {Kayitmazer}}, \bibinfo {author} {\bibfnamefont {A.~F.}\ \bibnamefont
  {Koksal}},\ and\ \bibinfo {author} {\bibfnamefont {E.~K.}\ \bibnamefont
  {Iyilik}},\ }\bibfield  {title} {\bibinfo {title} {Complex coacervation of
  hyaluronic acid and chitosan: effects of {pH}, ionic strength, charge
  density, chain length and the charge ratio},\ }\href
  {https://doi.org/10.1039/c5sm01829c} {\bibfield  {journal} {\bibinfo
  {journal} {Soft Matter}\ }\textbf {\bibinfo {volume} {11}},\ \bibinfo {pages}
  {8605} (\bibinfo {year} {2015})}\BibitemShut {NoStop}%
\bibitem [{\citenamefont {Fu}\ and\ \citenamefont {Schlenoff}(2016)}]{fu2016}%
  \BibitemOpen
  \bibfield  {author} {\bibinfo {author} {\bibfnamefont {J.}~\bibnamefont
  {Fu}}\ and\ \bibinfo {author} {\bibfnamefont {J.~B.}\ \bibnamefont
  {Schlenoff}},\ }\bibfield  {title} {\bibinfo {title} {Driving forces for
  oppositely charged polyion association in aqueous solutions: Enthalpic,
  entropic, but not electrostatic},\ }\href
  {https://doi.org/10.1021/jacs.5b11878} {\bibfield  {journal} {\bibinfo
  {journal} {JACS}\ }\textbf {\bibinfo {volume} {138}},\ \bibinfo {pages} {980}
  (\bibinfo {year} {2016})},\ \bibinfo {note} {pMID: 26771205},\ \Eprint
  {https://arxiv.org/abs/https://doi.org/10.1021/jacs.5b11878}
  {https://doi.org/10.1021/jacs.5b11878} \BibitemShut {NoStop}%
\bibitem [{\citenamefont {Meka}\ \emph {et~al.}(2017)\citenamefont {Meka},
  \citenamefont {Sing}, \citenamefont {Pichika}, \citenamefont {Nali},
  \citenamefont {Kolapalli},\ and\ \citenamefont {Kesharwani}}]{meka2017}%
  \BibitemOpen
  \bibfield  {author} {\bibinfo {author} {\bibfnamefont {V.~S.}\ \bibnamefont
  {Meka}}, \bibinfo {author} {\bibfnamefont {M.~K.}\ \bibnamefont {Sing}},
  \bibinfo {author} {\bibfnamefont {M.~R.}\ \bibnamefont {Pichika}}, \bibinfo
  {author} {\bibfnamefont {S.~R.}\ \bibnamefont {Nali}}, \bibinfo {author}
  {\bibfnamefont {V.~R.}\ \bibnamefont {Kolapalli}},\ and\ \bibinfo {author}
  {\bibfnamefont {P.}~\bibnamefont {Kesharwani}},\ }\bibfield  {title}
  {\bibinfo {title} {A comprehensive review on polyelectrolyte complexes},\
  }\href {https://doi.org/10.1016/j.drudis.2017.06.008} {\bibfield  {journal}
  {\bibinfo  {journal} {Drug Discov. Today}\ }\textbf {\bibinfo {volume}
  {22}},\ \bibinfo {pages} {1697} (\bibinfo {year} {2017})}\BibitemShut
  {NoStop}%
\bibitem [{\citenamefont {Fu}\ \emph {et~al.}(2017)\citenamefont {Fu},
  \citenamefont {Fares},\ and\ \citenamefont {Schlenoff}}]{schlenoff2017}%
  \BibitemOpen
  \bibfield  {author} {\bibinfo {author} {\bibfnamefont {J.}~\bibnamefont
  {Fu}}, \bibinfo {author} {\bibfnamefont {H.~M.}\ \bibnamefont {Fares}},\ and\
  \bibinfo {author} {\bibfnamefont {J.~B.}\ \bibnamefont {Schlenoff}},\
  }\bibfield  {title} {\bibinfo {title} {Ion-pairing strength in
  polyelectrolyte complexes},\ }\href
  {https://doi.org/10.1021/acs.macromol.6b02445} {\bibfield  {journal}
  {\bibinfo  {journal} {Macromolecules}\ }\textbf {\bibinfo {volume} {50}},\
  \bibinfo {pages} {1066} (\bibinfo {year} {2017})}\BibitemShut {NoStop}%
\bibitem [{\citenamefont {Ali}\ and\ \citenamefont {Prabhu}(2018)}]{ali2018}%
  \BibitemOpen
  \bibfield  {author} {\bibinfo {author} {\bibfnamefont {S.}~\bibnamefont
  {Ali}}\ and\ \bibinfo {author} {\bibfnamefont {V.}~\bibnamefont {Prabhu}},\
  }\bibfield  {title} {\bibinfo {title} {Relaxation behavior by time-salt and
  time-temperature superpositions of polyelectrolyte complexes from coacervate
  to precipitate},\ }\href {https://doi.org/10.3390/gels4010011} {\bibfield
  {journal} {\bibinfo  {journal} {Gels}\ }\textbf {\bibinfo {volume} {4}},\
  \bibinfo {pages} {11} (\bibinfo {year} {2018})}\BibitemShut {NoStop}%
\bibitem [{\citenamefont {Li}\ \emph {et~al.}(2018)\citenamefont {Li},
  \citenamefont {Srivastava}, \citenamefont {Andreev}, \citenamefont {Marciel},
  \citenamefont {de~Pablo},\ and\ \citenamefont {Tirrell}}]{dePablo2018}%
  \BibitemOpen
  \bibfield  {author} {\bibinfo {author} {\bibfnamefont {L.}~\bibnamefont
  {Li}}, \bibinfo {author} {\bibfnamefont {S.}~\bibnamefont {Srivastava}},
  \bibinfo {author} {\bibfnamefont {M.}~\bibnamefont {Andreev}}, \bibinfo
  {author} {\bibfnamefont {A.~B.}\ \bibnamefont {Marciel}}, \bibinfo {author}
  {\bibfnamefont {J.~J.}\ \bibnamefont {de~Pablo}},\ and\ \bibinfo {author}
  {\bibfnamefont {M.~V.}\ \bibnamefont {Tirrell}},\ }\bibfield  {title}
  {\bibinfo {title} {Phase behavior and salt partitioning in polyelectrolyte
  complex coacervates},\ }\href {https://doi.org/10.1021/acs.macromol.8b00238}
  {\bibfield  {journal} {\bibinfo  {journal} {Macromolecules}\ }\textbf
  {\bibinfo {volume} {51}},\ \bibinfo {pages} {2988} (\bibinfo {year}
  {2018})}\BibitemShut {NoStop}%
\bibitem [{\citenamefont {Marciel}\ \emph {et~al.}(2018)\citenamefont
  {Marciel}, \citenamefont {Srivastava},\ and\ \citenamefont
  {Tirrell}}]{tirrell2018}%
  \BibitemOpen
  \bibfield  {author} {\bibinfo {author} {\bibfnamefont {A.~B.}\ \bibnamefont
  {Marciel}}, \bibinfo {author} {\bibfnamefont {S.}~\bibnamefont
  {Srivastava}},\ and\ \bibinfo {author} {\bibfnamefont {M.~V.}\ \bibnamefont
  {Tirrell}},\ }\bibfield  {title} {\bibinfo {title} {Structure and rheology of
  polyelectrolyte complex coacervates},\ }\href
  {https://doi.org/10.1039/C7SM02041D} {\bibfield  {journal} {\bibinfo
  {journal} {Soft Matter}\ }\textbf {\bibinfo {volume} {14}},\ \bibinfo {pages}
  {2454} (\bibinfo {year} {2018})}\BibitemShut {NoStop}%
\bibitem [{\citenamefont {Ali}\ \emph {et~al.}(2019)\citenamefont {Ali},
  \citenamefont {Bleuel},\ and\ \citenamefont {Prabhu}}]{ali2019}%
  \BibitemOpen
  \bibfield  {author} {\bibinfo {author} {\bibfnamefont {S.}~\bibnamefont
  {Ali}}, \bibinfo {author} {\bibfnamefont {M.}~\bibnamefont {Bleuel}},\ and\
  \bibinfo {author} {\bibfnamefont {V.~M.}\ \bibnamefont {Prabhu}},\ }\bibfield
   {title} {\bibinfo {title} {Lower critical solution temperature in
  polyelectrolyte complex coacervates},\ }\href
  {https://doi.org/10.1021/acsmacrolett.8b00952} {\bibfield  {journal}
  {\bibinfo  {journal} {ACS Macro Lett.}\ }\textbf {\bibinfo {volume} {8}},\
  \bibinfo {pages} {289} (\bibinfo {year} {2019})}\BibitemShut {NoStop}%
\bibitem [{\citenamefont {Schlenoff}\ \emph {et~al.}(2019)\citenamefont
  {Schlenoff}, \citenamefont {Yang}, \citenamefont {Digby},\ and\ \citenamefont
  {Wang}}]{wang2019}%
  \BibitemOpen
  \bibfield  {author} {\bibinfo {author} {\bibfnamefont {J.~B.}\ \bibnamefont
  {Schlenoff}}, \bibinfo {author} {\bibfnamefont {M.}~\bibnamefont {Yang}},
  \bibinfo {author} {\bibfnamefont {Z.~A.}\ \bibnamefont {Digby}},\ and\
  \bibinfo {author} {\bibfnamefont {Q.}~\bibnamefont {Wang}},\ }\bibfield
  {title} {\bibinfo {title} {Ion content of polyelectrolyte complex coacervates
  and the donnan equilibrium},\ }\href
  {https://doi.org/10.1021/acs.macromol.9b01755} {\bibfield  {journal}
  {\bibinfo  {journal} {Macromolecules}\ }\textbf {\bibinfo {volume} {52}},\
  \bibinfo {pages} {9149} (\bibinfo {year} {2019})}\BibitemShut {NoStop}%
\bibitem [{\citenamefont {Huang}\ \emph {et~al.}(2019)\citenamefont {Huang},
  \citenamefont {Morin},\ and\ \citenamefont {Laaser}}]{huang2019}%
  \BibitemOpen
  \bibfield  {author} {\bibinfo {author} {\bibfnamefont {J.}~\bibnamefont
  {Huang}}, \bibinfo {author} {\bibfnamefont {F.~J.}\ \bibnamefont {Morin}},\
  and\ \bibinfo {author} {\bibfnamefont {J.~E.}\ \bibnamefont {Laaser}},\
  }\bibfield  {title} {\bibinfo {title} {Charge-density-dominated phase
  behavior and viscoelasticity of polyelectrolyte complex coacervates},\ }\href
  {https://doi.org/10.1021/acs.macromol.9b00036} {\bibfield  {journal}
  {\bibinfo  {journal} {Macromolecules}\ }\textbf {\bibinfo {volume} {52}},\
  \bibinfo {pages} {4957} (\bibinfo {year} {2019})}\BibitemShut {NoStop}%
\bibitem [{\citenamefont {Saha}\ \emph {et~al.}(2020)\citenamefont {Saha},
  \citenamefont {Gordievskaya}, \citenamefont {De},\ and\ \citenamefont
  {Kramarenko}}]{pde2020}%
  \BibitemOpen
  \bibfield  {author} {\bibinfo {author} {\bibfnamefont {B.}~\bibnamefont
  {Saha}}, \bibinfo {author} {\bibfnamefont {Y.~D.}\ \bibnamefont
  {Gordievskaya}}, \bibinfo {author} {\bibfnamefont {P.}~\bibnamefont {De}},\
  and\ \bibinfo {author} {\bibfnamefont {E.~Y.}\ \bibnamefont {Kramarenko}},\
  }\bibfield  {title} {\bibinfo {title} {Unusual nanostructured morphologies
  enabled by interpolyelectrolyte complexation of polyions bearing incompatible
  nonionic segments},\ }\href {https://doi.org/10.1021/acs.macromol.0c02230}
  {\bibfield  {journal} {\bibinfo  {journal} {Macromolecules}\ }\textbf
  {\bibinfo {volume} {53}},\ \bibinfo {pages} {10754} (\bibinfo {year}
  {2020})}\BibitemShut {NoStop}%
\bibitem [{\citenamefont {Meng}\ \emph
  {et~al.}(2020{\natexlab{a}})\citenamefont {Meng}, \citenamefont {Liu},
  \citenamefont {Yeo}, \citenamefont {Ting},\ and\ \citenamefont
  {Tirrell}}]{tirrell2020}%
  \BibitemOpen
  \bibfield  {author} {\bibinfo {author} {\bibfnamefont {S.}~\bibnamefont
  {Meng}}, \bibinfo {author} {\bibfnamefont {Y.}~\bibnamefont {Liu}}, \bibinfo
  {author} {\bibfnamefont {J.}~\bibnamefont {Yeo}}, \bibinfo {author}
  {\bibfnamefont {J.~M.}\ \bibnamefont {Ting}},\ and\ \bibinfo {author}
  {\bibfnamefont {M.~V.}\ \bibnamefont {Tirrell}},\ }\bibfield  {title}
  {\bibinfo {title} {Effect of mixed solvents on polyelectrolyte complexes with
  salt},\ }\href {https://doi.org/10.1007/s00396-020-04637-0} {\bibfield
  {journal} {\bibinfo  {journal} {Colloid and Polymer Science}\ }\textbf
  {\bibinfo {volume} {298}},\ \bibinfo {pages} {887} (\bibinfo {year}
  {2020}{\natexlab{a}})}\BibitemShut {NoStop}%
\bibitem [{\citenamefont {Meng}\ \emph
  {et~al.}(2020{\natexlab{b}})\citenamefont {Meng}, \citenamefont {Ting},
  \citenamefont {Wu},\ and\ \citenamefont {Tirrell}}]{meng2020}%
  \BibitemOpen
  \bibfield  {author} {\bibinfo {author} {\bibfnamefont {S.}~\bibnamefont
  {Meng}}, \bibinfo {author} {\bibfnamefont {J.~M.}\ \bibnamefont {Ting}},
  \bibinfo {author} {\bibfnamefont {H.}~\bibnamefont {Wu}},\ and\ \bibinfo
  {author} {\bibfnamefont {M.~V.}\ \bibnamefont {Tirrell}},\ }\bibfield
  {title} {\bibinfo {title} {Solid-to-liquid phase transition in
  polyelectrolyte complexes},\ }\href
  {https://doi.org/10.1021/acs.macromol.0c00930} {\bibfield  {journal}
  {\bibinfo  {journal} {Macromolecules}\ }\textbf {\bibinfo {volume} {53}},\
  \bibinfo {pages} {7944} (\bibinfo {year} {2020}{\natexlab{b}})}\BibitemShut
  {NoStop}%
\bibitem [{\citenamefont {Li}\ \emph {et~al.}(2020)\citenamefont {Li},
  \citenamefont {Rumyantsev}, \citenamefont {Srivastava}, \citenamefont {Meng},
  \citenamefont {de~Pablo},\ and\ \citenamefont {Tirrell}}]{tirrell2021}%
  \BibitemOpen
  \bibfield  {author} {\bibinfo {author} {\bibfnamefont {L.}~\bibnamefont
  {Li}}, \bibinfo {author} {\bibfnamefont {A.~M.}\ \bibnamefont {Rumyantsev}},
  \bibinfo {author} {\bibfnamefont {S.}~\bibnamefont {Srivastava}}, \bibinfo
  {author} {\bibfnamefont {S.}~\bibnamefont {Meng}}, \bibinfo {author}
  {\bibfnamefont {J.~J.}\ \bibnamefont {de~Pablo}},\ and\ \bibinfo {author}
  {\bibfnamefont {M.~V.}\ \bibnamefont {Tirrell}},\ }\bibfield  {title}
  {\bibinfo {title} {Effect of solvent quality on the phase behavior of
  polyelectrolyte complexes},\ }\href
  {https://doi.org/10.1021/acs.macromol.0c01000} {\bibfield  {journal}
  {\bibinfo  {journal} {Macromolecules}\ }\textbf {\bibinfo {volume} {54}},\
  \bibinfo {pages} {105} (\bibinfo {year} {2020})}\BibitemShut {NoStop}%
\bibitem [{\citenamefont {Chen}\ \emph
  {et~al.}(2021{\natexlab{b}})\citenamefont {Chen}, \citenamefont {Yang},
  \citenamefont {Shaheen},\ and\ \citenamefont {Schlenoff}}]{chen2021}%
  \BibitemOpen
  \bibfield  {author} {\bibinfo {author} {\bibfnamefont {Y.}~\bibnamefont
  {Chen}}, \bibinfo {author} {\bibfnamefont {M.}~\bibnamefont {Yang}}, \bibinfo
  {author} {\bibfnamefont {S.~A.}\ \bibnamefont {Shaheen}},\ and\ \bibinfo
  {author} {\bibfnamefont {J.~B.}\ \bibnamefont {Schlenoff}},\ }\bibfield
  {title} {\bibinfo {title} {Influence of nonstoichiometry on the viscoelastic
  properties of a polyelectrolyte complex},\ }\href
  {https://doi.org/10.1021/acs.macromol.1c01154} {\bibfield  {journal}
  {\bibinfo  {journal} {Macromolecules}\ }\textbf {\bibinfo {volume} {54}},\
  \bibinfo {pages} {7890} (\bibinfo {year} {2021}{\natexlab{b}})}\BibitemShut
  {NoStop}%
\bibitem [{\citenamefont {Neitzel}\ \emph {et~al.}(2021)\citenamefont
  {Neitzel}, \citenamefont {Fang}, \citenamefont {Yu}, \citenamefont
  {Rumyantsev}, \citenamefont {de~Pablo},\ and\ \citenamefont
  {Tirrell}}]{neitzel2021}%
  \BibitemOpen
  \bibfield  {author} {\bibinfo {author} {\bibfnamefont {A.~E.}\ \bibnamefont
  {Neitzel}}, \bibinfo {author} {\bibfnamefont {Y.~N.}\ \bibnamefont {Fang}},
  \bibinfo {author} {\bibfnamefont {B.}~\bibnamefont {Yu}}, \bibinfo {author}
  {\bibfnamefont {A.~M.}\ \bibnamefont {Rumyantsev}}, \bibinfo {author}
  {\bibfnamefont {J.~J.}\ \bibnamefont {de~Pablo}},\ and\ \bibinfo {author}
  {\bibfnamefont {M.~V.}\ \bibnamefont {Tirrell}},\ }\bibfield  {title}
  {\bibinfo {title} {Polyelectrolyte complex coacervation across a broad range
  of charge densities},\ }\href {https://doi.org/10.1021/acs.macromol.1c00703}
  {\bibfield  {journal} {\bibinfo  {journal} {Macromolecules}\ }\textbf
  {\bibinfo {volume} {54}},\ \bibinfo {pages} {6878} (\bibinfo {year}
  {2021})}\BibitemShut {NoStop}%
\bibitem [{\citenamefont {Priftis}\ and\ \citenamefont
  {Tirrell}(2012{\natexlab{b}})}]{priftis2012}%
  \BibitemOpen
  \bibfield  {author} {\bibinfo {author} {\bibfnamefont {D.}~\bibnamefont
  {Priftis}}\ and\ \bibinfo {author} {\bibfnamefont {M.}~\bibnamefont
  {Tirrell}},\ }\bibfield  {title} {\bibinfo {title} {Phase behaviour and
  complex coacervation of aqueous polypeptide solutions},\ }\href
  {https://doi.org/10.1039/c2sm25604e} {\ \textbf {\bibinfo {volume} {8}},\
  \bibinfo {pages} {9396} (\bibinfo {year} {2012}{\natexlab{b}})}\BibitemShut
  {NoStop}%
\bibitem [{\citenamefont {Priftis}\ and\ \citenamefont
  {Tirrell}(2012{\natexlab{c}})}]{priftis2012-softmatter}%
  \BibitemOpen
  \bibfield  {author} {\bibinfo {author} {\bibfnamefont {D.}~\bibnamefont
  {Priftis}}\ and\ \bibinfo {author} {\bibfnamefont {M.}~\bibnamefont
  {Tirrell}},\ }\bibfield  {title} {\bibinfo {title} {Phase behaviour and
  complex coacervation of aqueous polypeptide solutions},\ }\href
  {https://doi.org/10.1039/c2sm25604e} {\bibfield  {journal} {\bibinfo
  {journal} {Soft Matter}\ }\textbf {\bibinfo {volume} {8}},\ \bibinfo {pages}
  {9396} (\bibinfo {year} {2012}{\natexlab{c}})}\BibitemShut {NoStop}%
\bibitem [{\citenamefont {Subbotin}\ and\ \citenamefont
  {Semenov}(2021)}]{subbotin2021}%
  \BibitemOpen
  \bibfield  {author} {\bibinfo {author} {\bibfnamefont {A.~V.}\ \bibnamefont
  {Subbotin}}\ and\ \bibinfo {author} {\bibfnamefont {A.~N.}\ \bibnamefont
  {Semenov}},\ }\bibfield  {title} {\bibinfo {title} {The structure of
  polyelectrolyte complex coacervates and multilayers},\ }\href
  {https://doi.org/10.1021/acs.macromol.0c02470} {\bibfield  {journal}
  {\bibinfo  {journal} {Macromolecules}\ }\textbf {\bibinfo {volume} {54}},\
  \bibinfo {pages} {1314} (\bibinfo {year} {2021})}\BibitemShut {NoStop}%
\bibitem [{\citenamefont {Friedowitz}\ \emph {et~al.}(2021)\citenamefont
  {Friedowitz}, \citenamefont {Lou}, \citenamefont {Barker}, \citenamefont
  {Will}, \citenamefont {Xia},\ and\ \citenamefont {Qin}}]{friedowitz2021}%
  \BibitemOpen
  \bibfield  {author} {\bibinfo {author} {\bibfnamefont {S.}~\bibnamefont
  {Friedowitz}}, \bibinfo {author} {\bibfnamefont {J.}~\bibnamefont {Lou}},
  \bibinfo {author} {\bibfnamefont {K.~P.}\ \bibnamefont {Barker}}, \bibinfo
  {author} {\bibfnamefont {K.}~\bibnamefont {Will}}, \bibinfo {author}
  {\bibfnamefont {Y.}~\bibnamefont {Xia}},\ and\ \bibinfo {author}
  {\bibfnamefont {J.}~\bibnamefont {Qin}},\ }\bibfield  {title} {\bibinfo
  {title} {Looping-in complexation and ion partitioning in nonstoichiometric
  polyelectrolyte mixtures},\ }\bibfield  {journal} {\bibinfo  {journal}
  {Science Advances}\ }\textbf {\bibinfo {volume} {7}},\ \href
  {https://doi.org/10.1126/sciadv.abg8654} {10.1126/sciadv.abg8654} (\bibinfo
  {year} {2021})\BibitemShut {NoStop}%
\bibitem [{\citenamefont {Ma}\ \emph {et~al.}(2021)\citenamefont {Ma},
  \citenamefont {Ali},\ and\ \citenamefont {Prabhu}}]{ma2021}%
  \BibitemOpen
  \bibfield  {author} {\bibinfo {author} {\bibfnamefont {Y.}~\bibnamefont
  {Ma}}, \bibinfo {author} {\bibfnamefont {S.}~\bibnamefont {Ali}},\ and\
  \bibinfo {author} {\bibfnamefont {V.~M.}\ \bibnamefont {Prabhu}},\ }\bibfield
   {title} {\bibinfo {title} {Enhanced concentration fluctuations in model
  polyelectrolyte coacervate mixtures along a salt isopleth phase diagram},\
  }\href {https://doi.org/10.1021/acs.macromol.1c02001} {\bibfield  {journal}
  {\bibinfo  {journal} {Macromolecules}\ }\textbf {\bibinfo {volume} {54}},\
  \bibinfo {pages} {11338} (\bibinfo {year} {2021})}\BibitemShut {NoStop}%
\bibitem [{\citenamefont {Lalwani}\ \emph {et~al.}(2021)\citenamefont
  {Lalwani}, \citenamefont {Batys}, \citenamefont {Sammalkorpi},\ and\
  \citenamefont {Lutkenhaus}}]{lalwani2021}%
  \BibitemOpen
  \bibfield  {author} {\bibinfo {author} {\bibfnamefont {S.~M.}\ \bibnamefont
  {Lalwani}}, \bibinfo {author} {\bibfnamefont {P.}~\bibnamefont {Batys}},
  \bibinfo {author} {\bibfnamefont {M.}~\bibnamefont {Sammalkorpi}},\ and\
  \bibinfo {author} {\bibfnamefont {J.~L.}\ \bibnamefont {Lutkenhaus}},\
  }\bibfield  {title} {\bibinfo {title} {Relaxation times of solid-like
  polyelectrolyte complexes of varying {pH} and water content},\ }\href
  {https://doi.org/10.1021/acs.macromol.1c00940} {\bibfield  {journal}
  {\bibinfo  {journal} {Macromolecules}\ }\textbf {\bibinfo {volume} {54}},\
  \bibinfo {pages} {7765} (\bibinfo {year} {2021})}\BibitemShut {NoStop}%
\bibitem [{\citenamefont {Digby}\ \emph {et~al.}(2022)\citenamefont {Digby},
  \citenamefont {Yang}, \citenamefont {Lteif},\ and\ \citenamefont
  {Schlenoff}}]{digby2022}%
  \BibitemOpen
  \bibfield  {author} {\bibinfo {author} {\bibfnamefont {Z.~A.}\ \bibnamefont
  {Digby}}, \bibinfo {author} {\bibfnamefont {M.}~\bibnamefont {Yang}},
  \bibinfo {author} {\bibfnamefont {S.}~\bibnamefont {Lteif}},\ and\ \bibinfo
  {author} {\bibfnamefont {J.~B.}\ \bibnamefont {Schlenoff}},\ }\bibfield
  {title} {\bibinfo {title} {Salt resistance as a measure of the strength of
  polyelectrolyte complexation},\ }\href
  {https://doi.org/10.1021/acs.macromol.1c02151} {\bibfield  {journal}
  {\bibinfo  {journal} {Macromolecules}\ }\textbf {\bibinfo {volume} {55}},\
  \bibinfo {pages} {978} (\bibinfo {year} {2022})}\BibitemShut {NoStop}%
\bibitem [{\citenamefont {Hayashi}\ \emph {et~al.}(2002)\citenamefont
  {Hayashi}, \citenamefont {Ullner},\ and\ \citenamefont
  {Linse}}]{hayashi2002}%
  \BibitemOpen
  \bibfield  {author} {\bibinfo {author} {\bibfnamefont {Y.}~\bibnamefont
  {Hayashi}}, \bibinfo {author} {\bibfnamefont {M.}~\bibnamefont {Ullner}},\
  and\ \bibinfo {author} {\bibfnamefont {P.}~\bibnamefont {Linse}},\ }\bibfield
   {title} {\bibinfo {title} {A monte carlo study of solutions of oppositely
  charged polyelectrolytes},\ }\href {https://doi.org/10.1063/1.1460859}
  {\bibfield  {journal} {\bibinfo  {journal} {JCP}\ }\textbf {\bibinfo {volume}
  {116}},\ \bibinfo {pages} {6836} (\bibinfo {year} {2002})}\BibitemShut
  {NoStop}%
\bibitem [{\citenamefont {Winkler}\ \emph {et~al.}(2002)\citenamefont
  {Winkler}, \citenamefont {Steinhauser},\ and\ \citenamefont
  {Reineker}}]{winkler2002}%
  \BibitemOpen
  \bibfield  {author} {\bibinfo {author} {\bibfnamefont {R.~G.}\ \bibnamefont
  {Winkler}}, \bibinfo {author} {\bibfnamefont {M.~O.}\ \bibnamefont
  {Steinhauser}},\ and\ \bibinfo {author} {\bibfnamefont {P.}~\bibnamefont
  {Reineker}},\ }\bibfield  {title} {\bibinfo {title} {Complex formation in
  systems of oppositely charged polyelectrolytes: A molecular dynamics
  simulation study},\ }\href {https://doi.org/10.1103/PhysRevE.66.021802}
  {\bibfield  {journal} {\bibinfo  {journal} {Phys. Rev. E}\ }\textbf {\bibinfo
  {volume} {66}},\ \bibinfo {pages} {021802} (\bibinfo {year}
  {2002})}\BibitemShut {NoStop}%
\bibitem [{\citenamefont {Hayashi}\ \emph {et~al.}(2003)\citenamefont
  {Hayashi}, \citenamefont {Ullner},\ and\ \citenamefont
  {Linse}}]{hayashi2003}%
  \BibitemOpen
  \bibfield  {author} {\bibinfo {author} {\bibfnamefont {Y.}~\bibnamefont
  {Hayashi}}, \bibinfo {author} {\bibfnamefont {M.}~\bibnamefont {Ullner}},\
  and\ \bibinfo {author} {\bibfnamefont {P.}~\bibnamefont {Linse}},\ }\bibfield
   {title} {\bibinfo {title} {Complex formation in solutions of oppositely
  charged polyelectrolytes at different polyion compositions and salt
  content},\ }\href {https://doi.org/10.1021/jp022491a} {\bibfield  {journal}
  {\bibinfo  {journal} {JPCB}\ }\textbf {\bibinfo {volume} {107}},\ \bibinfo
  {pages} {8198} (\bibinfo {year} {2003})}\BibitemShut {NoStop}%
\bibitem [{\citenamefont {Hayashi}\ \emph {et~al.}(2004)\citenamefont
  {Hayashi}, \citenamefont {Ullner},\ and\ \citenamefont
  {Linse}}]{hayashi2004}%
  \BibitemOpen
  \bibfield  {author} {\bibinfo {author} {\bibfnamefont {Y.}~\bibnamefont
  {Hayashi}}, \bibinfo {author} {\bibfnamefont {M.}~\bibnamefont {Ullner}},\
  and\ \bibinfo {author} {\bibfnamefont {P.}~\bibnamefont {Linse}},\ }\bibfield
   {title} {\bibinfo {title} {Oppositely charged polyelectrolytes. complex
  formation and effects of chain asymmetry},\ }\href
  {https://doi.org/10.1021/jp048267y} {\bibfield  {journal} {\bibinfo
  {journal} {JPCB}\ }\textbf {\bibinfo {volume} {108}},\ \bibinfo {pages}
  {15266} (\bibinfo {year} {2004})}\BibitemShut {NoStop}%
\bibitem [{\citenamefont {Zhang}\ and\ \citenamefont
  {Shklovskii}(2005)}]{zhang2005}%
  \BibitemOpen
  \bibfield  {author} {\bibinfo {author} {\bibfnamefont {R.}~\bibnamefont
  {Zhang}}\ and\ \bibinfo {author} {\bibfnamefont {B.}~\bibnamefont
  {Shklovskii}},\ }\bibfield  {title} {\bibinfo {title} {Phase diagram of
  solution of oppositely charged polyelectrolytes},\ }\href
  {https://doi.org/https://doi.org/10.1016/j.physa.2004.12.037} {\bibfield
  {journal} {\bibinfo  {journal} {Physica A: Statistical Mechanics and its
  Applications}\ }\textbf {\bibinfo {volume} {352}},\ \bibinfo {pages} {216 }
  (\bibinfo {year} {2005})},\ \bibinfo {note} {physics Applied to Biological
  Systems}\BibitemShut {NoStop}%
\bibitem [{\citenamefont {Ou}\ and\ \citenamefont
  {Muthukumar}(2006)}]{zhaoyang2006}%
  \BibitemOpen
  \bibfield  {author} {\bibinfo {author} {\bibfnamefont {Z.}~\bibnamefont
  {Ou}}\ and\ \bibinfo {author} {\bibfnamefont {M.}~\bibnamefont
  {Muthukumar}},\ }\bibfield  {title} {\bibinfo {title} {Entropy and enthalpy
  of polyelectrolyte complexation: Langevin dynamics simulations},\ }\href@noop
  {} {\bibfield  {journal} {\bibinfo  {journal} {JCP}\ }\textbf {\bibinfo
  {volume} {124}},\ \bibinfo {pages} {154902} (\bibinfo {year}
  {2006})}\BibitemShut {NoStop}%
\bibitem [{\citenamefont {Popov}\ \emph {et~al.}(2007)\citenamefont {Popov},
  \citenamefont {Lee},\ and\ \citenamefont {Fredrickson}}]{fredrickson2007}%
  \BibitemOpen
  \bibfield  {author} {\bibinfo {author} {\bibfnamefont {Y.~O.}\ \bibnamefont
  {Popov}}, \bibinfo {author} {\bibfnamefont {J.}~\bibnamefont {Lee}},\ and\
  \bibinfo {author} {\bibfnamefont {G.~H.}\ \bibnamefont {Fredrickson}},\
  }\bibfield  {title} {\bibinfo {title} {Field-theoretic simulations of
  polyelectrolyte complexation},\ }\href {https://doi.org/10.1002/polb.21334}
  {\bibfield  {journal} {\bibinfo  {journal} {Journal of Polymer Science Part
  B: Polymer Physics}\ }\textbf {\bibinfo {volume} {45}},\ \bibinfo {pages}
  {3223} (\bibinfo {year} {2007})},\ \Eprint
  {https://arxiv.org/abs/https://onlinelibrary.wiley.com/doi/pdf/10.1002/polb.21334}
  {https://onlinelibrary.wiley.com/doi/pdf/10.1002/polb.21334} \BibitemShut
  {NoStop}%
\bibitem [{\citenamefont {Hoda}\ and\ \citenamefont
  {Larson}(2009)}]{larson2009}%
  \BibitemOpen
  \bibfield  {author} {\bibinfo {author} {\bibfnamefont {N.}~\bibnamefont
  {Hoda}}\ and\ \bibinfo {author} {\bibfnamefont {R.~G.}\ \bibnamefont
  {Larson}},\ }\bibfield  {title} {\bibinfo {title} {Explicit- and
  implicit-solvent molecular dynamics simulations of complex formation between
  polycations and polyanions},\ }\href {https://doi.org/10.1021/ma901632c}
  {\bibfield  {journal} {\bibinfo  {journal} {Macromolecules}\ }\textbf
  {\bibinfo {volume} {42}},\ \bibinfo {pages} {8851} (\bibinfo {year}
  {2009})}\BibitemShut {NoStop}%
\bibitem [{\citenamefont {Lazutin}\ \emph {et~al.}(2012)\citenamefont
  {Lazutin}, \citenamefont {Semenov},\ and\ \citenamefont
  {Vasilevskaya}}]{semenov2012}%
  \BibitemOpen
  \bibfield  {author} {\bibinfo {author} {\bibfnamefont {A.~A.}\ \bibnamefont
  {Lazutin}}, \bibinfo {author} {\bibfnamefont {A.~N.}\ \bibnamefont
  {Semenov}},\ and\ \bibinfo {author} {\bibfnamefont {V.~V.}\ \bibnamefont
  {Vasilevskaya}},\ }\bibfield  {title} {\bibinfo {title} {Polyelectrolyte
  complexes consisting of macromolecules with varied stiffness: Computer
  simulation},\ }\href {https://doi.org/10.1002/mats.201100097} {\bibfield
  {journal} {\bibinfo  {journal} {Macromol. Theory Simul.}\ }\textbf {\bibinfo
  {volume} {21}},\ \bibinfo {pages} {328} (\bibinfo {year} {2012})},\ \Eprint
  {https://arxiv.org/abs/https://onlinelibrary.wiley.com/doi/pdf/10.1002/mats.201100097}
  {https://onlinelibrary.wiley.com/doi/pdf/10.1002/mats.201100097} \BibitemShut
  {NoStop}%
\bibitem [{\citenamefont {Riggleman}\ \emph {et~al.}(2012)\citenamefont
  {Riggleman}, \citenamefont {Kumar},\ and\ \citenamefont
  {Fredrickson}}]{fredrickson2012}%
  \BibitemOpen
  \bibfield  {author} {\bibinfo {author} {\bibfnamefont {R.~A.}\ \bibnamefont
  {Riggleman}}, \bibinfo {author} {\bibfnamefont {R.}~\bibnamefont {Kumar}},\
  and\ \bibinfo {author} {\bibfnamefont {G.~H.}\ \bibnamefont {Fredrickson}},\
  }\bibfield  {title} {\bibinfo {title} {Investigation of the interfacial
  tension of complex coacervates using field-theoretic simulations},\ }\href
  {https://doi.org/10.1063/1.3674305} {\bibfield  {journal} {\bibinfo
  {journal} {JCP}\ }\textbf {\bibinfo {volume} {136}},\ \bibinfo {pages}
  {024903} (\bibinfo {year} {2012})},\ \Eprint
  {https://arxiv.org/abs/https://doi.org/10.1063/1.3674305}
  {https://doi.org/10.1063/1.3674305} \BibitemShut {NoStop}%
\bibitem [{\citenamefont {Meneses-Ju{\'{a}}rez}\ \emph
  {et~al.}(2015)\citenamefont {Meneses-Ju{\'{a}}rez}, \citenamefont
  {M{\'{a}}rquez-Beltr{\'{a}}n}, \citenamefont {Rivas-Silva}, \citenamefont
  {Pal},\ and\ \citenamefont {Gonz{\'{a}}lez-Melchor}}]{juarez2015}%
  \BibitemOpen
  \bibfield  {author} {\bibinfo {author} {\bibfnamefont {E.}~\bibnamefont
  {Meneses-Ju{\'{a}}rez}}, \bibinfo {author} {\bibfnamefont {C.}~\bibnamefont
  {M{\'{a}}rquez-Beltr{\'{a}}n}}, \bibinfo {author} {\bibfnamefont {J.~F.}\
  \bibnamefont {Rivas-Silva}}, \bibinfo {author} {\bibfnamefont
  {U.}~\bibnamefont {Pal}},\ and\ \bibinfo {author} {\bibfnamefont
  {M.}~\bibnamefont {Gonz{\'{a}}lez-Melchor}},\ }\bibfield  {title} {\bibinfo
  {title} {The structure and interaction mechanism of a polyelectrolyte
  complex: a dissipative particle dynamics study},\ }\href
  {https://doi.org/10.1039/c5sm00911a} {\bibfield  {journal} {\bibinfo
  {journal} {Soft Matter}\ }\textbf {\bibinfo {volume} {11}},\ \bibinfo {pages}
  {5889} (\bibinfo {year} {2015})}\BibitemShut {NoStop}%
\bibitem [{\citenamefont {Peng}\ and\ \citenamefont
  {Muthukumar}(2015)}]{peng2015}%
  \BibitemOpen
  \bibfield  {author} {\bibinfo {author} {\bibfnamefont {B.}~\bibnamefont
  {Peng}}\ and\ \bibinfo {author} {\bibfnamefont {M.}~\bibnamefont
  {Muthukumar}},\ }\bibfield  {title} {\bibinfo {title} {Modeling competitive
  substitution in a polyelectrolyte complex},\ }\href
  {https://doi.org/10.1063/1.4936256} {\bibfield  {journal} {\bibinfo
  {journal} {JCP}\ }\textbf {\bibinfo {volume} {143}},\ \bibinfo {pages}
  {243133} (\bibinfo {year} {2015})}\BibitemShut {NoStop}%
\bibitem [{\citenamefont {Lytle}\ \emph {et~al.}(2016)\citenamefont {Lytle},
  \citenamefont {Radhakrishna},\ and\ \citenamefont {Sing}}]{lytle2016}%
  \BibitemOpen
  \bibfield  {author} {\bibinfo {author} {\bibfnamefont {T.~K.}\ \bibnamefont
  {Lytle}}, \bibinfo {author} {\bibfnamefont {M.}~\bibnamefont
  {Radhakrishna}},\ and\ \bibinfo {author} {\bibfnamefont {C.~E.}\ \bibnamefont
  {Sing}},\ }\bibfield  {title} {\bibinfo {title} {High charge density
  coacervate assembly via hybrid monte carlo single chain in mean field
  theory},\ }\href {https://doi.org/10.1021/acs.macromol.6b02159} {\bibfield
  {journal} {\bibinfo  {journal} {Macromolecules}\ }\textbf {\bibinfo {volume}
  {49}},\ \bibinfo {pages} {9693} (\bibinfo {year} {2016})}\BibitemShut
  {NoStop}%
\bibitem [{\citenamefont {Xu}\ \emph {et~al.}(2016)\citenamefont {Xu},
  \citenamefont {Kandu{\v{c}}}, \citenamefont {Wu},\ and\ \citenamefont
  {Dzubiella}}]{dzubiella2016}%
  \BibitemOpen
  \bibfield  {author} {\bibinfo {author} {\bibfnamefont {X.}~\bibnamefont
  {Xu}}, \bibinfo {author} {\bibfnamefont {M.}~\bibnamefont {Kandu{\v{c}}}},
  \bibinfo {author} {\bibfnamefont {J.}~\bibnamefont {Wu}},\ and\ \bibinfo
  {author} {\bibfnamefont {J.}~\bibnamefont {Dzubiella}},\ }\bibfield  {title}
  {\bibinfo {title} {Potential of mean force and transient states in
  polyelectrolyte pair complexation},\ }\href
  {https://doi.org/10.1063/1.4958675} {\bibfield  {journal} {\bibinfo
  {journal} {JCP}\ }\textbf {\bibinfo {volume} {145}},\ \bibinfo {pages}
  {034901} (\bibinfo {year} {2016})}\BibitemShut {NoStop}%
\bibitem [{\citenamefont {Delaney}\ and\ \citenamefont
  {Fredrickson}(2017)}]{fredrickson2017}%
  \BibitemOpen
  \bibfield  {author} {\bibinfo {author} {\bibfnamefont {K.~T.}\ \bibnamefont
  {Delaney}}\ and\ \bibinfo {author} {\bibfnamefont {G.~H.}\ \bibnamefont
  {Fredrickson}},\ }\bibfield  {title} {\bibinfo {title} {Theory of
  polyelectrolyte complexation-complex coacervates are self-coacervates},\
  }\href {https://doi.org/10.1063/1.4985568} {\bibfield  {journal} {\bibinfo
  {journal} {JCP}\ }\textbf {\bibinfo {volume} {146}},\ \bibinfo {pages}
  {224902} (\bibinfo {year} {2017})},\ \Eprint
  {https://arxiv.org/abs/https://doi.org/10.1063/1.4985568}
  {https://doi.org/10.1063/1.4985568} \BibitemShut {NoStop}%
\bibitem [{\citenamefont {Chang}\ \emph {et~al.}(2017)\citenamefont {Chang},
  \citenamefont {Lytle}, \citenamefont {Radhakrishna}, \citenamefont {Madinya},
  \citenamefont {V{\'{e}}lez}, \citenamefont {Sing},\ and\ \citenamefont
  {Perry}}]{chang2017}%
  \BibitemOpen
  \bibfield  {author} {\bibinfo {author} {\bibfnamefont {L.-W.}\ \bibnamefont
  {Chang}}, \bibinfo {author} {\bibfnamefont {T.~K.}\ \bibnamefont {Lytle}},
  \bibinfo {author} {\bibfnamefont {M.}~\bibnamefont {Radhakrishna}}, \bibinfo
  {author} {\bibfnamefont {J.~J.}\ \bibnamefont {Madinya}}, \bibinfo {author}
  {\bibfnamefont {J.}~\bibnamefont {V{\'{e}}lez}}, \bibinfo {author}
  {\bibfnamefont {C.~E.}\ \bibnamefont {Sing}},\ and\ \bibinfo {author}
  {\bibfnamefont {S.~L.}\ \bibnamefont {Perry}},\ }\bibfield  {title} {\bibinfo
  {title} {Sequence and entropy-based control of complex coacervates},\
  }\bibfield  {journal} {\bibinfo  {journal} {Nat. Comm.}\ }\textbf {\bibinfo
  {volume} {8}},\ \href {https://doi.org/10.1038/s41467-017-01249-1}
  {10.1038/s41467-017-01249-1} (\bibinfo {year} {2017})\BibitemShut {NoStop}%
\bibitem [{\citenamefont {Radhakrishna}\ \emph {et~al.}(2017)\citenamefont
  {Radhakrishna}, \citenamefont {Basu}, \citenamefont {Liu}, \citenamefont
  {Shamsi}, \citenamefont {Perry},\ and\ \citenamefont
  {Sing}}]{radhakrishna2017}%
  \BibitemOpen
  \bibfield  {author} {\bibinfo {author} {\bibfnamefont {M.}~\bibnamefont
  {Radhakrishna}}, \bibinfo {author} {\bibfnamefont {K.}~\bibnamefont {Basu}},
  \bibinfo {author} {\bibfnamefont {Y.}~\bibnamefont {Liu}}, \bibinfo {author}
  {\bibfnamefont {R.}~\bibnamefont {Shamsi}}, \bibinfo {author} {\bibfnamefont
  {S.~L.}\ \bibnamefont {Perry}},\ and\ \bibinfo {author} {\bibfnamefont
  {C.~E.}\ \bibnamefont {Sing}},\ }\bibfield  {title} {\bibinfo {title}
  {Molecular connectivity and correlation effects on polymer coacervation},\
  }\href {https://doi.org/10.1021/acs.macromol.6b02582} {\bibfield  {journal}
  {\bibinfo  {journal} {Macromolecules}\ }\textbf {\bibinfo {volume} {50}},\
  \bibinfo {pages} {3030} (\bibinfo {year} {2017})}\BibitemShut {NoStop}%
\bibitem [{\citenamefont {Rathee}\ \emph
  {et~al.}(2018{\natexlab{a}})\citenamefont {Rathee}, \citenamefont
  {Zervoudakis}, \citenamefont {Sidky}, \citenamefont {Sikora},\ and\
  \citenamefont {Whitmer}}]{whitmer2018}%
  \BibitemOpen
  \bibfield  {author} {\bibinfo {author} {\bibfnamefont {V.~S.}\ \bibnamefont
  {Rathee}}, \bibinfo {author} {\bibfnamefont {A.~J.}\ \bibnamefont
  {Zervoudakis}}, \bibinfo {author} {\bibfnamefont {H.}~\bibnamefont {Sidky}},
  \bibinfo {author} {\bibfnamefont {B.~J.}\ \bibnamefont {Sikora}},\ and\
  \bibinfo {author} {\bibfnamefont {J.~K.}\ \bibnamefont {Whitmer}},\
  }\bibfield  {title} {\bibinfo {title} {Weak polyelectrolyte complexation
  driven by associative charging},\ }\href {https://doi.org/10.1063/1.5017941}
  {\bibfield  {journal} {\bibinfo  {journal} {JCP}\ }\textbf {\bibinfo {volume}
  {148}},\ \bibinfo {pages} {114901} (\bibinfo {year}
  {2018}{\natexlab{a}})}\BibitemShut {NoStop}%
\bibitem [{\citenamefont {Rathee}\ \emph
  {et~al.}(2018{\natexlab{b}})\citenamefont {Rathee}, \citenamefont {Sidky},
  \citenamefont {Sikora},\ and\ \citenamefont {Whitmer}}]{whitmer2018macro}%
  \BibitemOpen
  \bibfield  {author} {\bibinfo {author} {\bibfnamefont {V.~S.}\ \bibnamefont
  {Rathee}}, \bibinfo {author} {\bibfnamefont {H.}~\bibnamefont {Sidky}},
  \bibinfo {author} {\bibfnamefont {B.~J.}\ \bibnamefont {Sikora}},\ and\
  \bibinfo {author} {\bibfnamefont {J.~K.}\ \bibnamefont {Whitmer}},\
  }\bibfield  {title} {\bibinfo {title} {Role of associative charging in the
  entropy{\textendash}energy balance of polyelectrolyte complexes},\ }\href
  {https://doi.org/10.1021/jacs.8b08649} {\bibfield  {journal} {\bibinfo
  {journal} {JACS}\ }\textbf {\bibinfo {volume} {140}},\ \bibinfo {pages}
  {15319} (\bibinfo {year} {2018}{\natexlab{b}})}\BibitemShut {NoStop}%
\bibitem [{\citenamefont {Rumyantsev}\ \emph
  {et~al.}(2019{\natexlab{b}})\citenamefont {Rumyantsev}, \citenamefont
  {Gavrilov},\ and\ \citenamefont {Kramarenko}}]{rumyantsev2019macro}%
  \BibitemOpen
  \bibfield  {author} {\bibinfo {author} {\bibfnamefont {A.~M.}\ \bibnamefont
  {Rumyantsev}}, \bibinfo {author} {\bibfnamefont {A.~A.}\ \bibnamefont
  {Gavrilov}},\ and\ \bibinfo {author} {\bibfnamefont {E.~Y.}\ \bibnamefont
  {Kramarenko}},\ }\bibfield  {title} {\bibinfo {title} {Electrostatically
  stabilized microphase separation in blends of oppositely charged
  polyelectrolytes},\ }\href {https://doi.org/10.1021/acs.macromol.9b00883}
  {\bibfield  {journal} {\bibinfo  {journal} {Macromolecules}\ }\textbf
  {\bibinfo {volume} {52}},\ \bibinfo {pages} {7167} (\bibinfo {year}
  {2019}{\natexlab{b}})}\BibitemShut {NoStop}%
\bibitem [{\citenamefont {Shakya}\ \emph {et~al.}(2020)\citenamefont {Shakya},
  \citenamefont {Girard}, \citenamefont {King},\ and\ \citenamefont {de~la
  Cruz}}]{shakya2020}%
  \BibitemOpen
  \bibfield  {author} {\bibinfo {author} {\bibfnamefont {A.}~\bibnamefont
  {Shakya}}, \bibinfo {author} {\bibfnamefont {M.}~\bibnamefont {Girard}},
  \bibinfo {author} {\bibfnamefont {J.~T.}\ \bibnamefont {King}},\ and\
  \bibinfo {author} {\bibfnamefont {M.~O.}\ \bibnamefont {de~la Cruz}},\
  }\bibfield  {title} {\bibinfo {title} {Role of chain flexibility in
  asymmetric polyelectrolyte complexation in salt solutions},\ }\href
  {https://doi.org/10.1021/acs.macromol.9b02355} {\bibfield  {journal}
  {\bibinfo  {journal} {Macromolecules}\ }\textbf {\bibinfo {volume} {53}},\
  \bibinfo {pages} {1258} (\bibinfo {year} {2020})}\BibitemShut {NoStop}%
\bibitem [{\citenamefont {Bobbili}\ and\ \citenamefont
  {Milner}(2022)}]{bobbili2022}%
  \BibitemOpen
  \bibfield  {author} {\bibinfo {author} {\bibfnamefont {S.~V.}\ \bibnamefont
  {Bobbili}}\ and\ \bibinfo {author} {\bibfnamefont {S.~T.}\ \bibnamefont
  {Milner}},\ }\bibfield  {title} {\bibinfo {title} {Closed-loop phase behavior
  of nonstoichiometric coacervates in coarse-grained simulations},\ }\href
  {https://doi.org/10.1021/acs.macromol.1c02115} {\bibfield  {journal}
  {\bibinfo  {journal} {Macromolecules}\ }\textbf {\bibinfo {volume} {55}},\
  \bibinfo {pages} {511} (\bibinfo {year} {2022})}\BibitemShut {NoStop}%
\bibitem [{\citenamefont {Chen}\ \emph {et~al.}(2022)\citenamefont {Chen},
  \citenamefont {Zhang},\ and\ \citenamefont {Wang}}]{chen2022}%
  \BibitemOpen
  \bibfield  {author} {\bibinfo {author} {\bibfnamefont {S.}~\bibnamefont
  {Chen}}, \bibinfo {author} {\bibfnamefont {P.}~\bibnamefont {Zhang}},\ and\
  \bibinfo {author} {\bibfnamefont {Z.-G.}\ \bibnamefont {Wang}},\ }\bibfield
  {title} {\bibinfo {title} {Complexation between oppositely charged
  polyelectrolytes in dilute solution: Effects of charge asymmetry},\ }\href
  {https://doi.org/10.1021/acs.macromol.2c00339} {\bibfield  {journal}
  {\bibinfo  {journal} {Macromolecules}\ }\textbf {\bibinfo {volume} {55}},\
  \bibinfo {pages} {3898} (\bibinfo {year} {2022})}\BibitemShut {NoStop}%
\bibitem [{\citenamefont {Michaels}(1965)}]{michaels1965}%
  \BibitemOpen
  \bibfield  {author} {\bibinfo {author} {\bibfnamefont {A.~S.}\ \bibnamefont
  {Michaels}},\ }\bibfield  {title} {\bibinfo {title} {{POLYELECTROLYTE}
  {COMPLEXES}},\ }\href {https://doi.org/10.1021/ie50670a007} {\bibfield
  {journal} {\bibinfo  {journal} {Industrial {\&} Engineering Chemistry}\
  }\textbf {\bibinfo {volume} {57}},\ \bibinfo {pages} {32} (\bibinfo {year}
  {1965})}\BibitemShut {NoStop}%
\bibitem [{\citenamefont {Tainaka}(1980)}]{tianaka1980}%
  \BibitemOpen
  \bibfield  {author} {\bibinfo {author} {\bibfnamefont {K.-I.}\ \bibnamefont
  {Tainaka}},\ }\bibfield  {title} {\bibinfo {title} {Effect of counterions on
  complex coacervation},\ }\href
  {https://doi.org/https://doi.org/10.1002/bip.1980.360190705} {\bibfield
  {journal} {\bibinfo  {journal} {Biopolymers}\ }\textbf {\bibinfo {volume}
  {19}},\ \bibinfo {pages} {1289} (\bibinfo {year} {1980})},\ \Eprint
  {https://arxiv.org/abs/https://onlinelibrary.wiley.com/doi/pdf/10.1002/bip.1980.360190705}
  {https://onlinelibrary.wiley.com/doi/pdf/10.1002/bip.1980.360190705}
  \BibitemShut {NoStop}%
\bibitem [{\citenamefont {M{\'{a}}rquez-Beltr{\'{a}}n}\ \emph
  {et~al.}(2012)\citenamefont {M{\'{a}}rquez-Beltr{\'{a}}n}, \citenamefont
  {Casta{\~{n}}eda}, \citenamefont {Enciso-Aguilar}, \citenamefont
  {Paredes-Quijada}, \citenamefont {Acu{\~{n}}a-Campa}, \citenamefont
  {Maldonado-Arce},\ and\ \citenamefont {Argillier}}]{beltran2012}%
  \BibitemOpen
  \bibfield  {author} {\bibinfo {author} {\bibfnamefont {C.}~\bibnamefont
  {M{\'{a}}rquez-Beltr{\'{a}}n}}, \bibinfo {author} {\bibfnamefont
  {L.}~\bibnamefont {Casta{\~{n}}eda}}, \bibinfo {author} {\bibfnamefont
  {M.}~\bibnamefont {Enciso-Aguilar}}, \bibinfo {author} {\bibfnamefont
  {G.}~\bibnamefont {Paredes-Quijada}}, \bibinfo {author} {\bibfnamefont
  {H.}~\bibnamefont {Acu{\~{n}}a-Campa}}, \bibinfo {author} {\bibfnamefont
  {A.}~\bibnamefont {Maldonado-Arce}},\ and\ \bibinfo {author} {\bibfnamefont
  {J.-F.}\ \bibnamefont {Argillier}},\ }\bibfield  {title} {\bibinfo {title}
  {Structure and mechanism formation of polyelectrolyte complex obtained from
  {PSS}/{PAH} system: effect of molar mixing ratio, base{\textendash}acid
  conditions, and ionic strength},\ }\href
  {https://doi.org/10.1007/s00396-012-2775-y} {\bibfield  {journal} {\bibinfo
  {journal} {Colloid and Polymer Science}\ }\textbf {\bibinfo {volume} {291}},\
  \bibinfo {pages} {683} (\bibinfo {year} {2012})}\BibitemShut {NoStop}%
\bibitem [{\citenamefont {Muthukumar}(2017)}]{muthu2017}%
  \BibitemOpen
  \bibfield  {author} {\bibinfo {author} {\bibfnamefont {M.}~\bibnamefont
  {Muthukumar}},\ }\bibfield  {title} {\bibinfo {title} {50th anniversary
  perspective: A perspective on polyelectrolyte solutions},\ }\href
  {https://doi.org/10.1021/acs.macromol.7b01929} {\bibfield  {journal}
  {\bibinfo  {journal} {Macromolecules}\ }\textbf {\bibinfo {volume} {50}},\
  \bibinfo {pages} {9528} (\bibinfo {year} {2017})},\ \bibinfo {note} {pMID:
  29296029},\ \Eprint
  {https://arxiv.org/abs/https://doi.org/10.1021/acs.macromol.7b01929}
  {https://doi.org/10.1021/acs.macromol.7b01929} \BibitemShut {NoStop}%
\bibitem [{\citenamefont {Muthukumar}(2023)}]{Muthukumar2023-book}%
  \BibitemOpen
  \bibfield  {author} {\bibinfo {author} {\bibfnamefont {M.}~\bibnamefont
  {Muthukumar}},\ }\href {https://doi.org/10.1017/9781139046749} {\emph
  {\bibinfo {title} {Physics of Charged Macromolecules: Synthetic and
  Biological Systems}}}\ (\bibinfo  {publisher} {Cambridge University Press},\
  \bibinfo {year} {2023})\BibitemShut {NoStop}%
\bibitem [{\citenamefont {Edwards}\ and\ \citenamefont
  {Singh}(1979)}]{edwards1979}%
  \BibitemOpen
  \bibfield  {author} {\bibinfo {author} {\bibfnamefont {S.~F.}\ \bibnamefont
  {Edwards}}\ and\ \bibinfo {author} {\bibfnamefont {P.}~\bibnamefont
  {Singh}},\ }\bibfield  {title} {\bibinfo {title} {Size of a polymer molecule
  in solution. part 1. {E}xcluded {v}olume problem},\ }\href@noop {} {\bibfield
   {journal} {\bibinfo  {journal} {Journal of the Chemical Society, Faraday
  Transactions 2: Molecular and Chemical Physics}\ }\textbf {\bibinfo {volume}
  {75}},\ \bibinfo {pages} {1001} (\bibinfo {year} {1979})}\BibitemShut
  {NoStop}%
\bibitem [{\citenamefont {Flory}\ and\ \citenamefont
  {Krigbaum}(1950)}]{flory1950}%
  \BibitemOpen
  \bibfield  {author} {\bibinfo {author} {\bibfnamefont {P.~J.}\ \bibnamefont
  {Flory}}\ and\ \bibinfo {author} {\bibfnamefont {W.~R.}\ \bibnamefont
  {Krigbaum}},\ }\bibfield  {title} {\bibinfo {title} {Statistical mechanics of
  dilute polymer solutions. {II}},\ }\href {https://doi.org/10.1063/1.1747866}
  {\bibfield  {journal} {\bibinfo  {journal} {JCP}\ }\textbf {\bibinfo {volume}
  {18}},\ \bibinfo {pages} {1086} (\bibinfo {year} {1950})}\BibitemShut
  {NoStop}%
\bibitem [{\citenamefont {Podgornik}(1993)}]{podgornik1993}%
  \BibitemOpen
  \bibfield  {author} {\bibinfo {author} {\bibfnamefont {R.}~\bibnamefont
  {Podgornik}},\ }\bibfield  {title} {\bibinfo {title} {A variational approach
  to charged polymer chains: Polymer mediated interactions},\ }\href
  {https://doi.org/10.1063/1.465439} {\bibfield  {journal} {\bibinfo  {journal}
  {JCP}\ }\textbf {\bibinfo {volume} {99}},\ \bibinfo {pages} {7221} (\bibinfo
  {year} {1993})}\BibitemShut {NoStop}%
\bibitem [{\citenamefont {Muthukumar}(2012)}]{Muthukumar2012}%
  \BibitemOpen
  \bibfield  {author} {\bibinfo {author} {\bibfnamefont {M.}~\bibnamefont
  {Muthukumar}},\ }\bibfield  {title} {\bibinfo {title} {Counterion adsorption
  theory of dilute polyelectrolyte solutions: Apparent molecular weight, second
  virial coefficient, and intermolecular structure factor},\ }\href
  {https://doi.org/10.1063/1.4736545} {\bibfield  {journal} {\bibinfo
  {journal} {The Journal of Chemical Physics}\ }\textbf {\bibinfo {volume}
  {137}},\ \bibinfo {pages} {034902} (\bibinfo {year} {2012})}\BibitemShut
  {NoStop}%
\bibitem [{\citenamefont {Laugel}\ \emph {et~al.}(2006)\citenamefont {Laugel},
  \citenamefont {Betscha}, \citenamefont {Winterhalter}, \citenamefont
  {Voegel}, \citenamefont {Schaaf},\ and\ \citenamefont {Ball}}]{Laugel2006}%
  \BibitemOpen
  \bibfield  {author} {\bibinfo {author} {\bibfnamefont {N.}~\bibnamefont
  {Laugel}}, \bibinfo {author} {\bibfnamefont {C.}~\bibnamefont {Betscha}},
  \bibinfo {author} {\bibfnamefont {M.}~\bibnamefont {Winterhalter}}, \bibinfo
  {author} {\bibfnamefont {J.-C.}\ \bibnamefont {Voegel}}, \bibinfo {author}
  {\bibfnamefont {P.}~\bibnamefont {Schaaf}},\ and\ \bibinfo {author}
  {\bibfnamefont {V.}~\bibnamefont {Ball}},\ }\bibfield  {title} {\bibinfo
  {title} {Relationship between the growth regime of polyelectrolyte
  multilayers and the polyanion/polycation complexation enthalpy},\ }\href
  {https://doi.org/10.1021/jp062264z} {\bibfield  {journal} {\bibinfo
  {journal} {The Journal of Physical Chemistry B}\ }\textbf {\bibinfo {volume}
  {110}},\ \bibinfo {pages} {19443–19449} (\bibinfo {year}
  {2006})}\BibitemShut {NoStop}%
\bibitem [{\citenamefont {Mehler}\ and\ \citenamefont
  {Eichele}(1984)}]{Mehler1984}%
  \BibitemOpen
  \bibfield  {author} {\bibinfo {author} {\bibfnamefont {E.~L.}\ \bibnamefont
  {Mehler}}\ and\ \bibinfo {author} {\bibfnamefont {G.}~\bibnamefont
  {Eichele}},\ }\bibfield  {title} {\bibinfo {title} {Electrostatic effects in
  water-accessible regions of proteins},\ }\href
  {https://doi.org/10.1021/bi00312a015} {\bibfield  {journal} {\bibinfo
  {journal} {Biochemistry}\ }\textbf {\bibinfo {volume} {23}},\ \bibinfo
  {pages} {3887} (\bibinfo {year} {1984})}\BibitemShut {NoStop}%
\bibitem [{\citenamefont {Lamm}\ and\ \citenamefont {Pack}(1997)}]{Lamm1997}%
  \BibitemOpen
  \bibfield  {author} {\bibinfo {author} {\bibfnamefont {G.}~\bibnamefont
  {Lamm}}\ and\ \bibinfo {author} {\bibfnamefont {G.~R.}\ \bibnamefont
  {Pack}},\ }\bibfield  {title} {\bibinfo {title} {Calculation of dielectric
  constants near polyelectrolytes in solution},\ }\href
  {https://doi.org/10.1021/jp9623453} {\bibfield  {journal} {\bibinfo
  {journal} {The Journal of Physical Chemistry B}\ }\textbf {\bibinfo {volume}
  {101}},\ \bibinfo {pages} {959} (\bibinfo {year} {1997})}\BibitemShut
  {NoStop}%
\bibitem [{\citenamefont {Rouzina}\ and\ \citenamefont
  {Bloomfield}(1998)}]{Rouzina1998}%
  \BibitemOpen
  \bibfield  {author} {\bibinfo {author} {\bibfnamefont {I.}~\bibnamefont
  {Rouzina}}\ and\ \bibinfo {author} {\bibfnamefont {V.~A.}\ \bibnamefont
  {Bloomfield}},\ }\bibfield  {title} {\bibinfo {title} {{DNA} bending by
  small, mobile multivalent cations},\ }\href
  {https://doi.org/10.1016/s0006-3495(98)78021-x} {\bibfield  {journal}
  {\bibinfo  {journal} {Biophysical Journal}\ }\textbf {\bibinfo {volume}
  {74}},\ \bibinfo {pages} {3152} (\bibinfo {year} {1998})}\BibitemShut
  {NoStop}%
\bibitem [{\citenamefont {Muthukumar}(2004)}]{muthu2004}%
  \BibitemOpen
  \bibfield  {author} {\bibinfo {author} {\bibfnamefont {M.}~\bibnamefont
  {Muthukumar}},\ }\bibfield  {title} {\bibinfo {title} {Theory of counter-ion
  condensation on flexible polyelectrolytes: Adsorption mechanism},\ }\href
  {https://doi.org/10.1063/1.1701839} {\bibfield  {journal} {\bibinfo
  {journal} {JCP}\ }\textbf {\bibinfo {volume} {120}},\ \bibinfo {pages} {9343}
  (\bibinfo {year} {2004})},\ \Eprint
  {https://arxiv.org/abs/https://doi.org/10.1063/1.1701839}
  {https://doi.org/10.1063/1.1701839} \BibitemShut {NoStop}%
\bibitem [{\citenamefont {Ghosh}\ and\ \citenamefont
  {Kundagrami}(2023)}]{Ghosh20231}%
  \BibitemOpen
  \bibfield  {author} {\bibinfo {author} {\bibfnamefont {S.}~\bibnamefont
  {Ghosh}}\ and\ \bibinfo {author} {\bibfnamefont {A.}~\bibnamefont
  {Kundagrami}},\ }\bibfield  {title} {\bibinfo {title} {Effect of counterion
  size on polyelectrolyte conformations and thermodynamics}\ }\href
  {https://doi.org/10.48550/ARXIV.2310.11250} {10.48550/ARXIV.2310.11250}
  (\bibinfo {year} {2023})\BibitemShut {NoStop}%
\bibitem [{\citenamefont {Dinpajooh}\ and\ \citenamefont
  {Matyushov}(2016)}]{Dinpajooh2016}%
  \BibitemOpen
  \bibfield  {author} {\bibinfo {author} {\bibfnamefont {M.}~\bibnamefont
  {Dinpajooh}}\ and\ \bibinfo {author} {\bibfnamefont {D.~V.}\ \bibnamefont
  {Matyushov}},\ }\bibfield  {title} {\bibinfo {title} {Dielectric constant of
  water in the interface},\ }\href {https://doi.org/10.1063/1.4955203}
  {\bibfield  {journal} {\bibinfo  {journal} {The Journal of Chemical Physics}\
  }\textbf {\bibinfo {volume} {145}},\ \bibinfo {pages} {014504} (\bibinfo
  {year} {2016})}\BibitemShut {NoStop}%
\bibitem [{\citenamefont {Roux}\ \emph {et~al.}(1990)\citenamefont {Roux},
  \citenamefont {Yu},\ and\ \citenamefont {Karplus}}]{Roux1990}%
  \BibitemOpen
  \bibfield  {author} {\bibinfo {author} {\bibfnamefont {B.}~\bibnamefont
  {Roux}}, \bibinfo {author} {\bibfnamefont {H.~A.}\ \bibnamefont {Yu}},\ and\
  \bibinfo {author} {\bibfnamefont {M.}~\bibnamefont {Karplus}},\ }\bibfield
  {title} {\bibinfo {title} {Molecular basis for the born model of ion
  solvation},\ }\href {https://doi.org/10.1021/j100374a057} {\bibfield
  {journal} {\bibinfo  {journal} {The Journal of Physical Chemistry}\ }\textbf
  {\bibinfo {volume} {94}},\ \bibinfo {pages} {4683} (\bibinfo {year}
  {1990})}\BibitemShut {NoStop}%
\bibitem [{\citenamefont {Rick}\ and\ \citenamefont {Berne}(1994)}]{Rick1994}%
  \BibitemOpen
  \bibfield  {author} {\bibinfo {author} {\bibfnamefont {S.~W.}\ \bibnamefont
  {Rick}}\ and\ \bibinfo {author} {\bibfnamefont {B.~J.}\ \bibnamefont
  {Berne}},\ }\bibfield  {title} {\bibinfo {title} {The aqueous solvation of
  water: A comparison of continuum methods with molecular dynamics},\ }\href
  {https://doi.org/10.1021/ja00088a034} {\bibfield  {journal} {\bibinfo
  {journal} {Journal of the American Chemical Society}\ }\textbf {\bibinfo
  {volume} {116}},\ \bibinfo {pages} {3949} (\bibinfo {year}
  {1994})}\BibitemShut {NoStop}%
\bibitem [{\citenamefont {Lynden-Bell}\ and\ \citenamefont
  {Rasaiah}(1997)}]{LyndenBell1997}%
  \BibitemOpen
  \bibfield  {author} {\bibinfo {author} {\bibfnamefont {R.~M.}\ \bibnamefont
  {Lynden-Bell}}\ and\ \bibinfo {author} {\bibfnamefont {J.~C.}\ \bibnamefont
  {Rasaiah}},\ }\bibfield  {title} {\bibinfo {title} {From hydrophobic to
  hydrophilic behaviour: A simulation study of solvation entropy and free
  energy of simple solutes},\ }\href {https://doi.org/10.1063/1.474550}
  {\bibfield  {journal} {\bibinfo  {journal} {The Journal of Chemical Physics}\
  }\textbf {\bibinfo {volume} {107}},\ \bibinfo {pages} {1981} (\bibinfo {year}
  {1997})}\BibitemShut {NoStop}%
\bibitem [{\citenamefont {Rajamani}\ \emph {et~al.}(2004)\citenamefont
  {Rajamani}, \citenamefont {Ghosh},\ and\ \citenamefont
  {Garde}}]{Rajamani2004}%
  \BibitemOpen
  \bibfield  {author} {\bibinfo {author} {\bibfnamefont {S.}~\bibnamefont
  {Rajamani}}, \bibinfo {author} {\bibfnamefont {T.}~\bibnamefont {Ghosh}},\
  and\ \bibinfo {author} {\bibfnamefont {S.}~\bibnamefont {Garde}},\ }\bibfield
   {title} {\bibinfo {title} {Size dependent ion hydration, its asymmetry, and
  convergence to macroscopic behavior},\ }\href
  {https://doi.org/10.1063/1.1644536} {\bibfield  {journal} {\bibinfo
  {journal} {The Journal of Chemical Physics}\ }\textbf {\bibinfo {volume}
  {120}},\ \bibinfo {pages} {4457} (\bibinfo {year} {2004})}\BibitemShut
  {NoStop}%
\bibitem [{\citenamefont {Kumar}\ \emph {et~al.}(2012)\citenamefont {Kumar},
  \citenamefont {Sumpter},\ and\ \citenamefont {Kilbey}}]{Kumar2012}%
  \BibitemOpen
  \bibfield  {author} {\bibinfo {author} {\bibfnamefont {R.}~\bibnamefont
  {Kumar}}, \bibinfo {author} {\bibfnamefont {B.~G.}\ \bibnamefont {Sumpter}},\
  and\ \bibinfo {author} {\bibfnamefont {S.~M.}\ \bibnamefont {Kilbey}},\
  }\bibfield  {title} {\bibinfo {title} {Charge regulation and local dielectric
  function in planar polyelectrolyte brushes},\ }\href
  {https://doi.org/10.1063/1.4729158} {\bibfield  {journal} {\bibinfo
  {journal} {The Journal of Chemical Physics}\ }\textbf {\bibinfo {volume}
  {136}},\ \bibinfo {pages} {234901} (\bibinfo {year} {2012})}\BibitemShut
  {NoStop}%
\bibitem [{\citenamefont {Grant}\ \emph {et~al.}(2001)\citenamefont {Grant},
  \citenamefont {Pickup},\ and\ \citenamefont {Nicholls}}]{Grant2001}%
  \BibitemOpen
  \bibfield  {author} {\bibinfo {author} {\bibfnamefont {J.~A.}\ \bibnamefont
  {Grant}}, \bibinfo {author} {\bibfnamefont {B.~T.}\ \bibnamefont {Pickup}},\
  and\ \bibinfo {author} {\bibfnamefont {A.}~\bibnamefont {Nicholls}},\
  }\bibfield  {title} {\bibinfo {title} {A smooth permittivity function for
  poisson-boltzmann solvation methods},\ }\href
  {https://doi.org/10.1002/jcc.1032} {\bibfield  {journal} {\bibinfo  {journal}
  {Journal of Computational Chemistry}\ }\textbf {\bibinfo {volume} {22}},\
  \bibinfo {pages} {608} (\bibinfo {year} {2001})}\BibitemShut {NoStop}%
\bibitem [{\citenamefont {Muthukumar}(1987)}]{muthu1987}%
  \BibitemOpen
  \bibfield  {author} {\bibinfo {author} {\bibfnamefont {M.}~\bibnamefont
  {Muthukumar}},\ }\bibfield  {title} {\bibinfo {title} {Adsorption of a
  polyelectrolyte chain to a charged surface},\ }\href
  {https://doi.org/10.1063/1.452763} {\bibfield  {journal} {\bibinfo  {journal}
  {JCP}\ }\textbf {\bibinfo {volume} {86}},\ \bibinfo {pages} {7230} (\bibinfo
  {year} {1987})}\BibitemShut {NoStop}%
\bibitem [{\citenamefont {Tanford}(1961)}]{tanford1961physical}%
  \BibitemOpen
  \bibfield  {author} {\bibinfo {author} {\bibfnamefont {C.}~\bibnamefont
  {Tanford}},\ }\href {https://books.google.co.in/books?id=gjlRAAAAMAAJ} {\emph
  {\bibinfo {title} {Physical Chemistry of Macromolecules}}},\ \bibinfo
  {series} {Physical Chemistry of Macromolecules}\ No.\ \bibinfo {number} {v.
  164, p. 1961}\ (\bibinfo  {publisher} {Wiley},\ \bibinfo {year}
  {1961})\BibitemShut {NoStop}%
\bibitem [{\citenamefont {Kundu}\ and\ \citenamefont {Dua}(2014)}]{kundu2014}%
  \BibitemOpen
  \bibfield  {author} {\bibinfo {author} {\bibfnamefont {P.}~\bibnamefont
  {Kundu}}\ and\ \bibinfo {author} {\bibfnamefont {A.}~\bibnamefont {Dua}},\
  }\bibfield  {title} {\bibinfo {title} {Weak polyelectrolytes in the presence
  of counterion condensation with ions of variable size and polarizability},\
  }\href {https://doi.org/10.1088/1742-5468/2014/07/p07023} {\bibfield
  {journal} {\bibinfo  {journal} {Journal of Statistical Mechanics: Theory and
  Experiment}\ }\textbf {\bibinfo {volume} {2014}},\ \bibinfo {pages} {P07023}
  (\bibinfo {year} {2014})}\BibitemShut {NoStop}%
\end{thebibliography}%

		\end{document}